\documentclass[a4paper,fleqn]{cas-sc}

\usepackage[numbers,sort&compress]{natbib} % compact citations
\usepackage{amsmath}
\usepackage{amssymb}
\usepackage{bm}
\usepackage{graphicx}
\usepackage{tabularx}
\usepackage{booktabs}
\usepackage{siunitx} %temperature degree
\usepackage{color}% color

\def\tsc#1{\csdef{#1}{\textsc{\lowercase{#1}}\xspace}}
\tsc{WGM}
\tsc{QE}
\begin{document}
\let\WriteBookmarks\relax
\def\floatpagepagefraction{1}
\def\textpagefraction{.001}

\shorttitle{}

\shortauthors{}

%\title[mode=title]{A Geometry-Induced Finite Warburg element for Microdisk Electrodes Consistent with the Nernstian Limit Implied by Saito's Steady-State Current}
%\title[mode=title]{A Geometry-Induced Finite Warburg element for Microdisk Electrodes Consistent with the Nernstian Saito Limit}
%\title[mode=title]{A Geometry-Induced Finite Warburg element for Microdisk Electrodes Consistent with the Reversible Saito Limit}
\title[mode=title]{A Geometry-Induced Finite Warburg element for Reversible Diffusion Impedance at Microdisk Electrodes}
%\title[mode=title]{A Geometry-Induced Finite Warburg element for Microdisk Electrodes Consistent with Saito's Reversible Limit}



\author[1]{Kazuhiko Seki}%[<options>]

\cormark[1]


\ead{k-seki@aist.go.jp}


\credit{Conceptualization, Investigation, Writing -- original draft, Validation, Writing -- review \& editing}

\affiliation[1]{organization={Global Zero Emission Research Center (GZR)
National Institute of Advanced Industrial Science and Technology (AIST)},
            addressline={305-8569 Onogawa 16-1 AIST West},
            city={Tsukuba},
%          citysep={}, % Uncomment if no comma needed between city and postcode
           postcode={305-8569},
           state={Ibaraki},
           country={Japan}}

\author[2]{Yuko Yokoyama}%[]

% Footnote of the second author

% Email id of the second author
%\ead{}


\credit{Conceptualization, Validation, Writing -- review \& editing}

\affiliation[2]{organization={Department of Chemical Science and Engineering, Graduate School of Engineering, Kyoto University},
            addressline={Kyotodaigaku-Katsura, Nishikyo-ku},
           city={Kyoto},
%          citysep={}, % Uncomment if no comma needed between city and postcode
           postcode={615-8510},
            state={Kyoto},
            country={Japan}}

%% Corresponding author text
%%\cortext[1]{Corresponding author}

\author[3]{Masahiro Yamamoto}%[]
\affiliation[3]{organization={Department of Environmental and Energy Engineering,
Konan University},
            addressline={Okamoto 8-9-1},
            city={Kobe},
%          citysep={}, % Uncomment if no comma needed between city and postcode
            postcode={658-8501},
            state={Hyogo},
            country={Japan}}

%% Credit authorship
\credit{Conceptualization, Validation, Writing -- review \& editing}



\begin{abstract}
A compact analytical representation is developed for the diffusion impedance of a disk electrode under reversible conditions, obtained by transforming the chronoamperometric relaxation function into the frequency domain. A Pad\'{e} approximant is constructed for the uniform-concentration boundary condition, preserving the finite low-frequency diffusion resistance of the steady-state disk current and the high-frequency Cottrell regime. It is extended to reactions in which both oxidized and reduced species contribute, with unequal diffusion coefficients and surface concentrations. Its accuracy is quantified against converged numerical solutions, including an exact solution of the coupled problem with finite-rate kinetics. The approximant reproduces the Saito resistance and the $45^{\circ}$ Warburg phase exactly; its modulus is within 0.6\% of the converged solution for $\omega a^2/D\le12$ and 7\% below the Cottrell amplitude at high frequency. The uniform-concentration and uniform-flux boundary conditions both give a diffusion impedance of geometric origin, but differ by up to 11\% in modulus because each couples differently to the spreading diffusion field; a criterion based on the rate constant and the disk radius delineates where each applies. The exact solution shows that this crossover is broad and that the decomposition into a charge-transfer resistance and a diffusion impedance is accurate to about 5\% for $1\le2k^0a/D\le100$. Fitted to a published spectrum of a platinum microdisk, it reproduces the reported diffusion coefficient within 1\%, while the two limiting elements lie 4.8\% below and 4.1\% above it. The element is a practical tool for microdisk impedance analysis.
\end{abstract}
%%%\includegraphics[width=13.28cm, height=5.31cm]{Saitoeq_graphabst-crop}
%\includegraphics[width=13.28cm]{Saitoeq_graphabst_v3-crop}

\begin{highlights}
\item A geometry-induced finite Warburg element is derived for microdisk electrodes.
\item Derived from an exact integral equation, it recovers the Saito and Warburg limits.
\item The Pad\'{e} impedance is within 0.6\% of the converged response up to $\omega a^2/D=12$.
\item Uniform flux overestimates the diffusion impedance by $+11$\% near $\omega a^2/D\approx2$.
\item An exact coupled solution shows the Randles decomposition is accurate to about 5\%.
\end{highlights}

% Keywords

\maketitle
%%%%%%%%%%%%%%%%%%%%%%%%%%%%%%%%%%%%%%%%%%%%%%%%%%%%%%%%%%%%%%%%%%%%%%
%\setcounter{equation}{0}
\section{Introduction}
%%%%%%%%%%%%%%%%%%%%%%%%%%%%%%%%%%%%%%%%%%%%%%%%%%%%%%%%%%%%%%%%%%%%%%
\label{sec:Intro}

\begin{figure*}[t]
\centering
\includegraphics[width=0.85\textwidth]{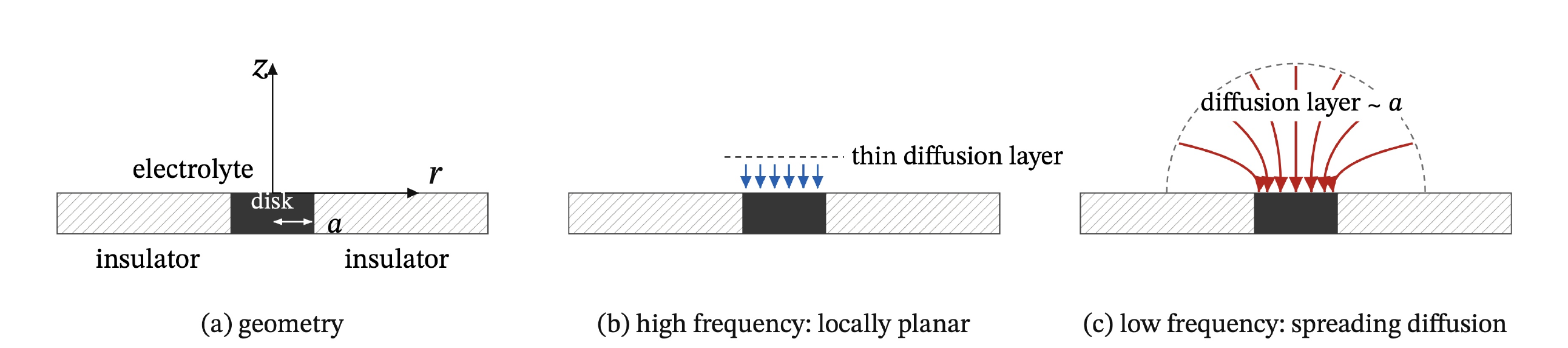}
\caption{
Schematic geometry and frequency-dependent diffusion regimes at a microdisk electrode.
(a) Cross-sectional geometry: a disk electrode of radius \(a\) is inlaid in an infinite insulating plane, with cylindrical coordinates \((r,z)\) and the electrode surface at \(z=0\).
(b) At high frequency, the diffusion layer is thin compared with the disk radius \(a\), and diffusion proceeds locally planar over the disk face, as in the Cottrell limit.
(c) At low frequency, the diffusion layer grows up to a size comparable to \(a\), and spreading diffusion from the disk edge into the bulk solution becomes significant, giving rise to the finite low-frequency diffusion resistance discussed throughout this work.
}
\label{fig:schematic}
\end{figure*}

Electrochemical impedance spectroscopy (EIS) is a powerful frequency-domain technique for separating charge-transfer, double-layer charging, and mass-transport processes at electrode interfaces.\cite{Lasia_14}
For diffusion-controlled reactions, the classical Warburg impedance, corresponding to the Cottrell equation in the time domain, provides the standard description of semi-infinite linear diffusion.\cite{Warburg}
However, this one-dimensional approximation is not generally applicable to microdisk electrodes, where the diffusion field changes from locally planar diffusion at high frequencies to radial, or spreading, diffusion at low frequencies.
This geometrical crossover leads to a finite steady-state current and, consequently, a finite low-frequency diffusion resistance.\cite{Jacobsen_95,Lasia_14}
This crossover between locally planar and spreading diffusion, illustrated schematically in Fig.~\ref{fig:schematic}, underlies both the finite low-frequency diffusion resistance and the boundary-condition dependence analyzed below.

Microdisk electrodes are particularly important because their small dimensions enhance mass transport, reduce ohmic losses, and enable steady-state measurements.
The steady-state diffusion-limited current at an inlaid disk electrode is given by Saito's equation,\cite{SAITO_68}
\begin{align}
I_\infty = 4DanFC_0,
\label{eq:SaitoIntro}
\end{align}
where \(a\) is the disk radius, \(D\) is the diffusion coefficient, \(C_0\) is the bulk concentration of the electroactive species, \(F\) is the Faraday constant, and \(n\) is the number of electrons transferred.
This result provides an exact benchmark for theoretical descriptions of diffusion at a disk electrode.

The impedance of microelectrodes has been studied extensively since the pioneering work of Fleischmann, Pons, and co-workers, who derived analytical expressions for spherical, cylindrical, disk, and ring microelectrodes.\cite{FLEISCHMANN_88,FLEISCHMANN_91,Gabrielli_06,VIVIER_22}
For disk electrodes, the expressions involving integral representations containing Bessel functions have been derived under a uniform-flux, or Neumann, boundary condition \cite{FLEISCHMANN_88,FLEISCHMANN_91,Gabrielli_06,VIVIER_22};
the analytical expressions have been widely reproduced in subsequent studies and reviews.\cite{MICHEL_10,VIVIER_22}

Using a finite element method with anisotropic mesh adaptation,
Michel et al. demonstrated that the classical Fleischmann--Pons expression has a limited range of validity in the reaction controlled limit,
which can be regarded as validating approximation of Dirichlet--Neumann boundary condition by a uniform-flux, or Neumann, boundary condition.\cite{MICHEL_10}
In this limit, the low-frequency Faradaic admittance predicted by the Fleischmann--Pons approximation is
\begin{align}
Y_{\rm FP}
=
\frac{3\pi^2}{16}
\frac{n^2F^2DC_ba}{RT},
\label{eq:AdFr}
\end{align}
where \(C_b\) denotes the bulk concentration of each redox species, and \(R\) and \(T\) are the gas constant and absolute temperature, respectively.

By contrast, the reversible limit corresponds to a Nernstian, or Dirichlet (uniform concentration), boundary condition.
For a symmetric reversible redox couple, in which the oxidized and reduced species (denoted by the subscripts O and R, respectively) have diffusion coefficients \(D_{\rm O}\) and \(D_{\rm R}\) and bulk concentrations \(C_{\rm O}^{*}\) and \(C_{\rm R}^{*}\), with
\(D_{\rm O}=D_{\rm R}=D\)
and
\(C_{\rm O}^{*}=C_{\rm R}^{*}=C_b\),
the low-frequency Faradaic admittance implied by Saito's steady-state current and the linearized Nernst relation is
\begin{align}
Y_{\infty,{\rm rev}}
=
\frac{2n^2F^2DC_ba}{RT}.
\label{eq:AdSaito}
\end{align}
The factor of two arises because both oxidized and reduced species contribute to the concentration-polarization response and are coupled through the Nernst boundary condition.
Equivalently, the corresponding dimensionless diffusion resistance, normalized by that of a hemispherical electrode, is \(R_d^*=\pi/2\),\cite{Alden_97}
whereas the uniform flux boundary condition gives \(R_d^*=16/(3\pi)\).\cite{MICHEL_10}
By considering that the admittance is proportional to the diffusion coefficient,
the diffusion coefficient under the uniform flux boundary condition $D_{\rm uf}$ (Fleischmann--Pons limit) against
that under the uniform concentration boundary condition $D_{\rm uc}$ (Saito limit) can be estimated as
\begin{align}
D_{\rm uf}/D_{\rm uc}=16/(3\pi)/(\pi/2)\approx1.081 .
\label{eq:Dratio}
\end{align}
if the steady state value of the admittance is the same.

The discrepancy between Eqs.~\eqref{eq:AdFr} and \eqref{eq:AdSaito} shows that the widely used Fleischmann--Pons microdisk impedance expression does not reproduce the reversible low-frequency limit associated with the Nernstian boundary condition.
An analytical expression that simultaneously reproduces the reversible Saito limit and the correct high-frequency Warburg behavior is therefore needed.
Such an expression would be useful not only for interpreting impedance spectra, but also for constructing physically meaningful equivalent-circuit models and extracting transport parameters from experimental data.

Recently, we developed an exact integral-equation formulation for the transient diffusion-limited current at a disk electrode.\cite{SEKI_26,SEKI_26c}
The mixed boundary-value problem reflecting circular disk geometry was reduced in the Laplace domain to a Fredholm integral equation for an auxiliary function that directly determines the Faradaic current
under the uniform concentration boundary condition.
The steady-state limit recovers Saito's equation, and the long-time relaxation was obtained systematically as an expansion in powers of \(a\sqrt{s/D}\), where \(s\) denotes the Laplace variable.
A Pad\'{e} approximant then yielded a compact analytical expression for chronoamperometric currents that showed excellent agreement with numerical simulations and experimental data.

The same Laplace-domain formulation provides a natural route to impedance spectroscopy.\cite{Jacobsen_95,Kubobook}
For a small harmonic perturbation, the frequency-domain response function is obtained by multiplying the Laplace transform of the relaxation function by \(s\) and performing the analytic continuation \(s=i\omega\),
where $\omega$ indicates the angular frequency and is related to the frequency $f$ (in Hz) by $\omega=2\pi f$.
Consequently, the asymptotic expansion derived for transient chronoamperometry is transformed into the diffusion admittance of a disk electrode.
The resulting response contains the steady-state admittance, the leading Warburg contribution, and higher-order corrections owing to disk geometry.

In the present work, we develop an analytical impedance formulation for reversible reactions at microdisk electrodes.
Starting from the disk-electrode response obtained from the Fredholm integral-equation formulation, we construct a Pad\'{e} representation of the corresponding frequency-domain response function.
The oxidized and reduced species are then combined through the linearized Nernst boundary condition to obtain the reversible diffusion impedance.
The resulting expression reproduces the reversible Saito limit at low frequencies and approaches the planar Warburg behavior at high frequencies, while remaining fully analytical over the frequency range considered.

We further compare the response under uniform-concentration boundary condition with that under the uniform-flux boundary condition associated with the Fleischmann--Pons formulation, clarifying how the imposed boundary condition affects the finite-disk impedance.
The resulting expressions and the exact solution of the coupled problem with finite-rate kinetics are then tested against a previously published experimental impedance spectrum for a Pt microdisk electrode, providing an experimental cross-check of the difference between the two boundary conditions.
Finally, the Pad\'{e} diffusion impedance is incorporated into a Randles-type equivalent circuit to demonstrate its use as a geometry-induced finite Warburg-type element for
fast reversible Nernstian reactions on microdisk electrodes.
Thus, the present formulation provides a physically grounded and compact alternative to empirical diffusion elements for quantitative interpretation of diffusion-controlled impedance at microdisk electrodes.

%%%%%%%%%%%%%%%%%%%%%%%%%%%%%%%%%%%%%%%
Existing analytical expressions for the diffusion impedance of a disk are the uniform-flux integral expression of Fleischmann et al.~\cite{FLEISCHMANN_88,FLEISCHMANN_91}, whose low-frequency resistance exceeds the Saito resistance of a reversible couple by the factor $32/(3\pi^2)$, and the time-domain interpolation of Shoup and Szabo~\cite{SHOUP_82}, which has no known Laplace transform and cannot be continued to $s=i\omega$ directly. Numerical methods, such as finite elements or the numerical solution of integral equations, can treat any boundary condition but give no closed form. The present work provides (i) a closed algebraic impedance with the uniform-concentration boundary condition, (ii) its extension to unequal diffusion coefficients and surface concentrations, which is exact for equal diffusion coefficients, and (iii) quantitative benchmarks against numerically exact solutions of the uniform-concentration problem (Appendix~C.1) and of the coupled problem with finite-rate kinetics (Sec.~\ref{sec:exact} and Appendix~D), which give the accuracy of the Pad\'e element and of the Randles decomposition. The exact solution is also fitted to a published experimental spectrum (Sec.~\ref{sec:Compex}). Table~\ref{tab:models} compares the models.

\begin{table}[h]
\centering
\small
\caption{Comparison of analytical models of the diffusion impedance of a disk. The errors of $|Z|$ are relative to the converged uniform-concentration response (Appendix~C.1); for the Shoup--Szabo formula the deviation in the time domain from the numerically exact current of Ref.~\cite{BIENIASZ_16} is given.}
\label{tab:models}
\begin{tabular}{p{3.7cm}p{2.5cm}p{2.7cm}p{4.6cm}}
\toprule
Model & Boundary condition & Form & Error of $|Z|$\\
\midrule
Fleischmann--Pons & uniform flux & integral with $J_1^2$ & $+8.4$\% ($\omega\to0$), $+11$\% near $\omega a^2/D\approx2$, $+0.8$\% at $10^3$ (Fig.~\ref{fig:uf_uc})\\
Shoup--Szabo & uniform concentration & time domain only & 0.65\% maximum in the time domain (Table~\ref{tab:appB_bieniasz})\\
Pad\'e, Eq.~\eqref{eq:Pade_dimensionless} & uniform concentration & algebraic in $\sqrt{i\omega}$ & $\le0.11$\% for $\omega a^2/D\le4$; 0.6\% at 12; $-5.9$\% at $10^3$\\
$[2/1]$ variant matching Cottrell (Appendix~B.2) & uniform concentration & algebraic & $\le1.1$\% for $10^{-2}\le\omega a^2/D\le10^3$\\
\bottomrule
\end{tabular}
\end{table}

\section{Pad\'{e} representation of the disk-electrode response}
%%%%%%%%%%%%%%%%%%%%%%%%%%%%%%%%%%%%%%%
\label{sec:Pade}

We consider a reversible electrode reaction,
\begin{align}
{\rm O}+ne^- \rightleftharpoons {\rm R},
\end{align}
for which both oxidized and reduced species contribute to the concentration-polarization impedance. The diffusion impedance must therefore be formulated by treating the diffusional responses of the two species separately.

We consider small perturbations around the steady state. The Laplace-domain impedance is defined as the transfer function relating the perturbation of the electrode potential, \(\delta \hat{E}(s)\), to the perturbation of the Faradaic current, \(\delta \hat{I}(s)\):\cite{Jacobsen_95,Lasia_14}
\begin{align}
\hat{Z}(s)
=
\frac{\delta \hat{E}(s)}{\delta \hat{I}(s)} .
\label{eq:Zdef}
\end{align}

The current-continuity condition at the electrode surface requires that the fluxes of O and R are coupled by the electrode reaction. We use the convention that the Faradaic current is positive for the reduction reaction \({\rm O}+ne^- \rightarrow {\rm R}\). With this convention, the perturbation current satisfies
\begin{align}
\delta \hat{I}(s)
=
\delta \hat{I}_{\rm O}(s)
=
-\delta \hat{I}_{\rm R}(s),
\end{align}
where \(\delta \hat{I}_{\rm O}(s)\) and \(\delta \hat{I}_{\rm R}(s)\) denote the current contributions associated with the fluxes of the oxidized and reduced species, respectively.

Let \(C_{\rm O}^{*}\) and \(C_{\rm R}^{*}\) be the steady-state surface concentrations of the oxidized and reduced species. Small deviations from these steady-state values are denoted by
\begin{align}
\delta C_{\rm O}(t)
=
C_{\rm O,s}(t)-C_{\rm O}^{*},
\qquad
\delta C_{\rm R}(t)
=
C_{\rm R,s}(t)-C_{\rm R}^{*},
\end{align}
and their Laplace transforms are denoted by \(\delta \hat{C}_{\rm O}(s)\) and \(\delta \hat{C}_{\rm R}(s)\), respectively.

For a reversible reaction, the perturbation of the electrode potential is obtained by subtracting the steady-state Nernst relation from the instantaneous Nernst relation. In terms of the instantaneous surface-concentration perturbations, this gives
\begin{align}
\delta E(t)
=
-\frac{RT}{nF}
\ln
\left[
\frac{
1+\delta C_{\rm R}(t)/C_{\rm R}^{*}
}{
1+\delta C_{\rm O}(t)/C_{\rm O}^{*}
}
\right].
\label{eq:Nernstianres}
\end{align}
For sufficiently small perturbations, this expression is linearized as
\begin{align}
\delta E(t)
\simeq
\frac{RT}{nF}
\left(
\frac{\delta C_{\rm O}(t)}{C_{\rm O}^{*}}
-
\frac{\delta C_{\rm R}(t)}{C_{\rm R}^{*}}
\right).
\label{eq:ECOR}
\end{align}
Taking the Laplace transform then yields
\begin{align}
\delta \hat{E}(s)
=
\frac{RT}{nF}
\left(
\frac{\delta \hat{C}_{\rm O}(s)}{C_{\rm O}^{*}}
-
\frac{\delta \hat{C}_{\rm R}(s)}{C_{\rm R}^{*}}
\right).
\label{eq:Nernst_linear}
\end{align}
Thus, the logarithmic form indicates the Nernstian origin of the relation, whereas only the linearized expression is required in the impedance derivation.

Within linear response theory, the current response to a small surface-concentration perturbation is written in terms of the diffusion response functions \(\hat{G}_{\rm O}(s)\) and \(\hat{G}_{\rm R}(s)\). For the oxidized species,
\begin{align}
\delta \hat{I}_{\rm O}(s)
=
nF \hat{G}_{\rm O}(s)
\frac{\delta \hat{C}_{\rm O}(s)}{C_{\rm O}^{*}} .
\label{eq:responseO}
\end{align}
Similarly, for the reduced species, taking account of the opposite flux direction,
\begin{align}
\delta \hat{I}_{\rm R}(s)
=
nF \hat{G}_{\rm R}(s)
\frac{\delta \hat{C}_{\rm R}(s)}{C_{\rm R}^{*}},
\qquad
\delta \hat{I}(s)
=
-\delta \hat{I}_{\rm R}(s).
\label{eq:responseR}
\end{align}
It follows that
\begin{align}
\frac{\delta \hat{C}_{\rm O}(s)}{C_{\rm O}^{*}}
=
\frac{\delta \hat{I}(s)}{nF\hat{G}_{\rm O}(s)},
\qquad
\frac{\delta \hat{C}_{\rm R}(s)}{C_{\rm R}^{*}}
=
-\frac{\delta \hat{I}(s)}{nF\hat{G}_{\rm R}(s)}.
\end{align}
Substitution into Eq.~\eqref{eq:Nernst_linear} gives
\begin{align}
\delta \hat{E}(s)
=
\frac{RT}{n^2F^2}
\left[
\frac{1}{\hat{G}_{\rm O}(s)}
+
\frac{1}{\hat{G}_{\rm R}(s)}
\right]
\delta \hat{I}(s).
\label{eq:EI}
\end{align}
Thus, the Laplace-domain diffusion impedance for the reversible reaction is \cite{Jacobsen_95,Lasia_14}
\begin{align}
\hat{Z}_{\rm W}(s)
=
\frac{RT}{n^2F^2}
\left[
\frac{1}{\hat{G}_{\rm O}(s)}
+
\frac{1}{\hat{G}_{\rm R}(s)}
\right].
\label{eq:zw_reversible_laplace}
\end{align}

For application to impedance spectroscopy, the Laplace-domain response function is analytically continued to the frequency domain by setting \(s=i\omega\). By construction, \(\hat{G}(s)\) [and, through Eq.~\eqref{eq:zw_reversible_laplace}, \(\hat{Z}_{\rm W}(s)\)] is the Laplace-domain transfer function relating a small perturbation to its Faradaic-current response; such a transfer function already fully specifies the linear response of the system to an arbitrary input. Because of the linearity, explicitly reconstructing the time-domain relaxation and Fourier-analyzing it would yield the same result, but is unnecessary: the analytic continuation of the already-known Laplace-domain expression is sufficient.
In contrast, Eq.~\eqref{eq:Nernstianres} in the time domain is required before linearization.
After analytical continuation, \(s=i\omega\), Eq.~\eqref{eq:zw_reversible_laplace} gives the frequency-domain diffusion impedance, \cite{Jacobsen_95,Lasia_14}
\begin{align}
Z_{\rm W}(\omega)
=
\frac{RT}{n^2F^2}
\left[
\frac{1}{G_{\rm O}(\omega)}
+
\frac{1}{G_{\rm R}(\omega)}
\right].
\label{eq:zw_reversible_general}
\end{align}
Equation~\eqref{eq:zw_reversible_general} has been widely applied to study the impedance associated with both infinite and finite planar linear diffusion, as well as cylindrical and spherical diffusion \cite{Jacobsen_95,Lasia_14}.
However, its application to disk electrodes has required careful finite-element methods under the uniform-concentration boundary condition \cite{Gabrielli_06,VIVIER_22}, even though microdisk geometry is well suited to studying diffusion impedance, since it suppresses the influence of the electric double layer.
To date, an analytical expression has been derived only under the uniform-flux boundary condition \cite{FLEISCHMANN_88,FLEISCHMANN_91}.
Below, we derive a compact analytical expression for the diffusion impedance of a disk electrode under the uniform-concentration boundary condition, using Eq.~\eqref{eq:zw_reversible_general}; as shown later, this expression can be incorporated into a Randles-type equivalent circuit to help extract diffusion coefficients from experimental impedance spectra.

To evaluate \(\hat{G}_{\rm O}(s)\) and \(\hat{G}_{\rm R}(s)\), we use the disk-electrode response obtained from the chronoamperometric step problem for a circular disk electrode of radius \(a\) embedded in an insulating plane, as shown schematically in Fig.~\ref{fig:schematic}(a); the diffusion impedance for the same geometry is considered below.
Denoting the angular frequency by \(\omega\) (\(\omega=2\pi f\)), the diffusion-layer thickness, characterized by \(\sqrt{D_{\rm O}/\omega}\), is thin at high frequency and grows as the frequency decreases.
Because the flux is normal to the disk surface, diffusion is effectively one-dimensional at high frequency, as illustrated in Fig.~\ref{fig:schematic}(b), and the diffusion impedance can be modeled by a Warburg element as long as the diffusion layer continues to grow with decreasing frequency.
This approximation breaks down once the diffusion-layer thickness becomes comparable to the disk radius \(a\): the diffusion layer can no longer grow indefinitely, because spreading diffusion from the disk edge into the surrounding solution becomes significant, as illustrated in Fig.~\ref{fig:schematic}(c).
Consequently, the diffusion impedance is expected to approach a finite low-frequency value, a direct consequence of the disk geometry, rather than diverging as in the classical semi-infinite Warburg case.

In the previous disk-electrode analysis, the interfacial concentration was changed stepwise at \(t=0\), and the resulting diffusion-limited current relaxed toward the steady-state limiting current. For the oxidized species, let the step amplitude be \(\Delta C_{\rm O}\). The corresponding steady-state current scale is
\begin{align}
I_{\rm O}^{(\infty)}
=
4aD_{\rm O}nF\Delta C_{\rm O}.
\end{align}
For a step perturbation, \(\delta \hat{C}_{\rm O}(s)=\Delta C_{\rm O}/s\),
where the Laplace transform of any constant \(c\) is \(c/s\),
and Eq.~\eqref{eq:responseO} becomes
\begin{align}
\delta \hat{I}_{\rm O}(s)
=
nF \hat{G}_{\rm O}(s)
\frac{\Delta C_{\rm O}}{s C_{\rm O}^{*}} .
\label{eq:responseO_step}
\end{align}
Using \(I_{\rm O}^{(\infty)}=4aD_{\rm O}nF\Delta C_{\rm O}\), we obtain
\begin{align}
\hat{G}_{\rm O}(s)
=
4aD_{\rm O}C_{\rm O}^{*}
\frac{s\delta \hat{I}_{\rm O}(s)}{I_{\rm O}^{(\infty)}}.
\label{eq:responseO_G}
\end{align}
Therefore, while \(\delta \hat{I}_{\rm O}(s)/I_{\rm O}^{(\infty)}\) describes the relaxation kinetics following a concentration step, \(s\delta \hat{I}_{\rm O}(s)/I_{\rm O}^{(\infty)}\) is the corresponding response function required for impedance spectroscopy apart from a time-independent prefactor. This relation is the electrochemical counterpart of the general linear-response relation between relaxation and response functions~\cite{Kubobook}. An analogous relation holds for the reduced species.

For a single species with diffusion coefficient \(D\), we define the normalized disk-electrode response function as
\begin{align}
\hat{\phi}(x)
=
\frac{s\delta \hat{I}(s)}{I_{\infty}},
\qquad
x=\frac{sa^2}{D},
\end{align}
where \(I_{\infty}=4aDnF\Delta C\) is the steady-state limiting current for the corresponding concentration step. The response function can be expanded in powers of \(x^{1/2}\). The first terms are \cite{SEKI_26,SEKI_26c}
\begin{align}
\hat{\phi}(x)
=
1
+\frac{2}{\pi}x^{1/2}
+\left(\frac{4}{\pi^2}-\frac{1}{3}\right)x
-\frac{8}{\pi}\left(\frac{1}{9}-\frac{1}{\pi^2}\right)x^{3/2}
+\cdots .
\label{eq:series_main}
\end{align}
The full symbolic expansion and its numerical form, retained up to order \(x^{21/2}\), are given in Appendix~A. These expressions are included to make the construction and comparison of the Pad\'{e} approximant reproducible and to allow direct use of the series expansion when needed.

To obtain a closed-form expression that preserves the low-frequency finite-resistance response while connecting smoothly to the high-frequency Cottrell limiting behavior, a Pad\'{e} approximant was constructed from the series expansion. Using \(x=sa^2/D\), the result is written as \cite{SEKI_26,SEKI_26c}
\begin{align}
\hat{\phi}_{\rm Pade}(x)
=
\frac{
3\pi\left(12-\pi^2\right)
+2\pi^2 x^{1/2}
+\pi\left(\pi^2-8\right)x
}{
3\pi\left(12-\pi^2\right)
+8\left(\pi^2-9\right)x^{1/2}
}.
\label{eq:Pade_dimensionless}
\end{align}
The four independent coefficients of Eq.~\eqref{eq:Pade_dimensionless} are fixed by the low-frequency series alone, reproducing Eq.~\eqref{eq:series_main} through order \(x^{3/2}\); the resulting high-frequency behavior is then a consequence rather than an imposed condition. Appendix~B examines this allocation, constructs the alternative member of the same \([2/1]\) family that matches the Cottrell limit exactly instead, and shows that Eq.~\eqref{eq:Pade_dimensionless} is the more accurate of the two over the frequency range in which the diffusion impedance is resolved. The approximant reproduces the Warburg phase ($45^{\circ}$) and the low-frequency resistance exactly but not the exact high-frequency amplitude: its response function exceeds the Cottrell amplitude by about 7.5\%, i.e., $|Z|$ is lower by about 7.0\% (Appendix~B.1). Appendices~C.2 and C.4 benchmark both members of the $[2/1]$ family against a converged numerical solution: Eq.~\eqref{eq:Pade_dimensionless} is more accurate in $|Z|$ up to $\omega a^2/D\approx17$, whereas the Cottrell-matching member is more uniform over $10^{-2}\le\omega a^2/D\le10^{3}$ (maximum error of $|Z|$ of 1.1\%).
After applying the inverse Laplace transform to
the corresponding Pad\'{e} approximant expression, $\hat{\phi}_{\rm Pade}(x)/s$,
the diffusion-limited current relaxation at a disk electrode
is recovered with an accuracy comparable to that of the widely used Shoup--Szabo equation \cite{SEKI_26,SEKI_26c}.
The Shoup--Szabo equation is obtained by interpolation between the short- and long-time limits of the relaxation function \cite{SHOUP_82}.
The Laplace transform of the Shoup--Szabo equation is not known in closed form,
whereas $\hat{\phi}_{\rm Pade}(x)$ is derived directly in the Laplace domain,
making it directly suitable for impedance analysis.

The value of $\hat{\phi}_{\rm Pade}(x)$ on the imaginary axis, \(s=i\omega\), directly gives the steady-state response to a harmonic perturbation of frequency \(\omega\), which is precisely the impedance~\cite{Kubobook}. Since the oxidized and reduced species may have different diffusion coefficients, the response function must be evaluated with separate dimensionless frequency variables,
$i\omega a^2/D_{\rm O}$ and $i\omega a^2/D_{\rm R}$.
Using Eq.~\eqref{eq:responseO_G} and the corresponding equation for the reduced species, the frequency-dependent diffusion responses are
\begin{align}
G_{\rm O}(\omega)
=
4aD_{\rm O}C_{\rm O}^{*}
\hat{\phi}_{\rm Pade}\left(\frac{i\omega a^2}{D_{\rm O}}\right),
\qquad
G_{\rm R}(\omega)
=
4aD_{\rm R}C_{\rm R}^{*}
\hat{\phi}_{\rm Pade}\left(\frac{i\omega a^2}{D_{\rm R}}\right).
\label{eq:G_OR}
\end{align}
Here, \(4aD_jC_j^{*}\) is the steady-state disk-electrode scale for species \(j\), while \(\hat{\phi}_{\rm Pade}\) represents the frequency-dependent response function obtained from the Pad\'{e} approximation, now evaluated directly at \(s=i\omega\). The quantity $G_j(\omega)$ has the dimension of a molar flow rate (${\rm mol~s^{-1}}$), so that the admittance $Y_{\rm W}$ below has the dimension of a conductance.

Substitution of Eq.~\eqref{eq:G_OR} into Eq.~\eqref{eq:zw_reversible_general} gives
\begin{align}
Z_{\rm W}(\omega)
=
\frac{RT}{n^2F^2}
\left[
\frac{1}{
4aD_{\rm O}C_{\rm O}^{*}
\hat{\phi}_{\rm Pade}\left(i\omega a^2/D_{\rm O}\right)}
+
\frac{1}{
4aD_{\rm R}C_{\rm R}^{*}
\hat{\phi}_{\rm Pade}\left(i\omega a^2/D_{\rm R}\right)}
\right].
\label{eq:zw_reversible_pade}
\end{align}
Equation~\eqref{eq:zw_reversible_pade} is the dimensional Pad\'{e} diffusion impedance for a reversible reaction with different diffusion coefficients for the oxidized and reduced species. Here $C_{\rm O}^{*}$ and $C_{\rm R}^{*}$ are the local-equilibrium surface concentrations at the applied dc potential;
the dc potential are assumed to be known and the surface concentrations are effectively governed by the Nernst equation. Equation~\eqref{eq:zw_reversible_pade} treats each species as if its surface concentration were uniform over the disk. For $D_{\rm O}=D_{\rm R}$ this is exact at all frequencies and for any ratio $C_{\rm O}^{*}/C_{\rm R}^{*}$: the sum $\delta C_{\rm O}+\delta C_{\rm R}$ then obeys the same diffusion equation as each species, has zero normal derivative on the whole plane $z=0$ (flux continuity on the disk and zero flux outside it), and decays at infinity, so that it vanishes identically; the linearized Nernst condition then forces both surface concentrations to be uniform (Appendix~D.1). For $D_{\rm O}\neq D_{\rm R}$ the two species obey different diffusion equations at finite frequency, and Eq.~\eqref{eq:zw_reversible_pade} is an approximation in which flux continuity is satisfied only for the total current. Comparison with the exact coupled solution gives errors in $|Z|$ below 0.1\% for $D_{\rm R}/D_{\rm O}\ge0.5$ and below 0.31\% for $D_{\rm R}/D_{\rm O}\ge0.25$ (Appendix~D.4). The corresponding admittance is
\begin{align}
Y_{\rm W}(\omega)
=
\frac{1}{Z_{\rm W}(\omega)}
=
\frac{n^2F^2}{RT}
\frac{
G_{\rm O}(\omega)G_{\rm R}(\omega)
}{
G_{\rm O}(\omega)+G_{\rm R}(\omega)
}.
\end{align}

When \(D_{\rm O}=D_{\rm R}=D\) and \(C_{\rm O}^{*}=C_{\rm R}^{*}=C_b\), Eq.~\eqref{eq:zw_reversible_pade} reduces to
\begin{align}
Z_{\rm W}(\omega)
=
\frac{RT}{2n^2F^2DaC_b}
\frac{1}{
\hat{\phi}_{\rm Pade}\left(i\omega a^2/D\right)
}.
\label{eq:Z}
\end{align}
Only in this symmetric case can the normalized impedance be written simply as
\begin{align}
Z_{\rm norm}(\omega)
=
\frac{Z_{\rm W}(\omega)}{RT/(2n^2F^2DaC_b)}
=
\frac{1}{
\hat{\phi}_{\rm Pade}\left(i\omega a^2/D\right)
}.
\label{eq:nZ}
\end{align}
Therefore, the compact normalized form is valid only for the symmetric case with equal diffusion coefficients and equal bulk concentrations. 
For the general reversible reaction, the dimensional expression in Eq.~\eqref{eq:zw_reversible_pade} should be used, though 
Eq.~\eqref{eq:zw_reversible_pade} is an approximation for $D_{\rm O}\neq D_{\rm R}$ (Appendix~D.4). 
%%%%%%%%%%%%%%%%%%%%%%%%%%%%%%%%%%%%%%%
\section{Behavior of the compact representation}
%%%%%%%%%%%%%%%%%%%%%%%%%%%%%%%%%%%%%%%
\label{sec:behavior}

\begin{figure*}[t]
\centering
\includegraphics[width=0.85\textwidth]{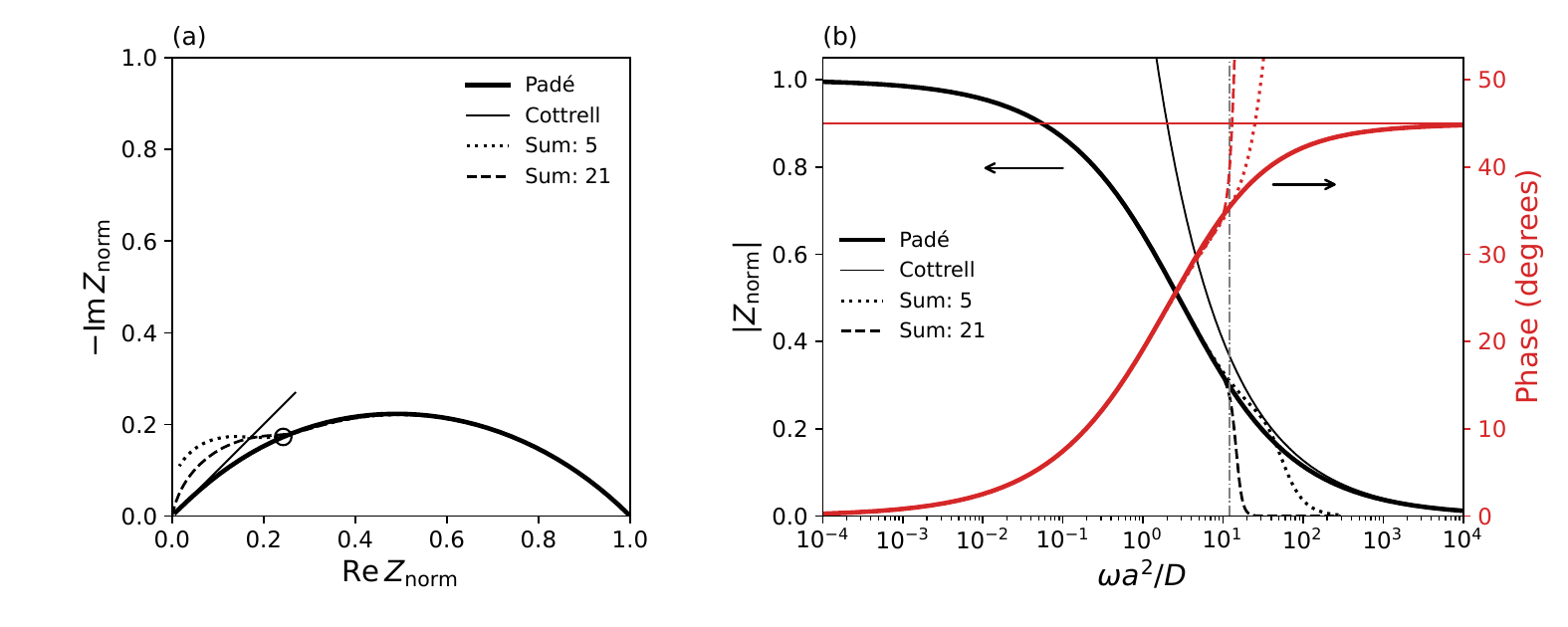}
\caption{
Comparison of the Pad\'{e} approximant with the Cottrell limiting response and truncated response-function expansions.
The impedance is normalized by the low-frequency limiting resistance, and the frequency is represented by the dimensionless variable \(\omega a^2/D\).
The thick solid curve shows the Pad\'{e} approximant [Eq.~\eqref{eq:nZ}], the thin solid curve shows the Cottrell response [Eqs.~\eqref{eq:Cottrell1} and \eqref{eq:Cottrell2}], the dotted curve shows the response-function expansion truncated at fifth order (up to the term \(x^{5/2}\)), and the dashed curve shows the expansion retained up to the term \(x^{21/2}\), where \(x=sa^2/D\).
(a) Cole--Cole plot of \(-{\rm Im}(Z)\) versus \({\rm Re}(Z)\). The open circle marks the point \(\omega a^2/D\simeq 12\). Up to this point, the original response-function expansion accurately describes the Cole--Cole arc and is reproduced by the Pad\'{e} approximant.
At higher frequencies, where \({\rm Re}(Z)\) becomes small, the Pad\'{e} approximant interpolates the response toward the Cottrell limiting behavior.
(b) Bode magnitude and phase plots. The arrows indicate the axes corresponding to \(|Z|\) and phase. The dash-dotted line marks \(\omega a^2/D=12\).
The point near \(\omega a^2/D\simeq 12\) can already be estimated from the fifth-order expansion, and increasing the number of retained terms does not substantially improve this estimate.
The Pad\'{e} approximant remains well behaved across the transition region, whereas the direct truncated series representations have limited practical ranges of validity. The phase angle approaches the classical Warburg value of \(45^\circ\) in the high-frequency limit, consistent with the Cottrell response.
}
\label{fig:pade_cottrell_series}
\end{figure*}

Figure~\ref{fig:pade_cottrell_series} compares the Pad\'{e} approximant with the Cottrell limiting response and with truncated response-function expansions. The impedance is normalized by the low-frequency limiting resistance, and the frequency is represented by the dimensionless variable \(\omega a^2/D\). The Pad\'{e} approximant was constructed from the response-function expansion in the Laplace domain.

In Fig.~\ref{fig:pade_cottrell_series}(a), the Pad\'{e} approximant gives a smooth finite arc in the Cole--Cole representation. The Cottrell response represents the high-frequency limiting behavior. In the present normalization, it is written as~\cite{Lasia_14}
\begin{align}
Z_{\rm Cottrell}(\omega)
=
\frac{RT}{2n^2F^2DaC_b}
\frac{1}{
\hat{\phi}_{\rm Cottrell}\left(i\omega a^2/D\right)
},
\label{eq:Cottrell1}
\end{align}
where
\begin{align}
\hat{\phi}_{\rm Cottrell}\left(i\omega a^2/D\right)
=
\frac{\pi}{4}
\left(
\frac{i\omega a^2}{D}
\right)^{1/2}.
\label{eq:Cottrell2}
\end{align}
This expression describes the semi-infinite planar-diffusion limit (Fig.~\ref{fig:schematic}b) and therefore does not reproduce the finite low-frequency intercept associated with disk geometry.

The open circle indicates the point \(\omega a^2/D\simeq 12\).
By considering $\omega = 2 \pi f$,
the point can be expressed as $f a^2/(2D) \simeq 1$.
Up to this point, the original response-function expansion accurately describes the Cole--Cole arc, and the Pad\'{e} approximant reproduces this behavior. This crossover point can already be estimated from the fifth-order expansion. Increasing the truncation order to the term \(x^{21/2}\), where \(x=i \omega a^2/D\), does not substantially improve the estimate of this point.

In summary, at low frequency (\(\omega a^2/D\lesssim12\)) the truncated series expansion and the Pad\'{e} approximant agree closely with each other, both describing the finite-disk Cole--Cole arc; the Cottrell response, which assumes semi-infinite planar diffusion, does not apply in this regime. At high frequency (\(\omega a^2/D\gtrsim12\)), the series expansion instead diverges from the true response because the underlying expansion in powers of \(x^{1/2}\) is only asymptotic, whereas the Pad\'{e} approximant continues to track the response and converges onto the Cottrell frequency dependence, since locally planar diffusion is recovered at sufficiently high frequency; the limiting amplitude approaches the Cottrell value to within $7\%$, as quantified in Appendix~B.1.

This behavior is also evident in the Bode plots in Fig.~\ref{fig:pade_cottrell_series}(b). The direct truncated series representations are reliable only over a limited frequency range and lose practical accuracy around the transition region. At higher frequencies, where the real part of the impedance becomes small, the Pad\'{e} approximant interpolates the response toward the high-frequency Cottrell limit; the accuracy of that interpolation, and the reason this particular member of the $[2/1]$ Pad\'{e} family was chosen, are examined in Appendices~B.2 and B.5. This comparison shows that the Pad\'{e} approximant provides a robust compact representation of the disk-electrode response, preserving the finite low-frequency response while capturing the crossover to the high-frequency behavior described by the Cottrell limiting response.

%%%%%%%%%%%%%%%%%%%%%%%%%%%%%%%%%%%%%%%
\section{The choice of boundary condition}
%%%%%%%%%%%%%%%%%%%%%%%%%%%%%%%%%%%%%%%
\label{sec:bc}

\begin{figure*}[t]
\centering
\includegraphics[width=0.85\textwidth]{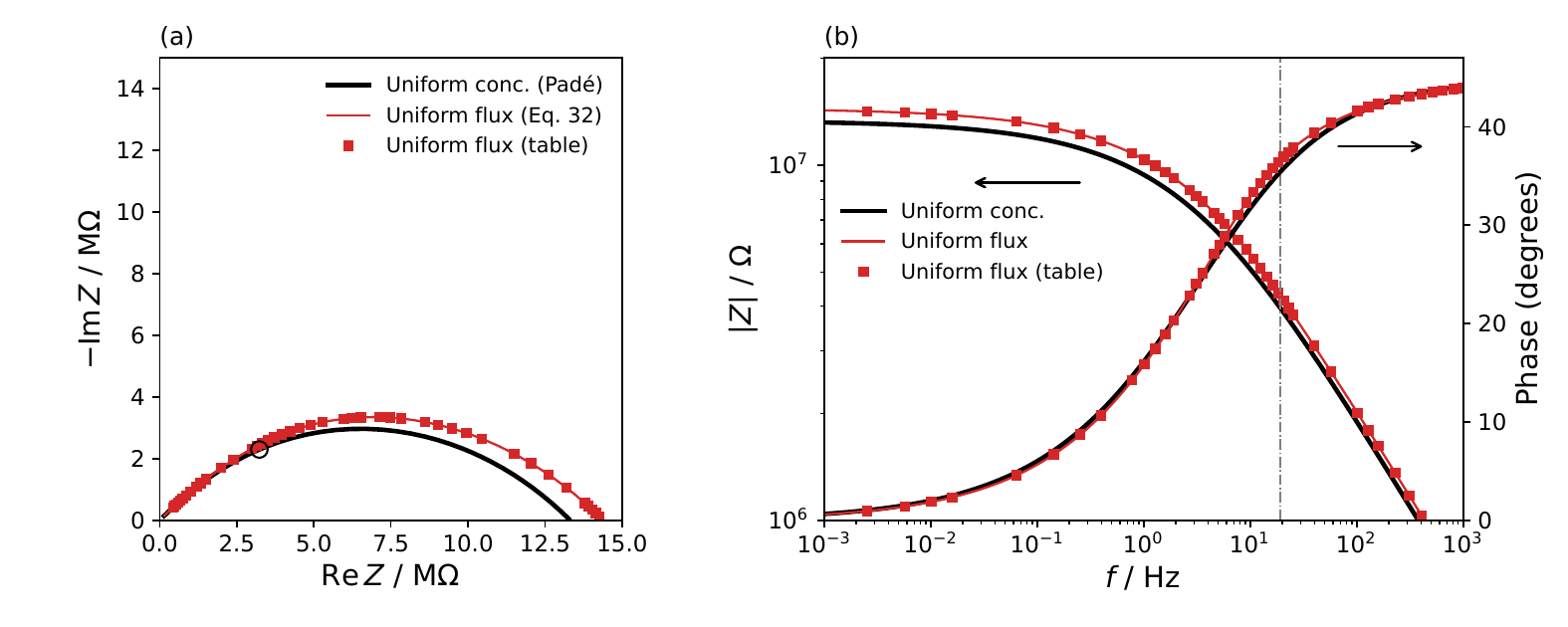}
\caption{
Comparison of impedance responses for disk electrodes under uniform-concentration and uniform-flux boundary conditions.
The thick black curves show the response under the uniform-concentration boundary condition calculated using the Pad\'{e} approximant [Eq.~\eqref{eq:Z}].
The thin red curves show the response under the uniform-flux boundary condition calculated from the analytical integral expression [Eq.~\eqref{eq:FP}]~\cite{FLEISCHMANN_91,Lasia_14}, and the red square symbols show the response under the uniform-flux boundary condition obtained from tabulated values of \(\Phi_4\) and \(\Phi_5\) in Ref.~\cite{FLEISCHMANN_88}.
The parameters are \(a=10~\mu{\rm m}\), \(D=1.0\times10^{-9}~{\rm m^2~s^{-1}}\), \(C_b=1.0~{\rm mol~m^{-3}}\), \(T=298.15~{\rm K}\), and \(n=1\), the same values used for the Randles-type equivalent circuit calculations in Figs.~\ref{fig:randles_pade} and \ref{fig:randles_pade_1M}.
(a) Cole--Cole plot of \(-{\rm Im}(Z)\) versus \({\rm Re}(Z)\). The open circle marks \(\omega a^2/D=12\). The uniform-flux boundary condition yields a slightly larger arc and a larger low-frequency real-axis intercept than the uniform-concentration boundary condition.
The difference reflects the distinct coupling of the imposed boundary constraint to the three-dimensional spreading diffusion field around a disk electrode embedded in an insulating plane.
(b) Bode magnitude and phase plots as a function of frequency \(f\) (Hz). The arrows indicate the axes corresponding to \(|Z|\) and phase. The dash-dotted line marks \(\omega a^2/D=12\) (\(f\approx19\)~Hz). Both boundary conditions exhibit similar diffusion-controlled frequency dependence; however,
the response under the uniform-flux boundary condition is shifted in magnitude, particularly in the low-frequency finite-disk regime.
The agreement between the red solid curve and the red square symbols confirms the consistency between the integral calculation and the tabulated representation for the uniform-flux boundary condition.
}
\label{fig:bc_comparison}
\end{figure*}

Figure~\ref{fig:bc_comparison} compares the impedance responses obtained under uniform-concentration and uniform-flux boundary conditions at a disk electrode.
The uniform-concentration boundary condition corresponds to fast reversible reactions on the electrode surface and is evaluated using the Pad\'{e} approximant. The uniform-flux boundary condition corresponds to a prescribed normal flux over the disk surface. For the uniform-flux case, two results are shown: one calculated directly from the analytical integral expression and the other obtained from tabulated values of the functions \(\Phi_4\) and \(\Phi_5\).~\cite{FLEISCHMANN_88} The close agreement between the red solid curve and the red square symbols confirms the consistency between the integral calculation and the tabulated representation.

For the uniform-concentration boundary condition,
the dimensional impedance and
the normalized impedance are given by Eqs.~\eqref{eq:Z} and \eqref{eq:nZ}, respectively.
Taking the reciprocal of the normalized impedance gives the normalized admittance,
\begin{align}
\frac{Y_{\rm cc}(\omega)}{Y_{\infty,{\rm rev}}}
=
\frac{ 3\pi\left(12-\pi^2\right) +2\pi^2 x^{1/2} +\pi\left(\pi^2-8\right)x }{ 3\pi\left(12-\pi^2\right) +8\left(\pi^2-9\right)x^{1/2} }, \qquad x=\frac{i\omega a^2}{D},
\end{align}
where \(Y_{\infty,{\rm rev}}\) is given by Eq.~\eqref{eq:AdSaito}.

For the uniform-flux boundary condition, the impedance is expressed as~\cite{FLEISCHMANN_88,FLEISCHMANN_91,Lasia_14}
\begin{align}
Z_{\rm cf}(\omega)
=
\frac{4RT}{n^2F^2\pi D^{1/2}a^2\omega^{1/2}C_b}
\left[
\Phi_4\left(\frac{a^2\omega}{D}\right)
-
i\Phi_5\left(\frac{a^2\omega}{D}\right)
\right],
\label{eq:FP}
\end{align}
where \(\Phi_4\) and \(\Phi_5\) are dimensionless functions describing the real and imaginary components of the response under uniform-flux boundary condition on the disk electrode:
\begin{align}
\Phi_4\left(\frac{a^2\omega}{D}\right)
&=
\int_0^\infty d\beta\,
J_1\left(\beta \frac{a\sqrt{\omega}}{\sqrt{D}}\right)^2
\frac{\cos\left(\Theta/2 \right)}
{\beta \left(1+\beta^4\right)^{1/4}},
\label{eq:Phi4}
\\
\Phi_5\left(\frac{a^2\omega}{D}\right)
&=
\int_0^\infty d\beta\,
J_1\left(\beta \frac{a\sqrt{\omega}}{\sqrt{D}}\right)^2
\frac{\sin\left(\Theta/2 \right)}
{\beta \left(1+\beta^4\right)^{1/4}},
\label{eq:Phi5}
\end{align}
with \(\tan\Theta=1/\beta^2\), and \(J_1(z)\) denoting the Bessel function of the first kind of order one.

As shown in Fig.~\ref{fig:bc_comparison}(a), both boundary conditions produce a depressed Cole--Cole arc arising from the disk geometry. The uniform-flux boundary condition produces a larger arc and a larger low-frequency real-axis intercept than the uniform-concentration boundary condition, particularly in the range \(\omega a^2/D < 12\).

From a physical standpoint, the uniform-concentration boundary condition is more relevant to reversible electrochemical systems with sufficiently fast interfacial kinetics. In such systems, the electrode surface remains close to local equilibrium, and the surface concentrations are effectively governed by the Nernst equation.

The difference between the two responses reflects the distinct ways in which the boundary conditions couple to the diffusion field around a circular disk embedded in an insulating plane. Physically, \(a^2/D\) is, up to a numerical prefactor, the characteristic time required for a diffusing species to spread over a distance comparable to the disk radius \(a\); this follows from the scaling of the diffusion equation, in which a length scale \(\ell\) and a time scale \(\tau\) are related by \(\ell^2\sim D\tau\). More precisely, using the mean-square displacement, \(\langle r^2(t)\rangle=2dDt\), where \(d\) is the dimensionality, the time required for \(\langle r^2(t)\rangle\) to reach \(a^2\) is \(\tau=a^2/(2dD)\); for one-dimensional spreading (\(d=1\)), this gives \(\tau=a^2/(2D)\) as the precise diffusion time needed for the diffusion layer to grow to the size of the disk itself, although the simpler estimate \(\tau\sim a^2/D\) remains roughly correct. When \(\omega a^2/D < 12\) ($f a^2/(2D) <1$),
the oscillation period, \(f^{-1}\), is comparable to or longer than this diffusion time \(a^2/(2D)\), so that within a single period the diffusion layer has time to spread over a distance comparable to \(a\), and the diffusion field extends over the disk scale (Fig.~\ref{fig:schematic} (c)). In this regime, spreading diffusion from the disk edge into the bulk solution contributes significantly to the finite-disk impedance response. This edge contribution is precisely the geometric effect that distinguishes the disk response from the one-dimensional, semi-infinite planar diffusion described by
the Cottrell equation [Eqs.~\eqref{eq:Cottrell1} and \eqref{eq:Cottrell2}]: whereas the Cottrell limit assumes diffusion confined to directions normal to the electrode surface, spreading diffusion at the disk edge introduces an additional, purely geometric relaxation channel, which is ultimately responsible for the finite low-frequency diffusion resistance and the departure from the classical Warburg behavior noted in Fig.~\ref{fig:pade_cottrell_series}.

Under the uniform-concentration boundary condition, the surface concentration is spatially uniform, and the flux distribution adjusts to the disk geometry and the associated spreading diffusion field. In contrast, under the uniform-flux boundary condition, the normal flux is prescribed over the disk, and the surface concentration must adjust to satisfy the diffusion field. The difference in impedance therefore reflects the different coupling between the imposed boundary constraint and the three-dimensional diffusion field.

Figure~\ref{fig:bc_comparison}(b) shows the corresponding Bode magnitude and phase plots. Both magnitude responses exhibit similar overall frequency dependence,
but the response under the uniform-flux boundary condition is shifted relative to that under the uniform-concentration boundary condition, most noticeably in the range \(\omega a^2/D < 12\). The phase spectra also show similar diffusion-controlled behavior, with small quantitative differences in the transition region.

These results indicate that the boundary condition has a measurable effect on the impedance response, particularly in the frequency range where the finite-disk spreading diffusion field is developed, even though both cases exhibit the characteristic finite-diffusion behavior of disk electrodes.

The difference between the two responses can be quantified with the converged uniform-concentration solution of Appendix~C.1 as reference. Figure~\ref{fig:uf_uc} gives the relative differences of $|Z|$, $\mathrm{Re}\,Z$, and $\mathrm{Im}\,Z$ and the phase difference; the reference values of both responses are listed in Table~\ref{tab:C_ref} of Appendix~C.5. The modulus of the uniform-flux impedance is higher by $8.4\%$ at $\omega a^2/D=0.01$ (the exact zero-frequency value is $32/(3\pi^2)-1=8.1\%$), reaches a maximum of $11.3\%$ near $\omega a^2/D\approx1.8$, and decreases to $7.5\%$ at 12 and $0.8\%$ at $10^3$. At low frequency the difference resides in the real part (up to about $+10.5\%$), and near the crossover it resides in the imaginary part (up to about $-8\%$ near $\omega a^2/D\approx8$), so that it is largest where the diffusion field extends over the disk; the phases differ by between $-2.1^{\circ}$ and $+0.5^{\circ}$. At high frequency the two descriptions merge, because the diffusion is locally planar and the boundary condition at the disk surface no longer matters (Cottrell regime). These differences set the scale of the bias in the diffusion coefficient when a spectrum is fitted with the wrong boundary condition (Sec.~\ref{sec:Compex}).

\begin{figure*}[t]
\centering
\includegraphics[width=0.85\textwidth]{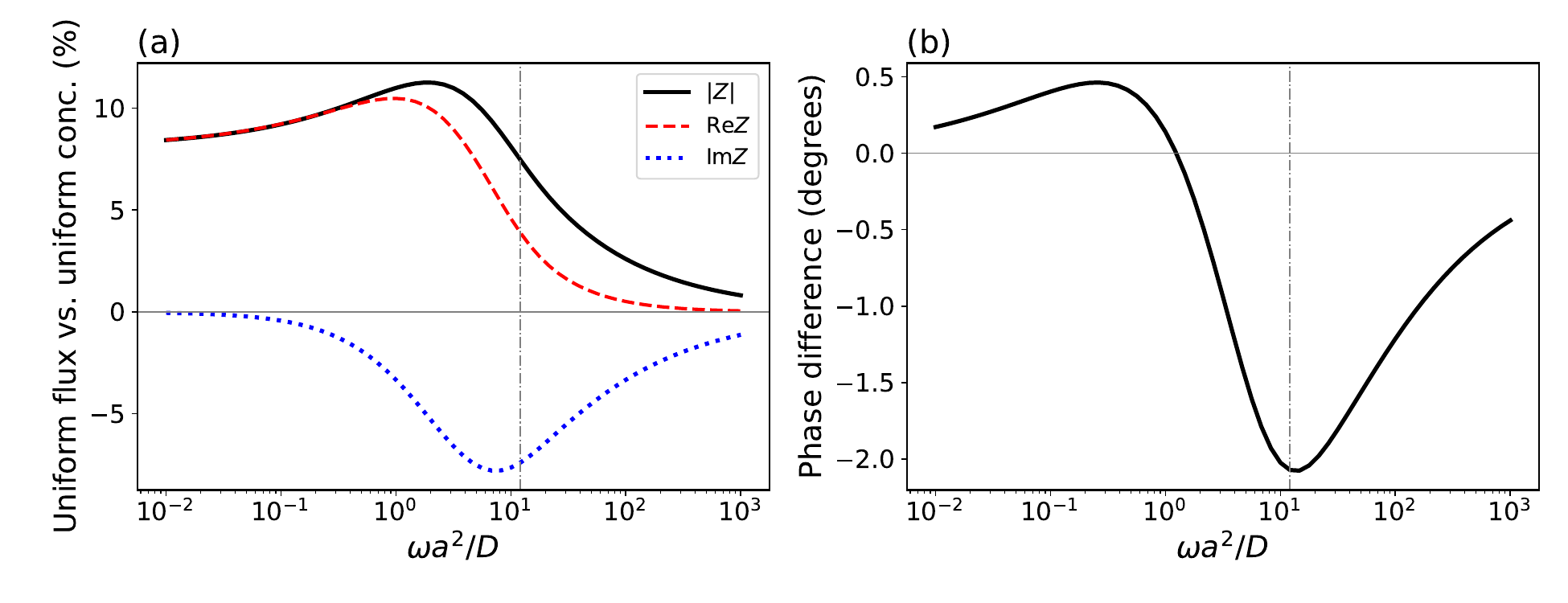}
\caption{Difference between the impedance under the uniform-flux boundary condition [Eq.~\eqref{eq:FP}] and under the uniform-concentration boundary condition (converged solution, Appendix~C.1) as a function of the dimensionless frequency $\omega a^2/D$. (a) Relative differences of $|Z|$, $\mathrm{Re}\,Z$, and $\mathrm{Im}\,Z$ in percent of $|Z|$ of the uniform-concentration response. (b) Phase difference. The dash-dotted line marks $\omega a^2/D=12$.}
\label{fig:uf_uc}
\end{figure*}

Under the presence of spreading flow associated with disk geometry,
$Z(\omega)$ in $\delta E(\omega)=Z(\omega)\delta I (\omega)$ cannot be decomposed as $Z(\omega)= R_{\rm ct}+Z_{\rm W}(\omega)$;
instead, $Z(\omega)$
should be obtained by solving the
Dirichlet--Neumann (Robin) boundary conditions.
By linearizing the time-dependent Butler--Volmer equation for $k_{\rm ox}$ and $k_{\rm red}$ around
$\bar{k}_{\rm ox}$ and $\bar{k}_{\rm red}$, respectively,
and
introducing $\delta C_{\rm O}(r,0,t) = C_{\rm O}(r,0,t) -C_{\rm O}^{*}$ and $\delta C_{\rm R}(r,0,t) = C_{\rm R}(r,0,t) -C_{\rm R}^{*}$,
Dirichlet--Neumann (Robin) boundary conditions can be expressed, following the same linearization approach as in Ref.~\cite{Lasia_14}, as
\begin{align}
\left. D_{\rm O} \frac{\partial}{\partial z} \delta C_{\rm O}(r,z,t) \right|_{z=0}&=\bar{k}_{\rm red} C_{\rm O}^{*}
\left(
\frac{\delta C_{\rm O}(r,0,t)}{C_{\rm O}^{*}}- \frac{\delta C_{\rm R}(r,0,t)}{C_{\rm R}^{*}}
- \frac{nF}{RT} \delta E(t) \right), \mbox{ for } r \leq a,
\label{eq:DNO}\\
\left. D_{\rm R} \frac{\partial}{\partial z} \delta C_{\rm R}(r,z,t)\right|_{z=0}&=\bar{k}_{\rm ox} C_{\rm R}^{*}
\left(
\frac{\delta C_{\rm R}(r,0,t)}{C_{\rm R}^{*}}- \frac{\delta C_{\rm O}(r,0,t)}{C_{\rm O}^{*}}
+ \frac{nF}{RT} \delta E(t) \right), \mbox{ for } r \leq a,
\label{eq:DNR}
\end{align}
where $\bar{k}_{\rm red} C_{\rm O}^{*}=\bar{k}_{\rm ox} C_{\rm R}^{*}$ is used
to obtain $\bar{k}_{\rm red} C_{\rm O}^{*}\,\delta C_{\rm R}(r,0,t)/C_{\rm R}^{*}=\bar{k}_{\rm ox} C_{\rm R}^{*}\,\delta C_{\rm R}(r,0,t)/C_{\rm R}^{*}$.
Both $\delta C_{\rm O}(r,z,t)$ and $\delta C_{\rm R}(r,z,t)$  satisfy the perfectly reflecting boundary condition for $r>a$.
Equations~\eqref{eq:DNO} and \eqref{eq:DNR} can be equivalently rewritten as
\begin{align}
\left. D_{\rm O} \frac{\partial}{\partial z} \delta C_{\rm O}(r,z,t) \right|_{z=0}&=\bar{k}_{\rm red} C_{\rm O}^{*}
\left(
\frac{\delta C_{\rm O}(r,0,t)-\delta C_{\rm O}(t)}{C_{\rm O}^{*}}- \frac{\delta C_{\rm R}(r,0,t)-\delta C_{\rm R}(t)}{C_{\rm R}^{*}}
 \right), \mbox{ for } r \leq a
\label{eq:DNO1}\\
\left. D_{\rm R} \frac{\partial}{\partial z} \delta C_{\rm R}(r,z,t)\right|_{z=0}&=\bar{k}_{\rm ox} C_{\rm R}^{*}
\left(
\frac{\delta C_{\rm R}(r,0,t)-\delta C_{\rm R}(t)}{C_{\rm R}^{*}}- \frac{\delta C_{\rm O}(r,0,t)-\delta C_{\rm O}(t)}{C_{\rm O}^{*}}
\right), \mbox{ for } r \leq a
\label{eq:DNR1}
\end{align}
by using an auxiliary equation
\begin{align}
 \frac{nF}{RT} \delta E(t)=\frac{\delta C_{\rm O}(t)}{C_{\rm O}^{*}}-\frac{\delta C_{\rm R}(t)}{C_{\rm R}^{*}},
\label{eq:ECOR1}
\end{align}
which is nothing but Eq.~\eqref{eq:ECOR} derived using the Nernst equation.
The current can be obtained from
\begin{align}
\left. \delta I (t)=- 2\pi \int_0^a rdr nFD_{\rm O} \frac{\partial}{\partial z} \delta C_{\rm O}(r,z,t) \right|_{z=0}=\left. 2\pi \int_0^a r dr nFD_{\rm R} \frac{\partial}{\partial z} \delta C_{\rm R}(r,z,t)\right|_{z=0} .
\label{eq:current}
\end{align}
Equations~\eqref{eq:responseO} and \eqref{eq:responseR} can be obtained through Eq.~\eqref{eq:current}, where
$G_{\rm O}(\omega)$ and $G_{\rm R}(\omega)$ are introduced.
$Z(\omega)$ can be obtained from $\delta E(\omega)=Z(\omega)\delta I (\omega)$.
By considering that Eq.~\eqref{eq:ECOR} [Eq.~\eqref{eq:ECOR1}] can be derived either by
the Nernst equation or the Butler--Volmer equation, both approaches are equivalent.

Previously, we calculated $\delta I (t)$ from $\delta C_{\rm O}(r,z,t)$ by solving the diffusion equation under the boundary condition of
$\delta C_{\rm O}(r,0,t)=\delta C_{\rm O}(t)$ for $r \leq a$ and $\delta C_{\rm R}(r,0,t)=\delta C_{\rm R}(t)$ for $r \leq a$ instead of solving
the Dirichlet--Neumann (Robin) boundary condition [Eqs.~\eqref{eq:DNO1} and \eqref{eq:DNR1}].
In this paper, $Z_{\rm W}(\omega)$ is obtained from $\delta E(\omega)=Z_{\rm W}(\omega)\delta I (\omega)$ using
Eq.~\eqref{eq:ECOR} [Eq.~\eqref{eq:ECOR1}].

For simplicity, we again consider the case of $k^0=\bar{k}_{\rm ox}=\bar{k}_{\rm red}$ and \(C_{\rm O}^{*}=C_{\rm R}^{*}=C_b\).
The uniform-concentration boundary condition corresponds to the limit $k^0 \rightarrow \infty$, and
the uniform-flux boundary condition corresponds to the limit $k^0 \rightarrow 0$ of Eqs. (\ref{eq:DNO1}) and (\ref{eq:DNR1}).
The two limits can therefore be regarded, respectively, as the ideal diffusion-controlled limit (reversible, uniform-concentration boundary condition) and the reaction-controlled limit.
Decomposing the impedance into the separate elements $R_{\rm ct}$ and $Z_{\rm W}(\omega)$ implicitly assumes
that the influence of $k^0$ on the diffusion-controlled element [$Z_{\rm W}(\omega)$] can be neglected.
The decomposition assumption can be justified when reactions occur on the plane and
diffusional flow is only normal to the plane without spreading.
Under spreading flow, the decomposition should be regarded as an approximation. 
In Sec.~\ref{sec:exact}, we quantify the deviation from the exact solution. 

As shown in Fig.~\ref{fig:bc_comparison},
the difference between the two boundary conditions (uniform-flux against uniform-concentration) increases as frequency decreases.
In the limit $\omega \rightarrow 0$,
the diffusion-controlled limit ($k^0 \rightarrow \infty$) relevant to the uniform-concentration boundary condition can be expressed as $R_{\rm ct} < 1/Y_{\infty,{\rm rev}}$,
where $Y_{\infty,{\rm rev}}$ is the admittance implied by Saito's steady-state current [Eq.~\eqref{eq:AdSaito}];
this condition leads to
\begin{align}
\frac{D_{\rm O}+D_{\rm R}}{\pi a} < k^0 .
\label{eq:DL}
\end{align}
In the opposite, reaction-controlled limit, where $k^0$ approaches $ 0$, the condition instead becomes
\begin{align}
\frac{D_{\rm O}+D_{\rm R}}{\pi a} > k^0 ,
\label{eq:RL}
\end{align}
and $Z_{\rm W}(\omega)$ should be replaced by the Fleischmann--Pons diffusion element,
$Z_{\rm cf}(\omega)$ [Eq.~\eqref{eq:FP}],
obtained under the uniform-flux boundary condition, to construct the Randles elements.

Equations~\eqref{eq:DL} and \eqref{eq:RL} are estimates of the rate constant at which $R_{\rm ct}$ equals the Saito resistance $1/Y_{\infty,{\rm rev}}$ and not sharp boundaries. In terms of the Damk\"ohler number $\mathrm{Da}=2k^0a/D$, Eq.~\eqref{eq:DL} can be rewritten as $1<\mathrm{Da}\pi/4$.
The exact solution of the coupled problem shows that the crossover between reaction and diffusion control is broad, so that these estimates locate it only in order of magnitude (Sec.~\ref{sec:exact}).

Here, we study a reversible electrode reaction using Randles elements with the diffusion-controlled element [$Z_{\rm W}(\omega)$];
the condition for the diffusion-controlled limit is given by Eq.~\eqref{eq:DL}.
Using \(a=10~\mu{\rm m}\), \(D_{\rm O}=D_{\rm R}=D=1.0\times10^{-9}~{\rm m^2~s^{-1}}\), $k^0> 6.3 \times 10^{-5}$ m s$^{-1}$ is estimated.
Similarly, using \(a=5~\mu{\rm m}\), \(D=0.633\times10^{-9}~{\rm m^2~s^{-1}}\), $k^0> 8.1 \times 10^{-5}$ m s$^{-1}$ is estimated.
If these conditions are violated, the results should instead be interpreted in the reaction-controlled limit, where
the Fleischmann--Pons diffusion element, $Z_{\rm cf}(\omega)$ [Eq.~\eqref{eq:FP}], rather than $Z_{\rm W}(\omega)$, would be relevant.

The criteria in Eqs.~\eqref{eq:DL} and \eqref{eq:RL} are consistent with,
and place the present impedance analysis within, the well-established
framework of dimensionless kinetic parameters used in steady-state
microelectrode voltammetry, where reversibility is governed by a
parameter of the form $k^0a/D$ (equivalently, $D/(k^0a)$), with the fully
reversible limit corresponding to this parameter tending to infinity and
the fully irreversible limit to zero~\cite{BondOldhamZoski_89,OldhamMylandZoskiBond_89}.
Since this parameter scales with $a/D$ for fixed $k^0$, a smaller disk
radius shifts a given reaction toward the reaction-controlled
(irreversible) limit for the same $k^0$; it is precisely this shift that
makes decreasing $a$ the standard strategy for resolving fast
(large-$k^0$) charge-transfer kinetics that would otherwise appear fully
reversible, and hence kinetically featureless, at a larger
electrode~\cite{OldhamMylandZoskiBond_89}.
This is the basis for using a systematic range of microdisk radii to
scan through the reversible, quasi-reversible, and irreversible regimes
and thereby extract $k^0$ from steady-state or
near-steady-state voltammograms~\cite{OldhamMylandZoskiBond_89}.

%%%%%%%%%%%%%%%%%%%%%%%%%%%%%%%%%%%%%%%
\section{The compact representation in a Randles circuit}
%%%%%%%%%%%%%%%%%%%%%%%%%%%%%%%%%%%%%%%
\label{sec:randles}

\begin{figure*}[t]
\centering
\includegraphics[width=0.85\textwidth]{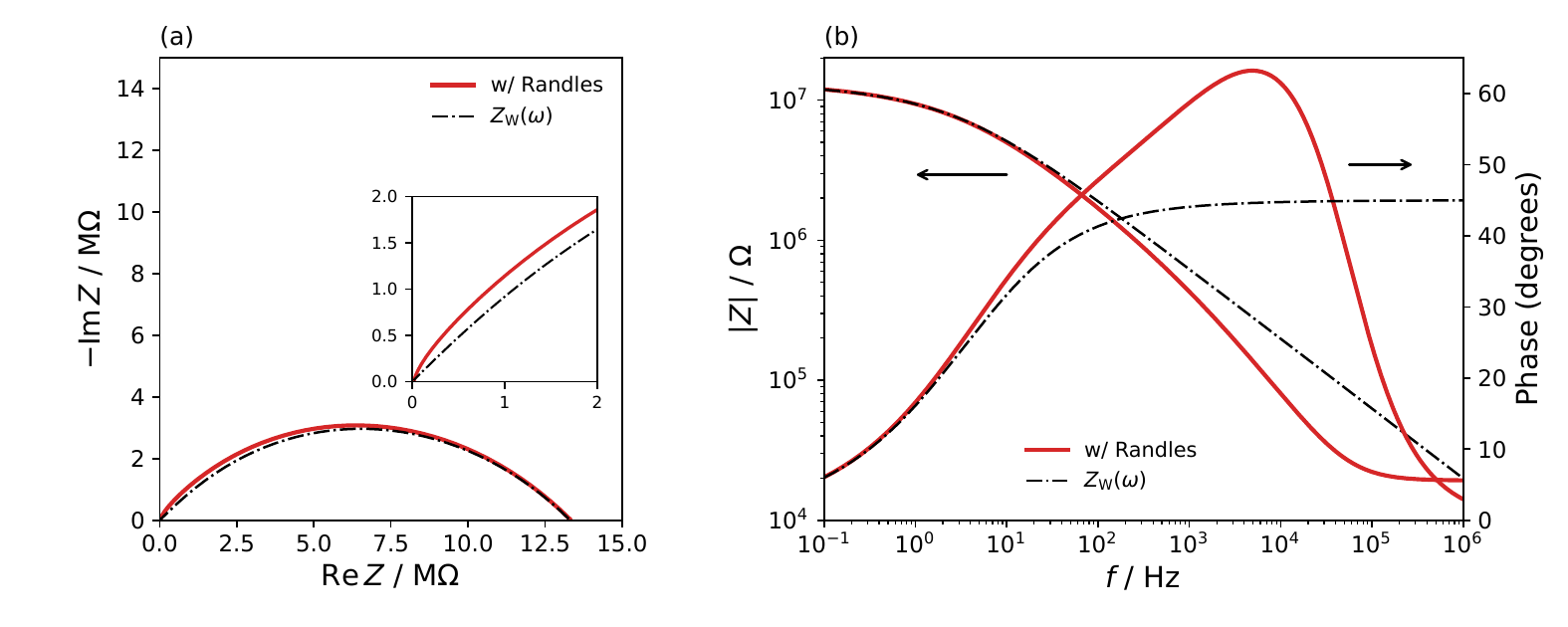}
\caption{
Impedance response calculated using the Pad\'{e} approximant as a geometry-induced finite Warburg-type diffusion element for a disk electrode.
The red solid curves show the full Randles-type response [Eq.~\eqref{eq:RW}],
\(Z(\omega)=R_{\Omega}+[i\omega C_{\rm dl}+1/(R_{\rm ct}+Z_{\rm W})]^{-1}\), whereas the black dash-dotted curves show the Pad\'{e} diffusion response without the other Randles elements [Eq.~\eqref{eq:Wp}].
(a) Cole--Cole plot of \(-{\rm Im}(Z)\) versus \({\rm Re}(Z)\). The inset magnifies the high-frequency region, where the contributions of \(R_{\Omega}\), \(R_{\rm ct}\), and \(C_{\rm dl}\) are more clearly visible.
(b) Bode magnitude and phase plots. The arrows indicate the axes corresponding to \(|Z|\) and phase.
The parameters are \(a=10~\mu{\rm m}\), \(D=1.0\times10^{-9}~{\rm m^2~s^{-1}}\), \(C_b=1.0~{\rm mol~m^{-3}}\), \(T=298.15~{\rm K}\), \(n=1\), \(R_{\rm ct}=1.0\times10^4~\Omega\), \(C_{\rm dl,total}=157~{\rm pF}\), and \(R_{\Omega}=1/(4\sigma_{\rm e} a)=1.9\times10^4~\Omega\), with the electrolyte conductivity \(\sigma_{\rm e}=1.3~{\rm S~m^{-1}}\).
}
\label{fig:randles_pade}
\end{figure*}

Figure~\ref{fig:randles_pade} shows the impedance response calculated by using the Pad\'{e} approximant for
the response under the uniform-concentration boundary condition on the disk electrode as a geometry-induced finite Warburg-type diffusion element. The total impedance was evaluated
using the Randles-type equivalent circuit,
\begin{align}
Z(\omega)
=
R_{\Omega}
+
\left[
i\omega C_{\rm dl}
+
\frac{1}{R_{\rm ct}+Z_{\rm W}(\omega)}
\right]^{-1},
\label{eq:RW}
\end{align}
where \(R_{\Omega}\) is the access resistance, \(R_{\rm ct}\) is the electron-transfer resistance, \(C_{\rm dl}\) is the total double-layer capacitance, and \(Z_{\rm W}(\omega)\) is the Pad\'{e} diffusion impedance. Using the normalized impedance in Eq.~\eqref{eq:nZ}, the dimensional diffusion impedance is written as
\begin{align}
Z_{\rm W}(\omega)
=
\frac{RT}{2n^2F^2DC_ba}
\left[ \frac{ 3\pi\left(12-\pi^2\right) +8\left(\pi^2-9\right)x^{1/2} }{ 3\pi\left(12-\pi^2\right) +2\pi^2 x^{1/2} +\pi\left(\pi^2-8\right)x } \right], \qquad x=\frac{i\omega a^2}{D}.
\label{eq:Wp}
\end{align}
Here, $R_{\rm ct}$ is given by,
\begin{align}
R_{\rm ct}
\simeq
\frac{RT}{n^2F^2 A k^0 C_b},
\qquad
A=\pi a^2 ,
\label{eq:Rct}
\end{align}
if the effect of the intrinsic rate constant $k^0$ on $Z_{\rm W}(\omega)$ is negligible.
Here, $k^0$ is the static part of the intrinsic rate constant for the electron-transfer reaction;
the static part of the intrinsic oxidation rate constant ($\bar{k}_{\rm ox}$) is equal to the static part of the intrinsic reduction rate constant ($\bar{k}_{\rm red}$)
since we have assumed \(C_{\rm O}^{*}=C_{\rm R}^{*}=C_b\) for simplicity;
under this condition, we have $\bar{k}_{\rm ox}=\bar{k}_{\rm red}$.
If we employ 1 dimensional diffusion without spreading flow to calculate
$\delta E(\omega)=Z(\omega)\delta I (\omega)$, we obtain
$Z(\omega)=R_{\rm ct}+Z_{\rm Cottrell}(\omega)$
with \cite{Lasia_14}
\begin{align}
k^0=\bar{k}_{\rm ox}=\bar{k}_{\rm red}.
\label{eq:k0}
\end{align}

\begin{figure*}[t]
\centering
\includegraphics[width=0.85\textwidth]{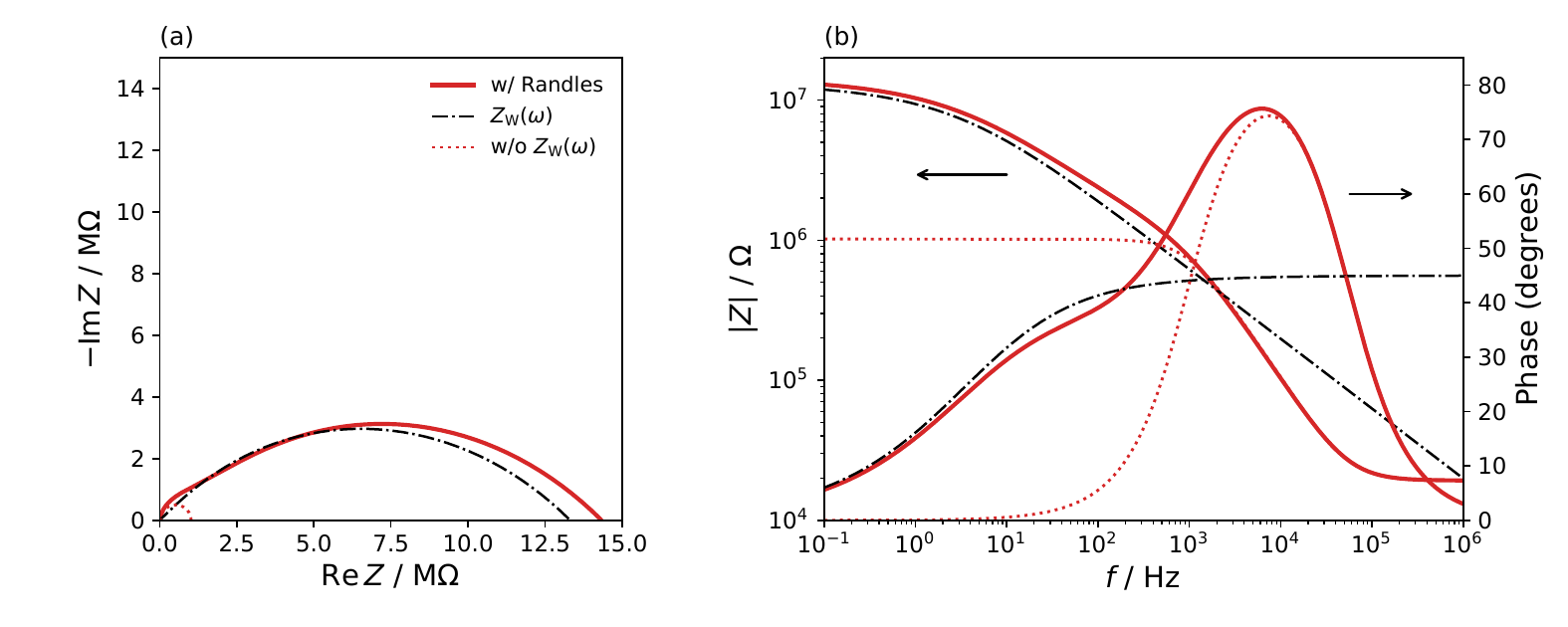}
\caption{
Impedance response calculated using the Pad\'{e} approximant as a geometry-induced finite Warburg-type diffusion element for a disk electrode.
The red solid curves show the full Randles-type response [Eq.~\eqref{eq:RW}], whereas the black dash-dotted curves show the Pad\'{e} diffusion response without the other Randles elements [Eq.~\eqref{eq:Wp}].
The red dotted curves show the response of the Randles elements without the Warburg diffusion impedance.
(a) Cole--Cole plot of \(-{\rm Im}(Z)\) versus \({\rm Re}(Z)\).
(b) Bode magnitude and phase plots. The arrows indicate the axes corresponding to \(|Z|\) and phase.
The parameters are the same as in Fig.~\ref{fig:randles_pade}, except that \(R_{\rm ct}=1.0\times10^6~\Omega\).
For \(a=10~\mu{\rm m}\) and \(C_b=1.0~{\rm mol~m^{-3}}\), this value corresponds to \(k^0\sim 10^{-3}~{\rm m~s^{-1}}\), which is a reasonable order-of-magnitude value for fast outer-sphere redox couples, such as ferrocene or \([{\rm Fe(CN)}_6]^{3-/4-}\), at a microdisk electrode.
The other parameters are \(D=1.0\times10^{-9}~{\rm m^2~s^{-1}}\), \(T=298.15~{\rm K}\), \(n=1\), \(C_{\rm dl,total}=157~{\rm pF}\), and \(R_{\Omega}=1/(4\sigma_{\rm e} a)=1.9\times10^4~\Omega\), with the electrolyte conductivity \(\sigma_{\rm e}=1.3~{\rm S~m^{-1}}\).
}
\label{fig:randles_pade_1M}
\end{figure*}

The Randles-type circuit of Eq.~\eqref{eq:RW} assumes a uniform interfacial potential over the disk, linear response, a symmetric reversible couple or equal diffusion coefficients (Eq.~\eqref{eq:zw_reversible_pade}), the access resistance $R_{\Omega}=1/(4\sigma_{\rm e} a)$ of a disk in an insulating plane in series, where $\sigma_{\rm e}$ is the electrolyte conductivity, a frequency-independent double-layer capacitance in parallel with the faradaic branch, and the decomposition of the faradaic impedance into $R_{\rm ct}$ and $Z_{\rm W}(\omega)$. The last assumption is an approximation, valid sufficiently close to the diffusion-controlled limit, as discussed below. Within the circuit, $R_{\rm ct}$ is an effective parameter: it equals the conventional small-signal charge-transfer resistance of Eq.~\eqref{eq:Rct} only if the decomposition is exact.

The parameters used in the calculation were \(a=10~\mu{\rm m}\), \(D=1.0\times10^{-9}~{\rm m^2~s^{-1}}\), \(C_b=1.0~{\rm mol~m^{-3}}\), \(T=298.15~{\rm K}\), and \(n=1\). For the Randles elements, \(R_{\Omega}=1/(4\sigma_{\rm e} a)\) was evaluated using \(\sigma_{\rm e}=1.3~{\rm S~m^{-1}}\), corresponding approximately to \(0.1~{\rm mol~dm^{-3}}\) KCl, giving \(R_{\Omega}\simeq1.9\times10^4~\Omega\). The total double-layer capacitance was estimated from \(C_{\rm dl}=50~\mu{\rm F~cm^{-2}}\) and the disk area, giving \(C_{\rm dl,total}\simeq157~{\rm pF}\).

The electron-transfer resistance was set to either \(R_{\rm ct}=1.0\times10^6~\Omega\) or \(R_{\rm ct}=1.0\times10^4~\Omega\). For comparison, \(R_{\rm ct}=1.0\times10^6~\Omega\) is a reasonable order-of-magnitude value for fast outer-sphere redox couples, such as ferrocene or \([{\rm Fe(CN)}_6]^{3-/4-}\), at a microdisk electrode. For a disk radius of \(a=10~\mu{\rm m}\) and a concentration of \(C_b=1.0~{\rm mol~m^{-3}}\), this value corresponds to an intrinsic rate constant of the order of \(k^0\sim 10^{-3}~{\rm m~s^{-1}}\), estimated from Eq.~\eqref{eq:Rct}
and satisfies $k^0> 6.3 \times 10^{-5}$ m s$^{-1}$.
Thus, the reaction is sufficiently fast that the electrode surface can be approximated by a reversible, Nernstian boundary condition. This value corresponds to $\mathrm{Da}=2k^0a/D\approx20$, for which the accuracy of the decomposition is quantified in Sec.~\ref{sec:exact}.

We also considered a smaller charge-transfer resistance, \(R_{\rm ct}=1.0\times10^4~\Omega\), corresponding to a standard heterogeneous rate constant of the order of \(k^0\sim 0.1~{\rm m~s^{-1}}\). 
This corresponds to $\mathrm{Da}=2k^0a/D\approx2\times10^{3}$. This case was included to examine the limiting behavior for very fast interfacial electron transfer and to make the effect of the double-layer capacitance more clearly visible.

The Cole--Cole plot in Fig.~\ref{fig:randles_pade}(a) shows that inclusion of the Randles elements other than $Z_{\rm W}(\omega)$
produces only a small modification of the Pad\'{e} diffusion response on the scale of the diffusion impedance for \(R_{\rm ct}=1.0\times10^4~\Omega\). This is because the characteristic diffusion impedance, \(1/Y_{\infty,{\rm rev}}\), is much larger than \(R_{\Omega}\) and \(R_{\rm ct}\) under the present conditions. The inset magnifies the high-frequency region, where the finite contributions of \(R_{\Omega}\), \(R_{\rm ct}\), and \(C_{\rm dl}\) are more clearly visible. In the Bode representation shown in Fig.~\ref{fig:randles_pade}(b),
the magnitude response is dominated by the Pad\'{e} diffusion impedance at low and intermediate frequencies,
whereas the high-frequency response approaches the resistive and capacitive response of the Randles elements other than $Z_{\rm W}(\omega)$.
The phase spectrum similarly shows the transition from diffusion-controlled behavior to the high-frequency response governed by \(R_{\Omega}\), \(R_{\rm ct}\), and \(C_{\rm dl}\).

Figure~\ref{fig:randles_pade_1M}(a) shows the corresponding result for the larger charge-transfer resistance, \(R_{\rm ct}=1.0\times10^6~\Omega\). In this case, the high-frequency semicircle associated with
the Randles elements other than $Z_{\rm W}(\omega)$ becomes more clearly visible, and the Cole--Cole response is shifted to larger real impedance compared with Fig.~\ref{fig:randles_pade}(a). In the Bode representation shown in Fig.~\ref{fig:randles_pade_1M}(b), the transition from the low- and intermediate-frequency regime dominated by the Pad\'{e} diffusion impedance to the high-frequency regime governed by
the Randles elements other than $Z_{\rm W}(\omega)$ becomes more pronounced in both the magnitude and phase spectra.

These results suggest that, for microdisk electrodes under conditions where the diffusion impedance is sufficiently large compared with
the other Randles elements, the Pad\'{e} representation can be used to estimate diffusion coefficients or effective electrode radii from impedance spectra. When charge-transfer resistance and double-layer capacitance become non-negligible, their influence can be accounted for by incorporating the Pad\'{e} diffusion impedance into a Randles-type equivalent circuit.

%%%%%%%%%%%%%%%%%%%%%%%%%%%%%%%%%%%%%%%
\section{Exact solution of the coupled problem}
%%%%%%%%%%%%%%%%%%%%%%%%%%%%%%%%%%%%%%%
\label{sec:exact}

The decomposition $Z_{\rm F}=R_{\rm ct}+Z_{\rm W}$ can be tested without approximation, because the coupled problem with the finite-rate boundary conditions of Eqs.~\eqref{eq:DNO1} and \eqref{eq:DNR1} can be solved numerically to high accuracy, as shown in Appendix~D.3. For $D_{\rm O}=D_{\rm R}$, flux continuity forces $\delta C_{\rm R}=-\delta C_{\rm O}$ everywhere (and, by current continuity, $\delta C_{\rm R}(t)=-\delta C_{\rm O}(t)$ for the auxiliary functions of Eq.~\eqref{eq:ECOR1}), and the two conditions collapse into a single Robin condition for the oxidized species,
\begin{align}
D\,\partial_z\delta C_{\rm O}\big|_{z=0}=\kappa\left[\delta C_{\rm O}(r,0,t)-\delta C_{\rm O}(t)\right],
\qquad
\kappa=\bar k_{\rm red}\left(1+\frac{C_{\rm O}^*}{C_{\rm R}^*}\right)=\bar k_{\rm ox}+\bar k_{\rm red},
\label{eq:D_robin}
\end{align}
with $\kappa=2k^0$ for $C_{\rm O}^{*}=C_{\rm R}^{*}$ (Appendix~D.1). The Damk\"ohler number is $\mathrm{Da}=\kappa a/D$.

Solving the diffusion equation for this condition at the Laplace variable $s=i\omega$, with $\Omega=\omega a^2/D$, gives the current $\delta I$ of the disk, which we express as $\psi(\mathrm{Da},i\Omega)=\delta I/I_{\rm S}$ in units of Saito's current $I_{\rm S}=Y_{\infty,{\rm rev}}\,\delta E_{\rm s}$, the steady-state current of the uniform-concentration disk ($\mathrm{Da}\to\infty$) for a potential perturbation $\delta E_{\rm s}$. The faradaic admittance of the coupled problem, $Y_{\rm F}(\omega)=\delta I/\delta E$, relates $\delta I$ to the potential perturbation $\delta E$. For $D_{\rm O}=D_{\rm R}$ the auxiliary function of Eq.~\eqref{eq:ECOR1} is related to $\delta E$ by [Eq.~\eqref{eq:D_dCR}]
\begin{align}
\delta C_{\rm O}(t)=-\delta C_{\rm R}(t)=\frac{nF}{RT}\,\delta E(t)\,\frac{C_{\rm O}^*C_{\rm R}^*}{C_{\rm O}^*+C_{\rm R}^*} ,
\label{eq:DOCRt}
\end{align}
so that, for $C_{\rm O}^*=C_{\rm R}^*=C_b$, Saito's current $I_{\rm S}=4nFDa|\delta C_{\rm O}(t)|$ used in Appendix~D.3 equals $Y_{\infty,{\rm rev}}\,\delta E_{\rm s}$ [Eq.~\eqref{eq:AdSaito}]. Within the linear response, $\psi(\mathrm{Da},i\Omega)$ is independent of $\delta E(t)$ and $\delta E_{\rm s}$,
\begin{equation}
\psi(\mathrm{Da},i\Omega)=Y_{\rm F}(\omega)/Y_{\infty,{\rm rev}}.
\label{eq:YF}
\end{equation}
The function $\psi$ is for the Robin boundary condition at the disk what $\hat\phi$ is for the uniform-concentration boundary condition at the disk:
$\psi\to\hat\phi(x)$ for $\mathrm{Da}\to\infty$. Correspondingly, the faradaic impedance of the coupled problem is
\begin{equation}
Z_{\rm F}(\omega)=\frac{R_\infty}{\psi(\mathrm{Da},i\Omega)},
\label{eq:ZF}
\end{equation}
where $R_\infty=1/Y_{\infty,{\rm rev}}$. The solution is converged in the number of basis functions (Appendix~D.3.3), and the method applies at $\omega=0$ as well as at finite frequency.

At zero frequency the linear response is the steady-state response to a small potential step. For $D_{\rm O}=D_{\rm R}$ and $C_{\rm O}^{*}=C_{\rm R}^{*}$, Eq.~\eqref{eq:YF} gives $Y_{\rm F}(0)=\chi(\mathrm{Da})\,Y_{\infty,{\rm rev}}$ with $\chi(\mathrm{Da})=\psi(\mathrm{Da},0)$, the steady-state current in units of Saito's current, so that
\begin{equation}
R_{\rm dc}=\frac{1}{Y_{\rm F}(0)}=\frac{R_\infty}{\chi(\mathrm{Da})}.
\label{eq:Rdc}
\end{equation}
To examine the decomposition we introduce $\mathcal{G}_{\rm s}(\mathrm{Da})$ by
\begin{align}
R_{\rm dc}=R_{\rm ct}+\mathcal{G}_{\rm s}(\mathrm{Da})\,R_\infty ,
\label{eq:decomp}
\end{align}
with $R_{\rm ct}/R_\infty=4/(\pi\mathrm{Da})$ [Eq.~\eqref{eq:Rct}]; Eq.~\eqref{eq:decomp} is the zero-frequency limit $Z_{\rm F}\to R_{\rm ct}+\mathcal{G}_{\rm s}\,R_\infty$. The decomposition is satisfied if $\kappa=\bar k_{\rm ox}+\bar k_{\rm red}$ is carried by $R_{\rm ct}$ alone, that is, if $\mathcal{G}_{\rm s}(\mathrm{Da})$ is independent of $\mathrm{Da}$. Comparing Eq.~\eqref{eq:Rdc} with Eq.~\eqref{eq:decomp} gives
\begin{align}
\mathcal{G}_{\rm s}(\mathrm{Da})=\frac1\chi-\frac{4}{\pi\mathrm{Da}}.
\label{eq:G}
\end{align}

The steady-state current $\chi$ is obtained from the solution of Appendix~D.3 at
$x=i\omega a^2/D=0$ ($\omega=0$); it tends to $(\pi/4)\mathrm{Da}$ for $\mathrm{Da}\to0$
(reaction controlled limit)
and to 1 for $\mathrm{Da}\to\infty$ (Saito's current in diffusion controlled limit),
and it reaches 0.5, 0.9, and 0.95 at $\mathrm{Da}=1.37$, 15.7, and 37.2.

The factor $\mathcal{G}_{\rm s}(\mathrm{Da})$ is the diffusion resistance of the coupled problem in units of $R_\infty$, and its two limits correspond to the two limiting diffusion elements:
$\mathcal{G}_{\rm s}(\mathrm{Da})\to1$ for $\mathrm{Da}\to\infty$ (uniform concentration in the diffusion-controlled limit, the element $Z_{\rm W}$ with the diffusion resistance $R_\infty$) and $\mathcal{G}_{\rm s}(\mathrm{Da})\to \mathcal{G}_0=32/(3\pi^2)=1.081$ for $\mathrm{Da}\to0$ (uniform flux in the reaction-controlled limit, the Fleischmann--Pons element with the diffusion resistance $\mathcal{G}_0R_\infty$). For intermediate $\mathrm{Da}$ neither element is exact. Table~\ref{tab:ss} tests each element in a Randles circuit against the exact solution at $\omega=0$ in two ways, in each case by the relative difference (approximate $-$ exact)/exact, where ``approximate'' refers to the circuit with the uniform-concentration element or with the uniform-flux element.

First, the dc resistance of the circuit, $R_{\rm ct}+R_\infty$ or $R_{\rm ct}+\mathcal{G}_{\rm s}(\mathrm{Da})R_\infty$, is compared with $R_{\rm dc}=R_\infty/\chi(\mathrm{Da})$;
the errors are $(1-\mathcal{G}_{\rm s}(\mathrm{Da}))\chi$ and $(\mathcal{G}_0-\mathcal{G}_{\rm s}(\mathrm{Da}))\chi$, which have the same normalization and can be compared directly. For the uniform-concentration element,
$(1-\mathcal{G}_{\rm s}(\mathrm{Da}))\chi$ is the zero-frequency limit of the error of the decomposition $Z_{\rm F}=R_{\rm ct}+Z_{\rm W}$ examined at finite frequency below (Fig.~\ref{fig:fd} of Appendix~D.3); it does not exceed 3.9\% in magnitude (at $\mathrm{Da}\approx3.3$).
The factor $\mathcal{G}_{\rm s}(\mathrm{Da})$ decreases monotonically with $\mathrm{Da}$, so that the deviation $|1-\mathcal{G}_{\rm s}(\mathrm{Da})|$ from the diffusion-controlled limit $\mathcal{G}_{\rm s}=1$ increases monotonically with decreasing $\mathrm{Da}$. The error $(1-\mathcal{G}_{\rm s}(\mathrm{Da}))\chi$ is nevertheless non-monotonic in Table~\ref{tab:ss}, because the deviation $1-\mathcal{G}_{\rm s}(\mathrm{Da})$ of the diffusion resistance is normalized by the total resistance $R_{\rm dc}$: for small $\mathrm{Da}$, $\chi\simeq(\pi/4)\mathrm{Da}$ (reaction-controlled limit), so that $R_{\rm dc}\simeq R_{\rm ct}$, which becomes much larger than the diffusion resistance and reduces $(1-\mathcal{G}_{\rm s}(\mathrm{Da}))\chi$.

Second, the diffusion resistance alone, $R_\infty$ or $\mathcal{G}_0R_\infty$, is compared with $\mathcal{G}_{\rm s}(\mathrm{Da})R_\infty$; the errors are $(1-\mathcal{G}_{\rm s}(\mathrm{Da}))/\mathcal{G}_{\rm s}(\mathrm{Da})$ and
$(\mathcal{G}_0-\mathcal{G}_{\rm s}(\mathrm{Da}))/\mathcal{G}_{\rm s}(\mathrm{Da})$. Because $R_\infty\propto1/D$, these are the relative errors of the diffusion coefficient deduced from the dc diffusion resistance when $R_{\rm ct}$ is known ($D_{\rm fit}/D=1/\mathcal{G}_{\rm s}$ for the uniform-concentration element and $\mathcal{G}_0/\mathcal{G}_{\rm s}(\mathrm{Da})$ for the uniform-flux element). Since the numerator is a difference of diffusion resistances, normalization by the exact diffusion resistance $\mathcal{G}_{\rm s}(\mathrm{Da})R_\infty$ is more appropriate for judging the diffusion element than the normalization by the exact total resistance $R_{\rm dc}$ in the first case;
these errors depend on $\mathcal{G}_{\rm s}(\mathrm{Da})$ alone and are therefore monotonic in $\mathrm{Da}$.
The two elements have equal errors of $(\mathcal{G}_0-1)/(\mathcal{G}_0+1)=3.9\%$ in the diffusion resistance ($3.4\%$ in $R_{\rm dc}$) at
$\mathcal{G}_{\rm s}(\mathrm{Da})=(1+\mathcal{G}_0)/2$, that is, at $\mathrm{Da}\approx8.3$; the uniform-flux element is the better one at smaller $\mathrm{Da}$ and the uniform-concentration element at larger $\mathrm{Da}$, and the error of the uniform-flux element grows to $\mathcal{G}_0-1=8.1\%$ for $\mathrm{Da}\to\infty$. The error is therefore at most about 4\% if the element is switched at $\mathrm{Da}\approx8.3$. The boundary between the reaction-limited and the diffusion-limited current, estimated by equating the limiting currents $(\pi/4)\mathrm{Da}$ and 1 [Eq.~\eqref{eq:DL}], is $\mathrm{Da}=4/\pi\approx1.3$. The boundary found here from the errors, $\mathrm{Da}\approx8.3$, is larger by a factor of about 6.5, but it is of the same order of magnitude, which is reasonable because the crossover between reaction and diffusion control is broad: at the nominal threshold $\mathrm{Da}=4/\pi$ of Eq.~\eqref{eq:DL} the exact steady-state current is already 0.48 of Saito's value, and $\chi$ rises from 0.5 to 0.95 only between $\mathrm{Da}=1.4$ and 37 ($k^0a/D=0.7$ and 19). The uniform-flux element is exact only for $\mathrm{Da}\to0$ and the uniform-concentration element only for $\mathrm{Da}\to\infty$; neither is an interchangeable model in the crossover region.

\begin{table}[h]
\centering
\caption{Steady-state quantities from the exact solution of the coupled problem at $\omega=0$ (Appendix~D.3 with $x=0$): $I/I_{\rm S}=\chi$ and $\mathcal{G}_{\rm s}(\mathrm{Da})=1/\chi-4/(\pi\mathrm{Da})$, with $R_{\rm dc}=R_{\rm ct}+\mathcal{G}_{\rm s}(\mathrm{Da})R_\infty$. Relative errors (approximate $-$ exact)/exact of a Randles circuit with the uniform-concentration element $Z_{\rm W}$ (diffusion resistance $R_\infty$) and with the uniform-flux (Fleischmann--Pons) element FP (diffusion resistance $\mathcal{G}_0R_\infty$, $\mathcal{G}_0=32/(3\pi^2)$): of the dc resistance $R_{\rm dc}$, $(1-\mathcal{G}_{\rm s}(\mathrm{Da}))\chi$ and $(\mathcal{G}_0-\mathcal{G}_{\rm s}(\mathrm{Da}))\chi$; of the diffusion resistance $\mathcal{G}_{\rm s}R_\infty$ alone, $(1-\mathcal{G}_{\rm s}(\mathrm{Da}))/\mathcal{G}_{\rm s}(\mathrm{Da})$ and $(\mathcal{G}_0-\mathcal{G}_{\rm s}(\mathrm{Da}))/\mathcal{G}_{\rm s}(\mathrm{Da})$.}
\label{tab:ss}
\begin{tabular}{rcccccc}
\toprule
 & & & \multicolumn{2}{c}{Error of $R_{\rm dc}$ (\%)} & \multicolumn{2}{c}{Error of $\mathcal{G}_{\rm s}R_\infty$ (\%)}\\
\cmidrule(lr){4-5}\cmidrule(lr){6-7}
$\mathrm{Da}$ & $I/I_{\rm S}$ & $\mathcal{G}_{\rm s}$ & $Z_{\rm W}$ & FP & $Z_{\rm W}$ & FP\\
\midrule
1   & 0.427 & 1.070 & $-3.0$ & $+0.4$ & $-6.6$ & $+1.0$\\
3.3 & 0.694 & 1.056 & $-3.9$ & $+1.7$ & $-5.3$ & $+2.4$\\
6   & 0.795 & 1.046 & $-3.7$ & $+2.8$ & $-4.4$ & $+3.3$\\
20  & 0.917 & 1.026 & $-2.4$ & $+5.0$ & $-2.6$ & $+5.3$\\
100 & 0.978 & 1.010 & $-0.9$ & $+7.0$ & $-0.9$ & $+7.1$\\
\bottomrule
\end{tabular}
\end{table}

\begin{figure*}[t]
\centering
\includegraphics[width=0.85\textwidth]{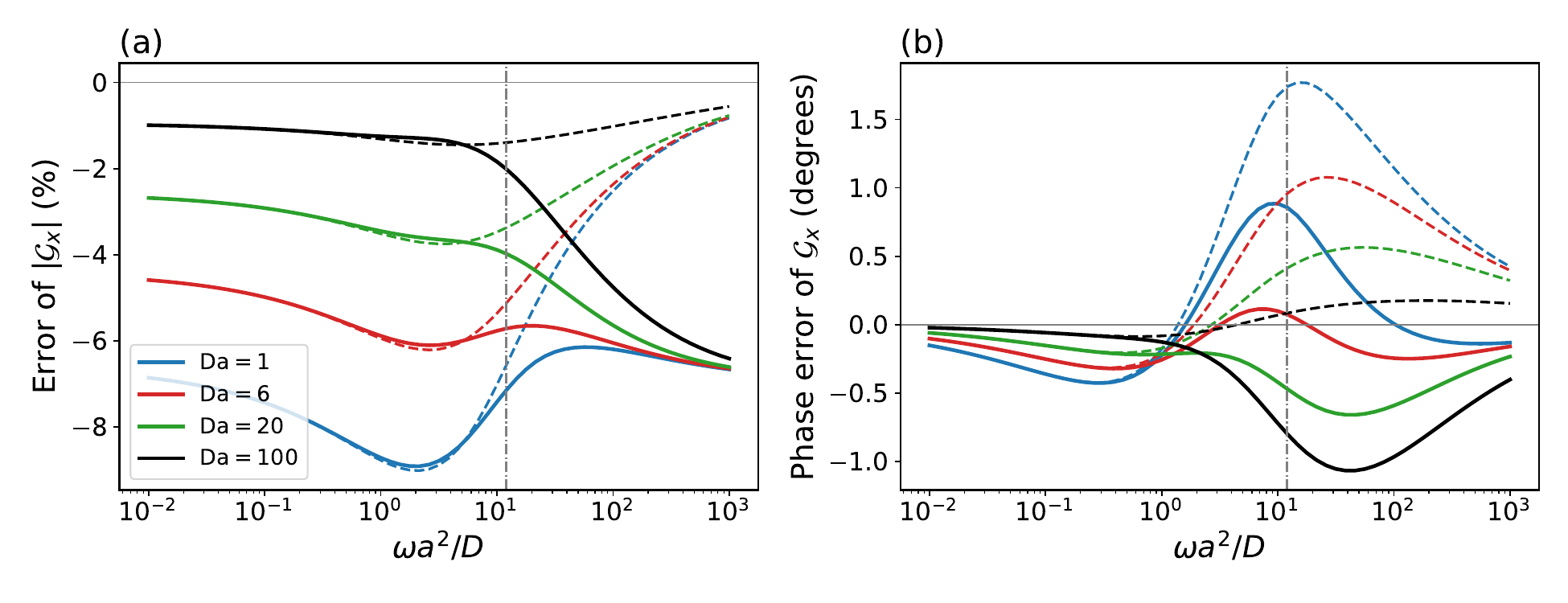}
\caption{Errors of the diffusion element $\mathcal{G}_x$ of the Randles circuit with respect to the diffusion resistance $\mathcal{G}(\mathrm{Da},\Omega)$ of the coupled problem, as functions of the dimensionless frequency $\omega a^2/D$ for $\mathrm{Da}=1$, 6, 20, and 100: (a) relative error of the modulus, $(|\mathcal{G}_x|-|\mathcal{G}(\mathrm{Da},\Omega)|)/|\mathcal{G}(\mathrm{Da},\Omega)|$; (b) error of the phase, $\arg \mathcal{G}_x-\arg \mathcal{G}(\mathrm{Da},\Omega)$. Solid curves: Pad\'e element $\mathcal{G}_x=\mathcal{G}_{\rm W}=Z_{\rm W}/R_\infty$, Eq.~\eqref{eq:Z}; dashed curves: converged element $\mathcal{G}_x=\mathcal{G}_\infty=Z_\infty/R_\infty$ (Appendix~C.1). The dash-dotted line marks $\omega a^2/D=12$.}
\label{fig:Gerr}
\end{figure*}

So far, we have studied the decomposition $R_{\rm dc}=R_{\rm ct}+\mathcal{G}_{\rm s}(\mathrm{Da})R_\infty$ in the steady state.
In the frequency domain we define the diffusion resistance of the coupled problem by
\begin{equation}
Z_{\rm F}(\mathrm{Da},\Omega)=R_{\rm ct}+R_\infty\,\mathcal{G}(\mathrm{Da},\Omega),
\end{equation}
so that $\mathcal{G}(\mathrm{Da},\Omega)=1/\psi(\mathrm{Da},i\Omega)-4/(\pi\mathrm{Da})$ is dimensionless and complex, with $\mathcal{G}(\mathrm{Da},0)=\mathcal{G}_{\rm s}(\mathrm{Da})$ of Eq.~\eqref{eq:G}. The two choices of the diffusion element in the Randles circuit are $\mathcal{G}_\infty(\Omega)=Z_\infty(\Omega)/R_\infty=1/\hat\phi(i\Omega)$, with the converged response function $\hat\phi$ of the uniform-concentration disk (Appendix~C.1), which is the limit of $\mathcal{G}(\mathrm{Da},\Omega)$ for $\mathrm{Da}\to\infty$ and isolates the error of the decomposition, and $\mathcal{G}_{\rm W}(\Omega)=Z_{\rm W}(\Omega)/R_\infty=1/\hat\phi_{\rm Pade}(i\Omega)$ [Eq.~\eqref{eq:Z}], which is obtained by approximating $\hat\phi$ with the Pad\'e approximant of Eq.~\eqref{eq:Pade_dimensionless} and adds the error of the approximant. Both tend to 1 for $\Omega\to0$. Figure~\ref{fig:Gerr} shows the relative error of the modulus, $(|\mathcal{G}_x|-|\mathcal{G}(\mathrm{Da},\Omega)|)/|\mathcal{G}(\mathrm{Da},\Omega)|$, and the error of the phase, $\arg \mathcal{G}_x-\arg \mathcal{G}(\mathrm{Da},\Omega)$, of the diffusion element $\mathcal{G}_x$ ($\mathcal{G}_\infty$ or $\mathcal{G}_{\rm W}$) with respect to the diffusion resistance of the coupled problem; these are the quantities used in Fig.~\ref{fig:fd} for the total impedance.
For the converged element, the error of the modulus tends to $(1-\mathcal{G}_{\rm s}(\mathrm{Da}))/\mathcal{G}_{\rm s}(\mathrm{Da})$ of Table~\ref{tab:ss} for $\Omega\to0$ and the error of the phase tends to zero. The error of the modulus is largest in magnitude at $\Omega\approx2$--5 (up to $-9.0$, $-6.2$, $-3.7$, and $-1.4\%$ for $\mathrm{Da}=1$, 6, 20, and 100) and below 0.9\% in magnitude at $\Omega=10^3$; the error of the phase does not exceed $1.8^{\circ}$ ($\mathrm{Da}=1$) and decreases to $0.2^{\circ}$ at $\mathrm{Da}=100$, with its maximum at $\Omega\approx15$--180. The Pad\'e element agrees with the converged one within 0.06 percentage points (modulus) and $0.04^{\circ}$ (phase) for $\Omega\le1$, but its error of the modulus remains between $-6.4$ and $-6.7\%$ at $\Omega=10^3$ because of its high-frequency deficit (Appendix~C.2), and its phase error stays within $\pm1.1^{\circ}$. The influence of these errors on the modulus and phase of $Z$ is shown by $Z_{\rm F}$, $Z_\infty$, and $Z_{\rm W}$ in Fig.~\ref{fig:fd} of Appendix~D.3. Over $10^{-2}\le\Omega\le10^3$ the maximum errors in $|Z|$ are 3.2, 4.6, 3.4, and 1.4\% for the converged element and 3.2, 4.6, 4.0, and 5.2\% for the Pad\'e element at $\mathrm{Da}=1$, 6, 20, and 100, with phase errors of at most $1.0^{\circ}$ and $0.9^{\circ}$, respectively; for $1\le\mathrm{Da}\le100$ the decomposition itself is therefore accurate to within 4.6\% in $|Z|$ and about $1^{\circ}$ in phase.
The error of $|Z|$ for the converged element is largest at low and intermediate frequencies, where it tends to the zero-frequency value $(1-\mathcal{G}_{\rm s}(\mathrm{Da}))\chi$ of Table~\ref{tab:ss}, and it vanishes at high frequency, where the diffusion is locally planar and the two descriptions coincide. For $\mathrm{Da}\gtrsim100$ the high-frequency deficit of the Pad\'e element (Appendix~C.2) dominates over the error of the decomposition. For the two parameter sets used in Sec.~\ref{sec:randles}, the decomposition error is at most 4.0\% in $|Z|$ at $\mathrm{Da}\approx20$, where the steady-state current is 92\% of Saito's limit (Table~\ref{tab:ss}), and below 0.1\% at zero frequency at $\mathrm{Da}\approx2\times10^{3}$.

We also examine the approximation of Eq.~\eqref{eq:zw_reversible_pade}, in which the local flux continuity at the disk surface is replaced by the continuity of the total current. It is exact for $D_{\rm O}=D_{\rm R}$; for $D_{\rm O}\neq D_{\rm R}$ the coupled solution (Appendix~D.4) shows that it deviates from the exact impedance by less than 0.1\% in $|Z|$ for $D_{\rm R}/D_{\rm O}\ge0.5$ and by less than 0.31\% for $D_{\rm R}/D_{\rm O}\ge0.25$.

The exact solution is applied to an experimental spectrum in Sec.~\ref{sec:Compex}.

%%%%%%%%%%%%%%%%%%%%%%%%%%%%%%%%%%%%%%%
\section{Comparison to experiments}
%%%%%%%%%%%%%%%%%%%%%%%%%%%%%%%%%%%%%%%
\label{sec:Compex}

To test the theoretical impedance expressions derived above against real
experimental data, we compare Eq.~\eqref{eq:RW}, combined with either the
uniform-concentration (Pad\'{e}) diffusion element of Eq.~\eqref{eq:Wp} or
the uniform-flux (Fleischmann--Pons) diffusion element of
Eq.~\eqref{eq:FP}~\cite{FLEISCHMANN_91,Lasia_14}, with the impedance
spectrum reported by Gabrielli et al.\ for a Pt microdisk electrode of
radius $a=5~\mu{\rm m}$~\cite{Gabrielli_06}. The exact solution of the coupled problem with finite-rate kinetics (Sec.~\ref{sec:exact}, Appendix~D.3), which assumes neither limiting boundary condition, is fitted as well.
This spectrum was obtained in
bulk solution and is explicitly distinguished, in the original
reference, from the thin-layer (confined) configuration also studied
there using a scanning electrochemical microscope.
The electrolyte consisted of
equimolar ($10~{\rm mol~m^{-3}}$) potassium ferricyanide and ferrocyanide
with a $5.0\times10^{2}~{\rm mol~m^{-3}}$ KCl supporting electrolyte, and
the impedance was measured at the equilibrium potential over the range
$0.01~{\rm Hz}$--$100~{\rm kHz}$.

The raw data are not available. The spectrum analyzed here consists of 46 points read by hand from the pixel coordinates of its Fig.~2 (macOS Preview), with the axes calibrated from the tick marks. One pixel of the figure ($500\times307$ pixels) corresponds to about 13~k$\Omega$, which is the digitization error; the points carry no frequency labels, and the source gives neither the temperature nor uncertainties of the radius and the concentration. A temperature of 298.15~K is assumed. For $a=5~\mu$m and $D=6.3\times10^{-10}~{\rm m^2~s^{-1}}$ the crossover $\omega a^2/D\approx12$ of the response corresponds to $f\approx48$~Hz, inside the measured window of 0.01~Hz--100~kHz.

\begin{figure*}[t]
\centering
\includegraphics[width=0.9\textwidth]{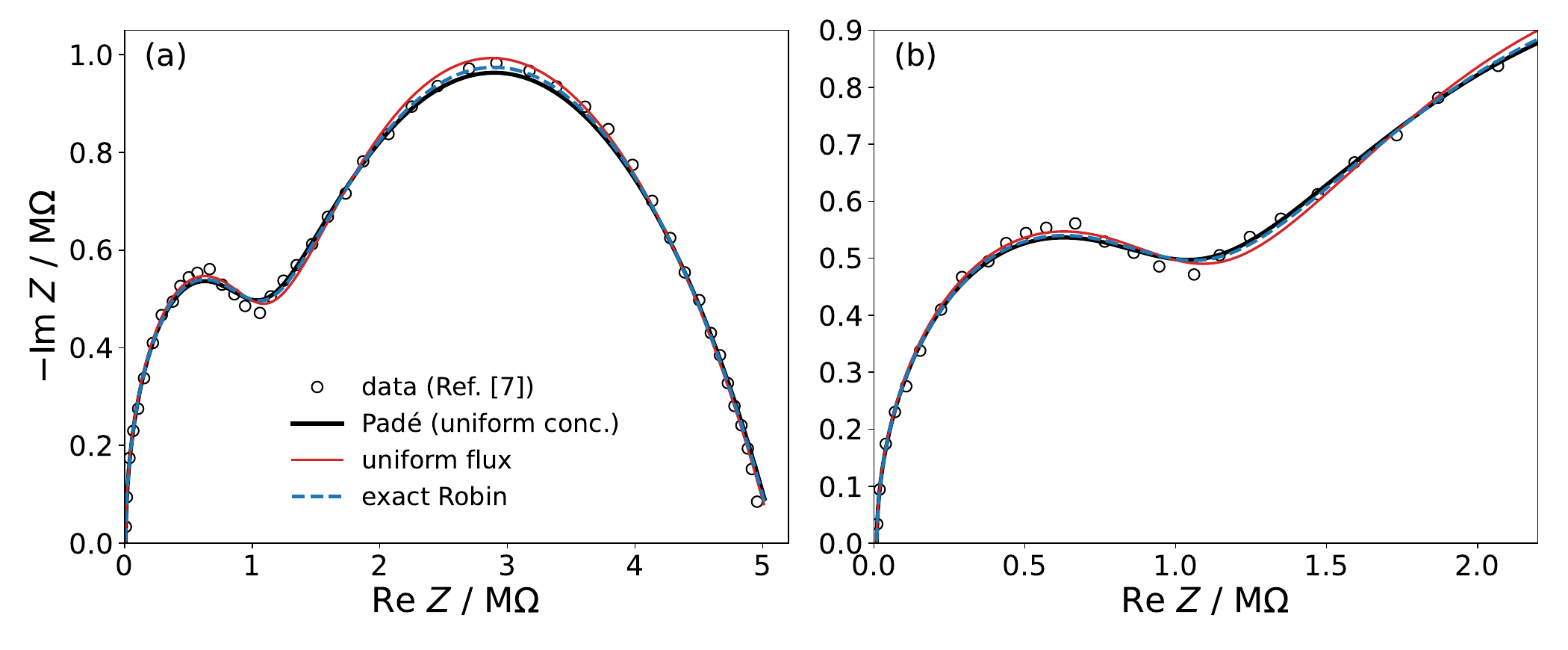}
\caption{
Fits to the experimental impedance data for a microdisk electrode: (a)~Nyquist plot; (b)~enlargement of the high-frequency arc.
Open circles indicate published data obtained with a $5~\mu\mathrm{m}$ radius Pt-microelectrode in a
$10~{\rm mol~m^{-3}}$ potassium ferricyanide (K$_3$Fe(CN)$_6$) / $10~{\rm mol~m^{-3}}$ potassium
ferrocyanide (K$_4$Fe(CN)$_6$) / $5.0\times10^{2}~{\rm mol~m^{-3}}$ KCl solution at the equilibrium
potential reported from Ref.~\cite{Gabrielli_06} (Fig. 2).
Thick black line, thin red line, and dashed blue line indicate the results of the uniform-concentration
boundary condition using the Pad\'{e} approximant [Eq.~\eqref{eq:RW} with Eq.~\eqref{eq:Wp}], of the
uniform-flux boundary condition [Eq.~\eqref{eq:RW} with Eq.~\eqref{eq:FP}]~\cite{FLEISCHMANN_91,Lasia_14},
and of the exact solution of the coupled problem with finite-rate kinetics (Sec.~\ref{sec:exact}, Appendix~D.3),
respectively. $R_{\Omega}$ is constrained to $R_{\Omega}=1/(4\sigma_{\rm e} a)=9.208\times10^{3}~\Omega$, with the electrolyte conductivity $\sigma_{\rm e}=5.43~{\rm S~m^{-1}}$ for all three; this value is the high-frequency intercept ($R_{\rm e}=9.2~{\rm k\Omega}$) measured in Ref.~\cite{Gabrielli_06}.
The common parameters are $a=5~\mu{\rm m}$, $C_b=10~{\rm mol~m^{-3}}$, $n=1$, $T=298.15$~K.
The remaining parameters are $R_{\rm ct}=8.9\times10^{5}~\Omega$ ($k^0=3.8\times10^{-4}~{\rm m~s^{-1}}$), $C_{\rm dl,total}=58.0~{\rm pF}$, and
$D_{\rm uc}=6.33\times10^{-10}~{\rm m^2~s^{-1}}$ (black); $R_{\rm ct}=9.2\times10^{5}~\Omega$ ($k^0=3.7\times10^{-4}~{\rm m~s^{-1}}$), $C_{\rm dl,total}=49.9~{\rm pF}$, and
$D_{\rm uf}=6.92\times10^{-10}~{\rm m^2~s^{-1}}$ (red); and $2k^0a/D=5.66$ ($k^0=3.8\times10^{-4}~{\rm m~s^{-1}}$, $R_{\rm ct}=9.0\times10^{5}~\Omega$), $C_{\rm dl,total}=52.4~{\rm pF}$, and
$D_{\rm ex}=6.65\times10^{-10}~{\rm m^2~s^{-1}}$ (blue); see Table~\ref{tab:fit_unc}.
Adapted with permission from Ref.~\cite{Gabrielli_06}. Copyright 2006 American Chemical Society.}
\label{fig:Vivier_comparison}
\end{figure*}

Because the electrolyte conductivity $\sigma_{\rm e}$ can be obtained independently
of the impedance fit itself, either from a separate conductivity
measurement or from the high-frequency intercept of the spectrum via the
disk access-resistance relation, $R_{\Omega}=1/(4\sigma_{\rm e} a)$, we fix
$R_{\Omega}$ at the high-frequency intercept $R_{\rm e}=9.2~{\rm k\Omega}$ measured in Ref.~\cite{Gabrielli_06} ($9.208\times10^{3}~\Omega$, equivalent to $\sigma_{\rm e}=5.43~{\rm S~m^{-1}}$ for $a=5~\mu{\rm m}$; the literature value given there is $9~{\rm k\Omega}$ and no conductivity is reported), rather than treating it as a free parameter of
the fit. The remaining parameters, $R_{\rm ct}$, $C_{\rm dl}$, and $D$, are
then obtained by least-squares fitting of Eq.~\eqref{eq:RW} to the
experimental Nyquist plot (by orthogonal-distance regression in the Nyquist plane and the real and imaginary residuals are weighted equally), using each of the two diffusion elements in turn, and then for the exact solution of the coupled problem, for which $Z_{\rm F}=R_\infty/\psi(\mathrm{Da},i\Omega)$ [Eq.~\eqref{eq:ZF}] replaces $R_{\rm ct}+Z_{\rm W}$ in Eq.~\eqref{eq:RW}. The free parameters of the latter fit are $\mathrm{Da}$, $C_{\rm dl}$, and $D$, i.e.\ as many as for the Randles fits, and $R_{\rm ct}=4R_\infty/(\pi\,\mathrm{Da})$ follows from the definition $\mathrm{Da}=2k^0a/D$. The response $\psi$ was computed with the Legendre-basis solution of Appendix~D.3 (100 basis functions), tabulated on a grid of $\mathrm{Da}$ and $\Omega$, and interpolated with cubic splines (relative interpolation error below $10^{-6}$); the model assumes equal diffusion coefficients and bulk concentrations of the two species, as appropriate for the equimolar couple, so that $\kappa=2k^0$ [Eq.~\eqref{eq:D_robin}]. Figure~\ref{fig:Vivier_comparison} shows the resulting fits together with the experimental data.

All three descriptions reproduce the experimental spectrum to within the digitization precision once $R_{\Omega}$ is fixed to its measured high-frequency value: the three curves are almost indistinguishable from each other and from the data over the entire frequency range (Fig.~\ref{fig:Vivier_comparison}). However, the fits require different diffusion coefficients, $D_{\rm uc}=6.33\times10^{-10}$, $D_{\rm uf}=6.92\times10^{-10}$, and $D_{\rm ex}=6.65\times10^{-10}~{\rm m^2~s^{-1}}$ for the uniform-concentration, uniform-flux, and exact solutions, respectively. The source quotes a voltammetric diffusion coefficient without an uncertainty and without describing how it was obtained, so that it is not used here as a reference.
The ratio $D_{\rm uf}/D_{\rm uc}\approx1.093$ agrees closely
with the ratio of the two boundary conditions'
zero-frequency diffusion resistances, $16/(3\pi)/(\pi/2)\approx1.081$,
presented in Sec.~\ref{sec:Intro} in the limit $\omega\rightarrow0$. The fitted ratio is 1.094 (95\% bootstrap interval 1.089--1.099), slightly above this limit because the difference between the two impedances depends on frequency (Appendix~C.5). This
agreement supports, using real experimental data rather than only the
analytic $\omega\to0$ and $\omega\to\infty$ limits, that the uniform-flux
boundary condition only slightly distorts the frequency dependence of the
impedance relative to the uniform-concentration case, through a slight
adjustment of $C_{\rm dl,total}$; it otherwise simply rescales
the apparent diffusion coefficient by the constant factor
$16/(3\pi)/(\pi/2)$. Consequently, a diffusion coefficient extracted from
impedance data using the conventional Fleischmann--Pons (uniform-flux)
expression for a reversible redox couple will
overestimate the diffusion coefficient of a reversible redox couple by up to about 8\% relative to the uniform-concentration element, which is the exact zero-frequency bound; for finite $\mathrm{Da}$ the bias is the smaller fraction $(\mathcal{G}_0-\mathcal{G}_{\rm s})/\mathcal{G}_{\rm s}$ and it tends to zero for $k^0\to0$ (Table~\ref{tab:ss}).

Relative to the exact solution, the uniform-concentration element underestimates $D$ by 4.8\% and the uniform-flux element overestimates it by 4.1\% ($D_{\rm ex}/D_{\rm uc}=1.051$ and $D_{\rm uf}/D_{\rm ex}=1.041$, from the values of $D$ in Table~\ref{tab:fit_unc}). This is what the steady-state analysis of Sec.~\ref{sec:exact} predicts: for $2k^0a/D=5.7$ the dc diffusion resistances give $D_{\rm ex}/D_{\rm uc}=\mathcal{G}_{\rm s}=1.047$ and $D_{\rm uf}/D_{\rm ex}=\mathcal{G}_0/\mathcal{G}_{\rm s}=1.032$ (Table~\ref{tab:ss}), so that the fitted shifts agree with the prediction to within one percentage point, although the prediction uses only the zero-frequency limit.

In the original study, a diffusion coefficient of
$6.60\times10^{-10}~{\rm m^2~s^{-1}}$ was also reported from numerical
simulations of the same spectrum~\cite{Gabrielli_06}. Those simulations
were performed with the Dirichlet--Neumann (Robin) boundary condition
obtained by linearizing the Butler--Volmer equation, which is
intermediate between the two limiting conditions compared here, and the
reported value lies accordingly between $D_{\rm uc}$ and $D_{\rm uf}$.
In particular, it is smaller than the value obtained here under the
uniform-flux boundary condition,
$D_{\rm uf}=6.92\times10^{-10}~{\rm m^2~s^{-1}}$, a reduction in the direction implied by the present analysis. It agrees with the exact value $D_{\rm ex}$ obtained here to within 0.8\%.

Table~\ref{tab:fit_unc} gives the fitted parameters with 95\% bootstrap intervals (659 resamples of the data points, paired for the three models), which describe the scatter of the digitized points only. All three fits give the same dc resistance, $R_{\rm dc}=5.08$--$5.10~{\rm M}\Omega$, so that the data fix the dc level and the models differ in how it is divided between $R_{\rm ct}$ and the diffusion resistance and in the frequency dependence. The experimental part is a reanalysis of one published spectrum; the quantitative benchmark of the diffusion elements remains the exact numerical solution of the coupled problem (Sec.~\ref{sec:exact}, Appendix~D).

\begin{table}[h]
\centering
\small
\caption{Fit parameters for the digitized spectrum of Ref.~\cite{Gabrielli_06} (Fig.~\ref{fig:Vivier_comparison}). Intervals are 95\% bootstrap intervals (659 resamples of the data points, paired for the three models) and describe the scatter of the digitized points only. For the Randles fits, $k^0=RT/(\pi F^2a^2C_bR_{\rm ct})$ and $2k^0a/D$ follow from the fitted $R_{\rm ct}$ and $D$; for the exact fit, $R_{\rm ct}=4R_\infty/(\pi\,\mathrm{Da})$ follows from the fitted $\mathrm{Da}=2k^0a/D$ and $D$. $R_{\rm dc}$ is the zero-frequency limit of the faradaic impedance.}
\label{tab:fit_unc}
\begin{tabular}{lccc}
\toprule
 & Uniform conc.\ (Pad\'{e}) & Uniform flux (F--P) & Exact (Robin)\\
\midrule
$R_{\rm ct}$ ($10^5~\Omega$) & 8.93 [8.78, 9.06] & 9.25 [9.10, 9.42] & 9.01 [8.88, 9.14]\\
$k^0$ ($10^{-4}$~m~s$^{-1}$) & 3.80 [3.74, 3.86] & 3.67 [3.60, 3.72] & 3.76 [3.71, 3.82]\\
$2k^0a/D$ & 6.00 [5.89, 6.12] & 5.30 [5.18, 5.40] & 5.66 [5.57, 5.76]\\
$C_{\rm dl,total}$ (pF) & 58.0 [54.7, 62.1] & 49.9 [45.1, 53.9] & 52.4 [49.1, 55.9]\\
$D$ ($10^{-10}$~m$^2$~s$^{-1}$) & 6.330 [6.300, 6.361] & 6.922 [6.894, 6.954] & 6.651 [6.624, 6.680]\\
$R_{\rm dc}$ (M$\Omega$) & 5.100 [5.084, 5.117] & 5.082 [5.071, 5.092] & 5.093 [5.079, 5.106]\\
\bottomrule
\end{tabular}
\end{table}

The result is consistent with the Bode plots in Fig.~\ref{fig:bc_comparison}(b),
where the influence of the boundary condition on phase is insignificant compared with its influence on magnitude;
the magnitude shifts higher under the uniform-flux boundary condition in the low-frequency range, $a^2 f/(2D) <1$.
For practical values of the Randles elements,
the diffusion coefficient can be estimated by fitting in this low-frequency range,
where the magnitude is approximately shifted by the constant factor
$16/(3\pi)/(\pi/2)$ while the phase is approximately retained.
Because the admittance is proportional to the diffusion coefficient and is the reciprocal of the impedance,
fitting the same experimental data under the assumption of uniform flux rather than uniform concentration
rescales the extracted diffusion coefficient by the same constant factor, $16/(3\pi)/(\pi/2)$ [Eq.~\eqref{eq:Dratio}].

Finally, the exact fit gives $k^0=3.76\times10^{-4}~{\rm m~s^{-1}}$ ($2k^0a/D=5.7$, $R_{\rm ct}=9.0\times10^{5}~\Omega$), which exceeds the nominal diffusion-controlled limit $k^0>8.1\times10^{-5}$ m s$^{-1}$ [Eq.~\eqref{eq:DL}] by a factor of 4.6. The steady-state current is 0.79 of Saito's value, so that the system lies in the broad crossover region (Sec.~\ref{sec:exact}), where neither limiting boundary condition is exact. Consistently, the exact fit shows that the uniform-concentration element underestimates $D$ by 4.8\% and the uniform-flux element overestimates it by 4.1\%, so that $D_{\rm uc}$ and $D_{\rm uf}$ bracket the exact value; the Randles decomposition itself has an error of up to 4.6\% in $|Z|$ at this $\mathrm{Da}$ (Fig.~\ref{fig:fd}). For a larger $\mathrm{Da}$ the uniform-concentration element approaches the exact result, and for a smaller $\mathrm{Da}$ the uniform-flux element does (Table~\ref{tab:ss}).

%%%%%%%%%%%%%%%%%%%%%%%%%%%%%%%%%%%%%%%
\section{Conclusion}
%%%%%%%%%%%%%%%%%%%%%%%%%%%%%%%%%%%%%%%
\label{sec:Conclusion}

We developed a compact analytical representation of the diffusion impedance at a disk electrode by converting the chronoamperometric disk response into the corresponding frequency-domain response function and constructing a Pad\'{e} approximant. The resulting expression preserves the finite low-frequency resistance associated with the steady-state disk current while connecting smoothly to the high-frequency Cottrell, or semi-infinite planar-diffusion, regime. Comparison with direct response-function expansions showed that the Pad\'{e} approximant remains well behaved across the transition region, whereas finite-order truncated series representations have a limited practical range of validity. A benchmark against a converged numerical solution (Appendix~C.2) gives errors in $|Z|$ below 0.1\% for $\omega a^2/D\le4$ and 0.6\% at 12, a phase error below $1.3^{\circ}$ at all frequencies, and a deficit approaching 7\% of the Cottrell amplitude at high frequency.

The formulation provides a geometry-induced finite Warburg-type element for microdisk electrodes. In contrast to conventional finite-length Warburg elements, whose finite resistance originates from an imposed diffusion-layer thickness,\cite{Jacobsen_95,Lasia_14} the finite diffusion resistance in the present theory arises from the disk geometry and the associated spreading diffusion field. Thus, the model gives a physically grounded impedance element for microdisk electrodes, rather than an empirical interpolation between limiting behaviors.

The analysis also clarifies the role of the electrode boundary condition. The uniform-concentration boundary condition, which is relevant to reversible systems with sufficiently fast interfacial kinetics, and the uniform-flux boundary condition both produce finite diffusion impedance responses arising from disk geometry. However, their Cole--Cole arcs, low-frequency intercepts, and Bode responses differ quantitatively because the imposed boundary constraint couples differently to the three-dimensional spreading diffusion field around a disk embedded in an insulating plane. Relative to the uniform-concentration response, the diffusion impedance is overestimated by $+11$\% in $|Z|$ near $\omega a^2/D\approx2$ under uniform-flux conditions (Fig.~\ref{fig:uf_uc}). This comparison shows that boundary conditions should be treated explicitly when interpreting disk-electrode impedance spectra.

The choice between the uniform-concentration element, $Z_{\rm W}(\omega)$, and the uniform-flux (Fleischmann--Pons) element, $Z_{\rm cf}(\omega)$, is not arbitrary: it is governed by the intrinsic rate constant relative to the disk radius and diffusion coefficients, with $Z_{\rm W}(\omega)$ appropriate in the diffusion-controlled limit and $Z_{\rm cf}(\omega)$ appropriate in the reaction-controlled limit, as made explicit by the criterion in Eqs.~\eqref{eq:DL} and \eqref{eq:RL}.

This distinction was illustrated against previously published experimental impedance data for a Pt microdisk electrode in a reversible ferri/ferrocyanide electrolyte. Once the electrolyte resistance was fixed at the measured high-frequency value, both the uniform-concentration (Pad\'{e}) and uniform-flux (Fleischmann--Pons) diffusion elements reproduced the digitized Cole--Cole plot to within the digitization precision, but only after adjusting the diffusion coefficient used in each fit. The two fitted diffusion coefficients differed by a factor ($1.094\pm0.005$) close to the theoretical ratio of the boundary conditions' zero-frequency diffusion resistances, \(16/(3\pi)/(\pi/2)\approx1.081\). The exact solution of the coupled problem, fitted to the same spectrum with the same number of free parameters, gives $D=6.65\times10^{-10}~{\rm m^2~s^{-1}}$ and $k^0=3.8\times10^{-4}~{\rm m~s^{-1}}$ ($2k^0a/D=5.7$), within 1\% of the value reported in the original study from simulations with the Robin boundary condition. The uniform-concentration and uniform-flux values lie 4.8\% below and 4.1\% above the exact one, as the dc diffusion resistances at this $\mathrm{Da}$ predict (Table~\ref{tab:ss}); in general the uniform-flux expression overestimates the diffusion coefficient of a reversible redox couple by up to about 8\% relative to the uniform-concentration element, the exact zero-frequency bound, with a smaller bias at finite $2k^0a/D$. The experimental part is a reanalysis of one published spectrum, digitized from the figure, and does not by itself establish the general applicability of either limiting element; it does show that the exact solution is consistent with the data without additional parameters.

Finally, incorporation of the Pad\'{e} diffusion impedance into a Randles-type equivalent circuit showed that, for representative microdisk-electrode parameters, the diffusion response can dominate over typical access resistance, charge-transfer resistance, and double-layer capacitance contributions over a broad frequency range. This result indicates that the present Pad\'{e} representation may be useful for estimating diffusion coefficients or effective electrode radii from impedance spectra, provided that the interfacial kinetics and capacitive contributions are either sufficiently small or properly included in the equivalent-circuit model. The decomposition of the faradaic impedance into $R_{\rm ct}$ and $Z_{\rm W}(\omega)$ is itself an approximation, accurate to about 5\% in $|Z|$ for $1\le2k^0a/D\le100$ (Fig.~\ref{fig:fd}), and the Pad\'{e} element underestimates the Cottrell amplitude by up to 7\% at high frequency (Appendix~C.2); the identifiability of the parameters should be checked for the frequency window used. The theory therefore offers a practical framework for equivalent-circuit modeling and quantitative interpretation of diffusion-controlled impedance at microdisk electrodes.

% Main text
%\section{}\label{}

% Use the style of numbering in square brackets.
% If nothing is used, default style will be taken.
%\begin{enumerate}[a)]
%\item
%\item
%\item
%\end{enumerate}

%\begin{itemize}
%\item
%\item
%\item
%\end{itemize}

%\begin{description}
%\item[]
%\item[]
%\item[]
%\end{description}

%\begin{table}%[]
%\caption{}\label{tbl1}
%\begin{tabular*}{\tblwidth}{@{}LL@{}}
%\toprule
% &  \\ % Table header row
%\midrule
% & \\
% & \\
% & \\
% & \\
%\bottomrule
%\end{tabular*}
%\end{table}

% Uncomment and use as the case may be
%\begin{theorem}
%\end{theorem}

% Uncomment and use as the case may be
%\begin{lemma}
%\end{lemma}

\renewcommand{\theequation}{A.\arabic{equation}}
\setcounter{equation}{0}  % reset counter
\section*{Appendix A. Series expansion of the response function}
%\section{Series expansion of the response function}
\label{app:series_expansion}

For reference and reproducibility, the response-function expansion used in this work is summarized here. Introducing \(x=sa^2/D\), the symbolic expansion is written explicitly up to order \(x^{13/2}\). The analytical coefficients beyond this order are too lengthy to be useful in the main printed form; therefore, the numerical coefficients are provided up to order \(x^{21/2}\), which was the highest order used in the comparison with the Pad\'{e} approximant.
Introducing the dimensionless Laplace variable
\begin{align}
x=\frac{sa^2}{D},
\end{align}
the normalized current response is expanded in powers of \(x^{1/2}\) as
\begin{multline}
\frac{s \hat{I}(s)}{ I_{\infty} } \approx
1+\frac{2}{\pi}x^{1/2}
+\left(\frac{4}{\pi ^2}-\frac{1}{3}\right)x
-\frac{8}{\pi}\left(\frac{1}{9}-\frac{1}{\pi^2}\right)x^{3/2}
+\left(\frac{16}{\pi^4}-\frac{20}{9\pi^2}+\frac{1}{15}
\right)x^2
\\
-\frac{1}{\pi} \left(\frac{-142}{675}
+\frac{16}{3 \pi^2}-\frac{32}{\pi^4}
\right)x^{5/2}
-\left(\frac{2}{315}-\frac{448}{675 \pi^2}
+\frac{112}{9\pi^4}-\frac{64}{\pi^6}
\right)x^3
\\
+\frac{1}{\pi} \left(
-\frac{1136}{33075}
+\frac{3872}{2025 \pi^2}
-\frac{256}{9 \pi^4}
+\frac{128}{\pi^6}
\right)x^{7/2}
\\
-\left(
-\frac{1}{2835}
+\frac{41428}{297675 \pi ^2}
-\frac{1168}{225 \pi ^4}
+\frac{64}{\pi ^6}
-\frac{256}{\pi ^8}
\right)x^4
\\
+\frac{1}{\pi}\left(
\frac{19046}{4465125}
-\frac{86336}{178605\pi^2}
+\frac{608}{45\pi^4}
-\frac{1280}{9\pi^6}
+\frac{512}{\pi^8}
\right)x^{9/2}
\\
+\left(
-\frac{2}{155925}
+\frac{505252}{22325625 \pi ^2}
-\frac{1363648}{893025 \pi ^4}
+\frac{68992}{2025 \pi ^6}
-\frac{2816}{9 \pi ^8}
+\frac{1024}{\pi ^{10}}
\right)x^5
\\
-\frac{1}{\pi}\left(
\frac{658136}{1620840375}
-\frac{1280408}{13395375 \pi^2}
+\frac{161408}{35721 \pi^4}
-\frac{56576}{675 \pi^6}
+\frac{2048}{3 \pi^8}
-\frac{2048}{\pi^{10}}
\right)x^{11/2}
\\
+\left(
\frac{2}{6081075}
-\frac{72782476}{24312605625 \pi ^2}
+\frac{70775744}{200930625 \pi ^4}
-\frac{3793088}{297675 \pi ^6}
+\frac{136448}{675 \pi ^8}
-\frac{13312}{9 \pi ^{10}}
+\frac{4096}{\pi ^{12}}
\right)x^6
\\
-\frac{1}{\pi}\left(
-\frac{57122836 }{1917454163625}
+\frac{400816}{25727625 \pi ^2}
-\frac{34023424}{28704375 \pi ^4}
+\frac{4417024}{127575 \pi ^6}
-\frac{971264}{2025 \pi ^8}
+\frac{28672}{9 \pi ^{10}}
-\frac{8192}{\pi ^{12}}
\right)x^{13/2}.
\label{eq:series_symbolic_appendix}
\end{multline}

The corresponding numerical form, retained up to order \(x^{21/2}\), is
\begin{multline}
\frac{s \hat{I}(s)}{ I_{\infty} } \approx
1
+0.63662 x^{1/2}
+0.0719514 x
-0.0249298 x^{3/2}
+0.0057642 x^2
-0.000476778 x^{5/2}
\\
-0.000286073 x^3
+0.000165659 x^{7/2}
-0.000046869 x^4
+5.86109\times10^{-6}x^{9/2}
+1.54037\times10^{-6}x^5
\\
-1.1752\times10^{-6}x^{11/2}
+3.7459\times10^{-7}x^6
-6.01208\times10^{-8}x^{13/2}
-6.45121\times10^{-9}x^7
\\
+7.99337\times10^{-9}x^{15/2}
-2.89631\times10^{-9}x^8
+5.59838\times10^{-10}x^{17/2}
+8.75713\times10^{-12}x^9
\\
-5.22477\times10^{-11}x^{19/2}
+2.18236\times10^{-11}x^{10}
-4.90934\times10^{-12}x^{21/2}.
\label{eq:series_numerical_appendix}
\end{multline}

Equations~\eqref{eq:series_symbolic_appendix} and
\eqref{eq:series_numerical_appendix} are included to allow direct use of the
series expansion and to make the construction and comparison of the Pad\'{e}
approximant reproducible.

%\renewcommand{\theequation}{A.\arabic{equation}}
%\setcounter{equation}{0}  % reset counter
%\section*{Appendix A. Derivation of Eq. (\ref{eq:T1})}
%\appendix
%\section{Derivation of Eq. (\ref{eq:T1})}
%\label{appA}

\renewcommand{\theequation}{B.\arabic{equation}}
\setcounter{equation}{0}  % reset counter for Appendix B
\renewcommand{\thetable}{B.\arabic{table}}
\setcounter{table}{0}     % reset counter for Appendix B

\section*{Appendix B. An alternative approximant matching the Cottrell
limit exactly, and why Eq.~\eqref{eq:Pade_dimensionless} is preferred}
\label{app:cottrell_variant}

\subsection*{B.1. The high-frequency limit of Eq.~\eqref{eq:Pade_dimensionless}}

In the main text, the Pad\'{e} approximant is described as interpolating
the response toward the high-frequency Cottrell limit. That statement is
asymptotic in form but not exact in magnitude. Here, we quantify it.

For semi-infinite planar diffusion the exact response function is
\begin{align}
\hat{\phi}_{\rm Cottrell}(x)=\frac{\pi}{4}x^{1/2},
\label{eq:appB_cottrell}
\end{align}
whereas Eq.~\eqref{eq:Pade_dimensionless} gives
\begin{align}
\hat{\phi}_{\rm Pade}(x)
\;\xrightarrow[x\to\infty]{}\;
\frac{\pi\left(\pi^2-8\right)}{8\left(\pi^2-9\right)}x^{1/2}
=0.844283\,x^{1/2},
\qquad
\frac{0.844283}{\pi/4}=1.07497 .
\label{eq:appB_hf}
\end{align}
The approximant therefore overestimates the limiting Cottrell response
function by $7.50\%$; equivalently, since $Z_{\rm norm}=1/\hat{\phi}$,
it underestimates the high-frequency impedance magnitude by $6.97\%$.
The phase is unaffected: both Eqs.~\eqref{eq:appB_cottrell}
and~\eqref{eq:appB_hf} are proportional to $x^{1/2}$, so on the
imaginary axis both approach the classical Warburg value of $45^{\circ}$
exactly, as seen in Fig.~\ref{fig:pade_cottrell_series}(b). The low-frequency limit is exact, $\hat{\phi}_{\rm Pade}(0)=1$,
which is the Saito steady-state disk current.

The origin of the discrepancy is a matter of counting. Writing the
approximant in the $[2/1]$ form
\begin{align}
\hat{\phi}(x)=\frac{A+B\,x^{1/2}+C\,x}{A'+B'\,x^{1/2}},
\label{eq:appB_general}
\end{align}
and setting $A=A'$ so that $\hat{\phi}(0)=1$, the expansion coefficients
of $\hat{\phi}=1+c_1x^{1/2}+c_2x+c_3x^{3/2}+\cdots$ are
\begin{align}
c_1=\frac{B-B'}{A},
\qquad
c_2=\frac{C}{A}-\frac{B B'}{A^2}+\frac{B'^2}{A^2},
\qquad
c_3=-\frac{B'}{A}c_2 ,
\label{eq:appB_coeffs}
\end{align}
while the high-frequency slope is $C/B'$. The relation between $c_2$ and
$c_3$ in Eq.~\eqref{eq:appB_coeffs} is structural, holding for any
$[2/1]$ form, so the approximant can satisfy exactly four independent
conditions, one of which is $A=A'$. Equation~\eqref{eq:Pade_dimensionless}
spends all four on the low-frequency series, reproducing $c_0=1$
together with $c_1$, $c_2$ and $c_3$ of Eq.~\eqref{eq:series_main}
exactly; the high-frequency slope is then
determined rather than imposed, and Eq.~\eqref{eq:appB_hf} is the value
it takes.

\subsection*{B.2. The alternative allocation}

The only other natural choice is to impose the Cottrell slope and
relinquish $c_3$. Requiring $A=A'$, $c_1=2/\pi$, $C/B'=\pi/4$ and
$c_2=(12-\pi^2)/(3\pi^2)$ fixes all four coefficients uniquely, and
after clearing common factors the result takes a form closely parallel
to Eq.~\eqref{eq:Pade_dimensionless}, with the roles of $(12-\pi^2)$ and
$(\pi^2-8)$ interchanged:
\begin{align}
\hat{\phi}_{\rm C}(x)
=
\frac{
3\pi\left(\pi^2-8\right)
+2\pi^2 x^{1/2}
+\pi\left(12-\pi^2\right)x
}{
3\pi\left(\pi^2-8\right)
+4\left(12-\pi^2\right)x^{1/2}
}.
\label{eq:appB_phiC}
\end{align}
Numerically the coefficients are $A=A'=17.620606$, $B=2\pi^2=19.739209$,
$C=6.692835$ and $B'=8.521582$. By construction
$\hat{\phi}_{\rm C}(0)=1$, the $x^{1/2}$ and $x$ terms of
Eq.~\eqref{eq:series_main} are reproduced exactly, and
$\hat{\phi}_{\rm C}(x)\to(\pi/4)x^{1/2}$ as $x\to\infty$. The price is
the $x^{3/2}$ coefficient, which becomes $-0.0347967$ against the exact
value $-0.0249298$, an overestimate of $40\%$.

\subsection*{B.3. Time-domain forms}

Both approximants can be inverted in closed form, which permits a direct
comparison against numerically exact chronoamperometric data. For any
approximant of the form Eq.~\eqref{eq:appB_general} with $A=A'$,
dividing out and inverting term by term gives
\begin{align}
\frac{I(t)}{I_{\infty}}
=
1
+\frac{C}{B'\sqrt{\pi}}\frac{a}{\sqrt{Dt}}
-\frac{R}{A}
\exp\!\left(c_{\rm p}^{2}\frac{Dt}{a^{2}}\right)
{\rm erfc}\!\left(c_{\rm p}\frac{\sqrt{Dt}}{a}\right),
\qquad
c_{\rm p}=\frac{A}{B'},
\quad
\frac{R}{A}=1-\frac{B}{B'}+\frac{CA}{B'^{2}},
\label{eq:appB_inversion}
\end{align}
where the leading constant is unity because $A=A'$, again the Saito
limit. Specializing Eq.~\eqref{eq:appB_inversion} to
Eq.~\eqref{eq:Pade_dimensionless} gives
\begin{align}
\frac{I(t)}{I_{\infty}}
=
1
+\frac{\sqrt{\pi}\left(\pi^{2}-8\right)}{8\left(\pi^{2}-9\right)}
\frac{a}{\sqrt{Dt}}
-\frac{3\left(12-\pi^{2}\right)^{3}}{64\left(\pi^{2}-9\right)^{2}}
\exp\!\left(c_{\rm pd}^{2}\frac{Dt}{a^{2}}\right)
{\rm erfc}\!\left(c_{\rm pd}\frac{\sqrt{Dt}}{a}\right),
\qquad
c_{\rm pd}=\frac{3\pi\left(12-\pi^{2}\right)}{8\left(\pi^{2}-9\right)},
\label{eq:appB_time_pade}
\end{align}
with $\sqrt{\pi}(\pi^{2}-8)/[8(\pi^{2}-9)]=0.476335$,
$3(12-\pi^{2})^{3}/[64(\pi^{2}-9)^{2}]=0.599347$ and
$c_{\rm pd}=2.886155$, while for Eq.~\eqref{eq:appB_phiC}
\begin{align}
\frac{I(t)}{I_{\infty}}
=
1
+\frac{\sqrt{\pi}}{4}\frac{a}{\sqrt{Dt}}
-\frac{3\left(\pi^{2}-8\right)^{2}}{16\left(12-\pi^{2}\right)}
\exp\!\left(c_{\rm C}^{2}\frac{Dt}{a^{2}}\right)
{\rm erfc}\!\left(c_{\rm C}\frac{\sqrt{Dt}}{a}\right),
\qquad
c_{\rm C}=\frac{3\pi\left(\pi^{2}-8\right)}{4\left(12-\pi^{2}\right)},
\label{eq:appB_time_C}
\end{align}
with $3(\pi^{2}-8)^{2}/[16(12-\pi^{2})]=0.307638$ and
$c_{\rm C}=2.067762$. The coefficient of $t^{-1/2}$ in
Eq.~\eqref{eq:appB_time_C} is $\sqrt{\pi}/4=0.443113$, the exact
Cottrell value; in Eq.~\eqref{eq:appB_time_pade} it is larger by the
same factor $1.07497$ appearing in Eq.~\eqref{eq:appB_hf}, so the
high-frequency excess of the response function and the short-time excess
of the relaxation current are the same statement.

\subsection*{B.4. Comparison with numerically exact values}

Table~\ref{tab:appB_bieniasz} compares Eqs.~\eqref{eq:appB_time_pade}
and~\eqref{eq:appB_time_C} with the high-accuracy numerical values of
Ref.~\cite{BIENIASZ_16}, tabulated over four decades in $Dt/a^{2}$. The
Shoup--Szabo interpolation formula~\cite{SHOUP_82} is included for
reference.

\begin{table}[h]
\centering
\caption{Relative deviation (\%) of the time-domain forms from the
numerically exact dimensionless current of Ref.~\cite{BIENIASZ_16}.
Entries marked $<10^{-4}$ fall below the third decimal place in percent.
The tabulated values of Ref.~\cite{BIENIASZ_16} are listed against
$\log_{10}(Dt/a^{2})$; they are given here in $Dt/a^{2}$ itself.}
\label{tab:appB_bieniasz}
\begin{tabular}{rrrrr}
\toprule
$Dt/a^{2}$ & $I/I_{\infty}$ (exact) &
Eq.~\eqref{eq:appB_time_pade} & Eq.~\eqref{eq:appB_time_C} &
Shoup--Szabo \\
\midrule
$0.01000$ & $5.237740$ & $+1.539$ & $-1.028$ & $-0.323$ \\
$0.01778$ & $4.136156$ & $+0.687$ & $-1.076$ & $-0.398$ \\
$0.03162$ & $3.313676$ & $-0.001$ & $-1.033$ & $-0.383$ \\
$0.05623$ & $2.701397$ & $-0.430$ & $-0.879$ & $-0.228$ \\
$0.10000$ & $2.247737$ & $-0.567$ & $-0.624$ & $+0.054$ \\
$0.17783$ & $1.913944$ & $-0.471$ & $-0.329$ & $+0.372$ \\
$0.31623$ & $1.670456$ & $-0.283$ & $-0.090$ & $+0.597$ \\
$0.56234$ & $1.493947$ & $-0.129$ & $+0.034$ & $+0.652$ \\
$1.0000$  & $1.365864$ & $-0.048$ & $+0.062$ & $+0.570$ \\
$1.7783$  & $1.272239$ & $-0.015$ & $+0.048$ & $+0.432$ \\
$3.1623$  & $1.203215$ & $-0.004$ & $+0.029$ & $+0.299$ \\
$5.6234$  & $1.151987$ & $-0.001$ & $+0.015$ & $+0.195$ \\
$10.00$   & $1.113803$ & $<10^{-4}$ & $+0.007$ & $+0.122$ \\
$17.78$   & $1.085267$ & $<10^{-4}$ & $+0.003$ & $+0.074$ \\
$31.62$   & $1.063911$ & $<10^{-4}$ & $+0.001$ & $+0.044$ \\
$56.23$   & $1.047913$ & $<10^{-4}$ & $+0.001$ & $+0.026$ \\
$100.0$   & $1.035924$ & $<10^{-4}$ & $<10^{-4}$ & $+0.015$ \\
$316.2$   & $1.020199$ & $<10^{-4}$ & $<10^{-4}$ & $+0.005$ \\
$1000$    & $1.011358$ & $<10^{-4}$ & $<10^{-4}$ & $+0.002$ \\
$3162$    & $1.006387$ & $<10^{-4}$ & $<10^{-4}$ & $+0.001$ \\
$10000$   & $1.003592$ & $<10^{-4}$ & $<10^{-4}$ & $<10^{-4}$ \\
\bottomrule
\end{tabular}
\end{table}

Equation~\eqref{eq:appB_time_pade} is essentially exact for
$Dt/a^{2}\gtrsim1$, remaining below $0.05\%$ and falling below $10^{-4}\%$
for $Dt/a^{2}\gtrsim10$, and it stays within $0.57\%$ for
$Dt/a^{2}\gtrsim0.03$. Its Cottrell excess becomes visible only below
$Dt/a^{2}\simeq0.03$, and even at $Dt/a^{2}=0.01$ it amounts to $1.5\%$
rather than the asymptotic $7.5\%$. Equation~\eqref{eq:appB_time_C},
although exact in both limits, deviates by about $1\%$ through the
intermediate region and is the less accurate of the two over the whole
range $Dt/a^{2}\gtrsim0.03$.

\subsection*{B.5. Choice of approximant for impedance analysis}

The two allocations therefore trade accuracy in the intermediate region
against exactness of the high-frequency asymptote, and the choice is
settled by the frequency range over which the diffusion impedance is
actually resolved.

Diffusion dominates the measured spectrum at low frequency. As noted in
Sec.~\ref{sec:Compex}, the diffusion coefficient is determined by
fitting in the range $a^{2}f/(2D)<1$, that is $\omega a^{2}/D\lesssim
4\pi$, which is also the transition region identified in
Fig.~\ref{fig:pade_cottrell_series} at $\omega a^{2}/D\simeq12$. For the
experimental parameters of Sec.~\ref{sec:Compex} ($a=5~\mu{\rm m}$,
$D_{\rm uc}=6.33\times10^{-10}~{\rm m^{2}~s^{-1}}$, hence
$a^{2}/D=39.5~{\rm ms}$) this corresponds to $f\lesssim50~{\rm Hz}$.
Within this window the two approximants can be compared directly
against the series expansion of Appendix~A, which is itself convergent
there. Evaluating Eq.~\eqref{eq:series_numerical_appendix} on the
imaginary axis, and using the change on truncating from $x^{21/2}$ to
$x^{19/2}$ as an internal convergence check (below $0.001\%$ for
$\omega a^{2}/D\le4$ and $0.2\%$ at $\omega a^{2}/D=8$, beyond which the
asymptotic series ceases to be usable), gives the deviations in
$\left|Z_{\rm norm}\right|$ collected in
Table~\ref{tab:appB_frequency}.

\begin{table}[h]
\centering
\caption{Deviation (\%) of $\left|Z_{\rm norm}\right|$ from the series
expansion of Appendix~A, evaluated at $x=i\omega a^{2}/D$, within the
frequency window where that series is convergent.}
\label{tab:appB_frequency}
\begin{tabular}{rrr}
\toprule
$\omega a^{2}/D$ & Eq.~\eqref{eq:Pade_dimensionless} &
Eq.~\eqref{eq:appB_phiC} \\
\midrule
$1$ & $+0.064$ & $+0.060$ \\
$2$ & $+0.110$ & $+0.248$ \\
$4$ & $+0.074$ & $+0.576$ \\
$8$ & $-0.143$ & $+1.018$ \\
\bottomrule
\end{tabular}
\end{table}

Equation~\eqref{eq:Pade_dimensionless} remains within $0.15\%$ across
this window, whereas Eq.~\eqref{eq:appB_phiC} drifts to about $1\%$,
consistent with the former reproducing Eq.~\eqref{eq:series_main} one
order further. The same ordering appears in the time domain: the
frequency window corresponds to $Dt/a^{2}\gtrsim0.08$, which is
precisely the region in which Table~\ref{tab:appB_bieniasz} shows
Eq.~\eqref{eq:Pade_dimensionless} to be the more accurate of the two,
and in which it becomes rapidly exact as the frequency is lowered
further toward the diffusion-resistance plateau.

At the opposite end, the two approximants differ in
$\left|Z_{\rm norm}\right|$ by $1.7\%$ at $\omega a^{2}/D=12$, $4.5\%$ at
$10^{2}$ and $6.1\%$ at $10^{3}$, approaching $6.97\%$ asymptotically.
However, at those frequencies the diffusion impedance is small compared
with the electrolyte access resistance $R_{\Omega}$ and is shunted by
the double-layer capacitance $C_{\rm dl}$, so it contributes little to
the measured spectrum and still less to the fitted value of $D$; the
two approximants also differ by at most $1.1^{\circ}$ in phase, at
$\omega a^{2}/D\simeq29$, and share the same exact $45^{\circ}$
high-frequency asymptote. The high-frequency
exactness of Eq.~\eqref{eq:appB_phiC} is thus of limited practical
benefit, while its intermediate-region deficit falls in the range that
matters.

The consequence for the analysis of Sec.~\ref{sec:Compex} is that the
diffusion coefficient obtained there is insensitive to this choice. A
deviation of $0.15\%$ in $\left|Z_{\rm W}\right|$ over the fitting
window propagates to a comparable relative shift in the fitted
$D_{\rm uc}$, since $Z_{\rm W}\propto1/D$ in the low-frequency limit.
That is nearly
two orders of magnitude below the $8\%$ discrepancy that distinguishes
the uniform-concentration and uniform-flux boundary conditions. The
conclusions of Sec.~\ref{sec:Compex} therefore do not depend on which
member of the $[2/1]$ family is used, and would be unchanged if
Eq.~\eqref{eq:appB_phiC} were substituted throughout.

Equation~\eqref{eq:Pade_dimensionless} is retained throughout this work
for the reasons given above. Appendix~C shows, using a converged numerical solution, that the alternative member is the more uniform over $10^{-2}\le\omega a^2/D\le10^{3}$ (maximum error of $|Z|$ of 1.1\% against 5.9\%), whereas Eq.~\eqref{eq:Pade_dimensionless} is the more accurate in $|Z|$ up to $\omega a^2/D\approx17$. Equation~\eqref{eq:appB_phiC} is recorded here because
it makes the structure of the approximation explicit: within the $[2/1]$
family the Cottrell limit and the $x^{3/2}$ term of the low-frequency
expansion cannot both be satisfied, and Eq.~\eqref{eq:Pade_dimensionless}
represents the allocation appropriate to the diffusion-controlled
frequency window.

\begingroup
\renewcommand{\theequation}{C.\arabic{equation}}
\setcounter{equation}{0}
\renewcommand{\thetable}{C.\arabic{table}}
\setcounter{table}{0}
\section*{Appendix C. Benchmark of the Pad\'{e} impedance against a converged numerical solution}

Appendix~B examined Eq.~\eqref{eq:Pade_dimensionless} against a series expansion and against time-domain values. Here the complex impedance itself is compared with an independent, converged solution of the uniform-concentration disk problem in the frequency domain.

\subsection*{C.1. Reference solution}

We use units $a=D=1$, with $x=sa^2/D=i\Omega$ and $\Omega=\omega a^2/D$. For the uniform-concentration (Dirichlet) condition on the disk, the response function is $\hat\phi(x)=x\int_0^1\hat h(y,x)\,dy$, where $\hat h$ obeys a Fredholm integral equation of the second kind (Eqs.~(36) and (37) of Ref.~\cite{SEKI_26}; the same structure is found in Ref.~\cite{Duffy_08}),
\begin{align}
\hat h(r,x)=\frac1x+\int_0^1\hat h(y,x)\,\hat K(r,y,x)\,dy,
\qquad 0\le r\le1,
\label{eq:C_fred}
\end{align}
with the kernel
\begin{align}
\hat K(r,y,x)=-\frac2\pi\int_0^\infty\left[\frac{k}{\sqrt{k^2+x}}-1\right]\cos(ky)\cos(kr)\,dk .
\label{eq:C_K}
\end{align}
Because the bracket decays as $k^{-2}$, the kernel is bounded; it has a kink at $y=r$. The integral can be evaluated in closed form, $\hat K=-[F(|r-y|)+F(r+y)]/\pi$ with $F(\xi)=\sqrt{x}\,\{(\pi/2)[I_1(\sqrt{x}\,\xi)-L_1(\sqrt{x}\,\xi)]-1\}$ and $F(0)=-\sqrt{x}$, where $I_1$ and $L_1$ are the modified Bessel and the modified Struve function of order one \cite{SEKI_26}.

We discretize Eq.~\eqref{eq:C_fred} directly with the trapezoidal rule on $N+1$ equidistant nodes $r_i=i/N$, without basis functions,
\begin{align}
\hat h_i=\frac1x+\sum_{j=0}^{N}w_j\,\hat K(r_i,r_j,x)\,\hat h_j,
\qquad
\hat\phi=x\sum_{j=0}^{N}w_j\,\hat h_j,
\end{align}
with $w_j=1/N$ ($1/(2N)$ for $j=0$ and $N$), which is a dense complex linear system of size $N+1$. The matrix elements $\hat K(r_i,r_j,x)=-[F(|i-j|/N)+F((i+j)/N)]/\pi$ require $F$ only at the $2N+1$ values $\xi=m/N$, $m=0,\ldots,2N$. For large $|z|=|\sqrt{x}\,\xi|$ the functions $I_1$ and $L_1$ are exponentially large and nearly cancel in $I_1-L_1$, so that $F$ is evaluated in arbitrary-precision arithmetic with $40+|z|_{\max}/2$ digits, $|z|_{\max}=2|\sqrt{x}|$. The kink of the kernel limits the error to $O(N^{-2})$, and the Richardson extrapolation $\hat\phi\approx(4\hat\phi_{2N}-\hat\phi_N)/3$ removes the leading term. Table~\ref{tab:C_conv} gives the relative error with respect to the independent basis-function solution described below: it decreases by a factor of four for each doubling of $N$, and the extrapolation reaches $10^{-10}$ at $\Omega=0.1$, $10^{-9}$ at $\Omega=1$, $2\times10^{-8}$ at $\Omega=12$, $4\times10^{-7}$ at $\Omega=100$, and $8\times10^{-7}$ at $\Omega=10^3$ (the extrapolation uses $N=200$ and 400 for $\Omega\le12$, $N=400$ and 800 for $\Omega=30$ and 100, and $N=800$ and 1600 for $\Omega=10^3$). At higher frequencies the boundary layer, of thickness proportional to $\Omega^{-1/2}$, must be resolved, so that the trapezoidal solution becomes impractical above $\Omega=10^3$. The errors of Table~\ref{tab:C_err} computed with the extrapolated trapezoidal solution coincide with the listed ones to within 0.001 percentage points.

\begin{table}[h]
\centering
\caption{Relative error $|\hat\phi_N-\hat\phi_{\rm Gal}|/|\hat\phi_{\rm Gal}|$ of the trapezoidal solution $\hat\phi_N$ of Eq.~\eqref{eq:C_fred} with $N+1$ nodes, with respect to the independent basis-function (Galerkin) solution $\hat\phi_{\rm Gal}$ of Appendix~D.4.2; the last column is the Richardson extrapolation from $N=200$ and 400.}
\label{tab:C_conv}
\begin{tabular}{rccccc}
\toprule
$\Omega$ & $N=50$ & $N=100$ & $N=200$ & $N=400$ & extrapolated\\
\midrule
0.1 & $1.3\times10^{-6}$ & $3.3\times10^{-7}$ & $8.3\times10^{-8}$ & $2.1\times10^{-8}$ & $1\times10^{-10}$\\
1   & $1.5\times10^{-5}$ & $3.7\times10^{-6}$ & $9.2\times10^{-7}$ & $2.3\times10^{-7}$ & $1\times10^{-9}$\\
12  & $6.3\times10^{-4}$ & $1.6\times10^{-4}$ & $3.9\times10^{-5}$ & $9.8\times10^{-6}$ & $2\times10^{-8}$\\
100 & $1.5\times10^{-2}$ & $3.7\times10^{-3}$ & $9.2\times10^{-4}$ & $2.3\times10^{-4}$ & $4\times10^{-7}$\\
\bottomrule
\end{tabular}
\end{table}

\textbf{Independent check.} The same response function was also computed by a second, independent method. The flux density on the disk is expanded in a small number $N$ of basis functions, which turns the integral equation into an $N\times N$ matrix equation; the formulas and the numerical details are given in Appendix~D.4.2, where the method is used for the coupled problem. It converges to nine digits or better with $N=20$--40 functions for $\Omega\le100$ and with $N=80$ for $\Omega=10^3$. It reproduces the series of Appendix~A at small $x$, the Cottrell limit $(\pi/4)x^{1/2}$ at large $x$, and the deviations of $|Z_{\rm norm}|$ in Table~\ref{tab:appB_frequency} (0.064, 0.110, and 0.074\%). The reference of Table~\ref{tab:C_err} and Fig.~\ref{fig:C_err} is this solution; its real and imaginary parts are listed in Table~\ref{tab:C_ref}.

\subsection*{C.2. Errors of the Pad\'{e} impedance}

Table~\ref{tab:C_err} gives the errors in the modulus and phase of $Z_{\rm norm}=1/\hat\phi_{\rm Pade}(i\Omega)$, Eq.~\eqref{eq:Pade_dimensionless}, and of the alternative member of the $[2/1]$ family, Eq.~\eqref{eq:appB_phiC} (Appendix~C.4), relative to the reference, and Fig.~\ref{fig:C_err} shows them as functions of frequency. The modulus is accurate to 0.11\% for $\Omega\le4$ and to 0.62\% at the crossover value $\Omega=12$. The error then grows to 3.8\% at $\Omega=100$ and approaches the asymptotic value $-6.97\%$ of Appendix~B ($-5.9\%$ at $\Omega=10^3$). The phase error does not exceed $1.22^{\circ}$ at any frequency and is $0.9^{\circ}$ at $\Omega=12$. The curves of Fig.~\ref{fig:C_err} were computed at 61 logarithmically spaced frequencies $10^{-2}\le\Omega\le10^3$, with the basis-function solution as reference ($N=40$ functions for $\Omega\le100$ and 80 above).

\begin{table}[h]
\centering
\caption{Errors of the Pad\'{e} impedance, $Z_{\rm norm}=1/\hat\phi_{\rm Pade}(i\Omega)$, Eq.~\eqref{eq:Pade_dimensionless}, and of the alternative member of the $[2/1]$ family, Eq.~\eqref{eq:appB_phiC}, relative to the converged reference: relative error of the modulus and error of the phase. Fig.~\ref{fig:C_err} shows these errors as functions of frequency.}
\label{tab:C_err}
\begin{tabular}{rrrrr}
\toprule
 & \multicolumn{2}{c}{Eq.~\eqref{eq:Pade_dimensionless}} & \multicolumn{2}{c}{Eq.~\eqref{eq:appB_phiC}}\\
\cmidrule(lr){2-3}\cmidrule(lr){4-5}
$\omega a^2/D$ & $\Delta|Z|/|Z|$ (\%) & $\Delta\varphi$ ($^\circ$) & $\Delta|Z|/|Z|$ (\%) & $\Delta\varphi$ ($^\circ$)\\
\midrule
0.1  & $+0.002$ & $-0.000$ & $-0.009$ & $+0.011$\\
1    & $+0.064$ & $-0.044$ & $+0.060$ & $+0.164$\\
4    & $+0.074$ & $-0.348$ & $+0.576$ & $+0.304$\\
12   & $-0.620$ & $-0.881$ & $+1.046$ & $+0.140$\\
30   & $-1.921$ & $-1.188$ & $+1.040$ & $-0.066$\\
100  & $-3.772$ & $-1.139$ & $+0.733$ & $-0.178$\\
1000 & $-5.888$ & $-0.557$ & $+0.261$ & $-0.116$\\
\bottomrule
\end{tabular}
\end{table}

\begin{figure}[h]
\centering
\includegraphics[width=0.95\linewidth]{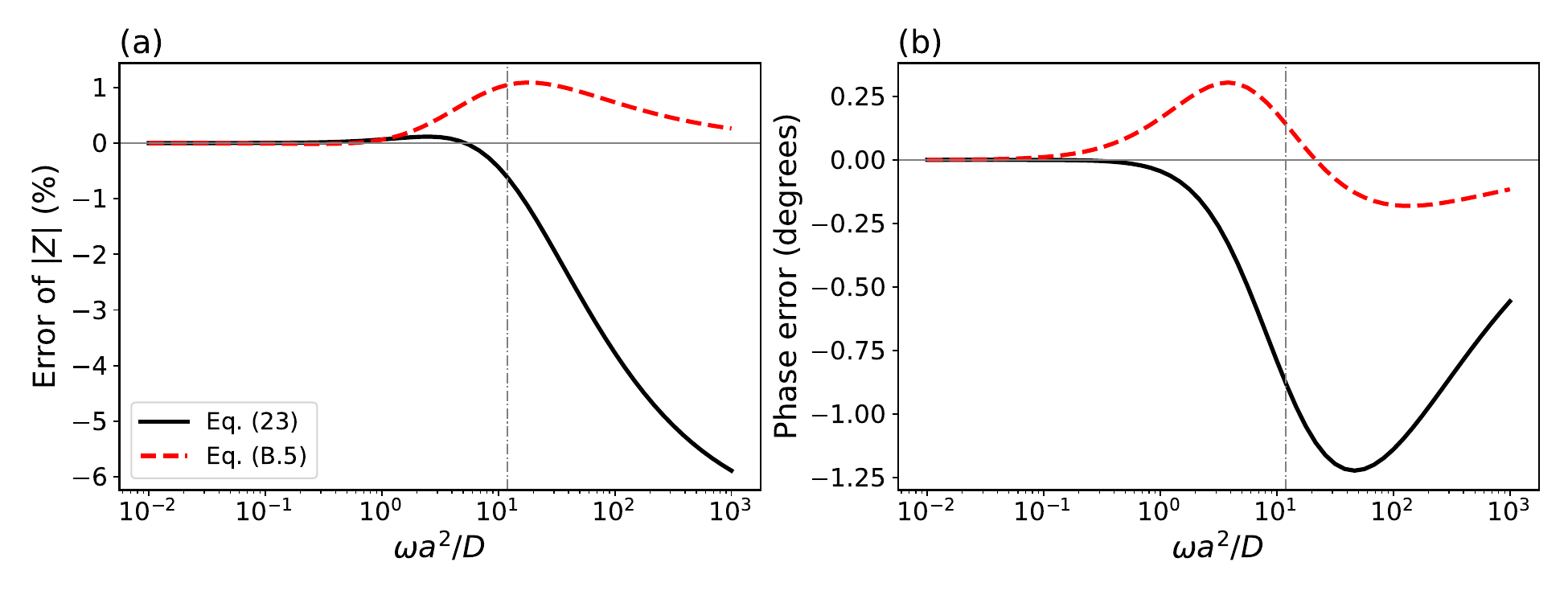}
\caption{Errors of the Pad\'{e} impedance $Z_{\rm norm}=1/\hat\phi_{\rm Pade}(i\Omega)$, Eq.~\eqref{eq:Pade_dimensionless} (solid), and of the alternative member of the $[2/1]$ family, Eq.~\eqref{eq:appB_phiC} (dashed), relative to the converged uniform-concentration solution. (a) Modulus $|Z|$. (b) Phase. The dash-dotted line marks $\omega a^2/D=12$.}
\label{fig:C_err}
\end{figure}

\subsection*{C.3. Poles, zeros, and passivity}

As polynomials in $x^{1/2}$, the numerator and the denominator of Eq.~\eqref{eq:Pade_dimensionless} have only positive coefficients. By the Routh--Hurwitz criterion no root lies in $\mathrm{Re}\,x^{1/2}\ge0$. The numerator roots are $x^{1/2}=-1.680\pm0.771i$ and the denominator root is $x^{1/2}=-2.886$. All of them are outside the physical sheet $|\arg x|<\pi$, i.e., $\mathrm{Re}\,x^{1/2}>0$, so that $Z_{\rm W}$ has neither poles nor zeros there. Moreover $\mathrm{Re}\,Z_{\rm norm}>0$ throughout $\mathrm{Re}\,x^{1/2}>0$, so that the element is passive and cannot generate spurious resonances in circuit fitting.

\subsection*{C.4. The alternative member of the $[2/1]$ family}

The member of Appendix~B that matches the Cottrell limit, Eq.~\eqref{eq:appB_phiC}, has a maximum error of $|Z|$ of 1.09\% and a maximum phase error of $0.30^{\circ}$ over $10^{-2}\le\Omega\le10^{3}$, against 5.9\% and $1.2^{\circ}$ for Eq.~\eqref{eq:Pade_dimensionless} (Fig.~\ref{fig:C_err} and Table~\ref{tab:C_err}). It is therefore the more uniform of the two. Eq.~\eqref{eq:Pade_dimensionless} is the more accurate in $|Z|$ up to $\Omega\approx17$, for example $+0.07\%$ against $+0.58\%$ at $\Omega=4$ (Table~\ref{tab:C_err}) and $-0.24\%$ against $+0.92\%$ at $\Omega=8$, but its error then grows to $-3.8\%$ at $\Omega=100$, against $+0.7\%$ for Eq.~\eqref{eq:appB_phiC}. In the phase, Eq.~\eqref{eq:Pade_dimensionless} is the more accurate below $\Omega\approx3.5$ and Eq.~\eqref{eq:appB_phiC} above it. The choice depends on the fitting window: for $\Omega\lesssim8$ (below about 30~Hz for $a=5~\mu$m and $D=6.3\times10^{-10}~{\rm m^2~s^{-1}}$) the error of $|Z|$ is below 0.25\% for Eq.~\eqref{eq:Pade_dimensionless} against up to 0.9\% for the alternative member, whereas the phase error is up to $0.7^{\circ}$ against $0.3^{\circ}$, so that Eq.~\eqref{eq:Pade_dimensionless} is preferable for the determination of $D$ from the modulus, and for wide-band fits the alternative member is preferable.

\subsection*{C.5. Comparison with the uniform-flux element}

The comparison of the Fleischmann--Pons impedance [Eq.~\eqref{eq:FP}] with the converged uniform-concentration impedance is shown in Fig.~\ref{fig:uf_uc} and discussed after Fig.~\ref{fig:bc_comparison}. The integrals $\Phi_4$ and $\Phi_5$ of Eqs.~\eqref{eq:Phi4} and \eqref{eq:Phi5} were evaluated by Simpson's rule on $0\le\beta\le1000$ with step $\min[0.004,\,0.02/(a\sqrt{\omega/D})]$ and the contribution of $\beta>1000$ to $\Phi_4$ was added analytically as $1/(2\pi a\sqrt{\omega/D}\,\beta_{\max}^2)$, $\beta_{\max}=1000$; refining the step and $\beta_{\max}$ (to 3000) changes the results by less than $10^{-6}$ in relative terms. The curves of Fig.~\ref{fig:uf_uc} were computed at the same 61 frequencies as Fig.~\ref{fig:C_err}, with the same converged uniform-concentration solution of Appendix~C.1 as reference. Table~\ref{tab:C_ref} lists the real and imaginary parts of both responses at selected frequencies.

Figure~\ref{fig:uf_uc} and Table~\ref{tab:C_ref} refer to equal diffusion coefficients. At zero frequency the ratio of the two responses is $32/(3\pi^2)$ for any $D_{\rm O}$, $D_{\rm R}$, $C_{\rm O}^{*}$, and $C_{\rm R}^{*}$. This follows from the general forms of Eqs.~\eqref{eq:AdSaito} and \eqref{eq:AdFr}. At zero frequency the linearized Nernst condition makes both surface concentrations uniform, so that each species solves a Dirichlet problem whose steady-state flux is Saito's, $G_j(0)=4aD_jC_j^{*}$; flux continuity for the total current then gives the series combination of Eq.~\eqref{eq:zw_reversible_general} exactly, and
\begin{align}
Y_{\infty,{\rm rev}}=\frac{4n^2F^2a}{RT}\left[\frac{1}{D_{\rm O}C_{\rm O}^{*}}+\frac{1}{D_{\rm R}C_{\rm R}^{*}}\right]^{-1},
\qquad
Y_{\rm FP}=\frac{3\pi^2}{8}\frac{n^2F^2a}{RT}\left[\frac{1}{D_{\rm O}C_{\rm O}^{*}}+\frac{1}{D_{\rm R}C_{\rm R}^{*}}\right]^{-1}.
\label{eq:C_general}
\end{align}
For $D_{\rm O}=D_{\rm R}=D$ and $C_{\rm O}^{*}=C_{\rm R}^{*}=C_b$ these reduce to Eqs.~\eqref{eq:AdSaito} and \eqref{eq:AdFr}, and their ratio is $32/(3\pi^2)$ whatever the diffusion coefficients and the bulk concentrations, because the bracket is common to both. The statement is therefore exact. At finite frequency no corresponding statement is available, because Eq.~\eqref{eq:FP} is derived for a single diffusion coefficient: combining the two species in series, each at its own dimensionless frequency $\omega a^2/D_j$, is itself an approximation of the same kind as Eq.~\eqref{eq:zw_reversible_pade}, so that the resulting ratio would compare two approximations rather than exact responses. The finite-frequency comparison of the two boundary conditions is therefore restricted here to equal diffusion coefficients, for which the uniform-concentration reference of Appendix~C.1 is exact.

\begin{table}[h]
\centering
\caption{Reference values of the normalized impedance $Z_{\rm norm}=Z/R_\infty$ ($R_\infty$ is the Saito resistance) as a function of $\omega a^2/D$: converged uniform-concentration response, $Z_{\rm norm}=1/\hat\phi(i\Omega)$ (Appendix~C.1), and uniform-flux (Fleischmann--Pons) response, Eq.~\eqref{eq:FP}. All digits shown are converged.}
\label{tab:C_ref}
\begin{tabular}{rrrrr}
\toprule
 & \multicolumn{2}{c}{Uniform concentration} & \multicolumn{2}{c}{Uniform flux}\\
\cmidrule(lr){2-3}\cmidrule(lr){4-5}
$\omega a^2/D$ & $\mathrm{Re}\,Z_{\rm norm}$ & $-\mathrm{Im}\,Z_{\rm norm}$ & $\mathrm{Re}\,Z_{\rm norm}$ & $-\mathrm{Im}\,Z_{\rm norm}$\\
\midrule
0.01 & 0.955080 & 0.041782 & 1.035815 & 0.042209\\
0.1 & 0.860380 & 0.112166 & 0.940470 & 0.115887\\
1 & 0.610943 & 0.210965 & 0.678704 & 0.232517\\
4 & 0.393544 & 0.212932 & 0.430351 & 0.244871\\
12 & 0.246926 & 0.169866 & 0.258654 & 0.192070\\
30 & 0.160793 & 0.126030 & 0.164149 & 0.137354\\
100 & 0.089404 & 0.077974 & 0.090011 & 0.081926\\
1000 & 0.028450 & 0.027219 & 0.028470 & 0.027660\\
\bottomrule
\end{tabular}
\end{table}

\renewcommand{\theequation}{D.\arabic{equation}}
\setcounter{equation}{0}
\section*{Appendix D. Exact solution of the coupled problem}

This appendix describes the numerical solution of the coupled problem with the boundary conditions of Eqs.~\eqref{eq:DNO1} and \eqref{eq:DNR1}, which is used in the Results to test the decomposition $Z_{\rm F}=R_{\rm ct}+Z_{\rm W}$ and Eq.~\eqref{eq:zw_reversible_pade}.

\subsection*{D.1. Equal diffusion coefficients: reduction to one species}

For $D_{\rm O}=D_{\rm R}=D$, flux continuity on the disk, $D\,\partial_z\delta C_{\rm O}=-D\,\partial_z\delta C_{\rm R}$, together with zero flux for $r>a$, means that the sum $w=\delta C_{\rm O}+\delta C_{\rm R}$ obeys the same diffusion equation as $\delta C_{\rm O}$ and $\delta C_{\rm R}$, has $\partial_z w=0$ on the whole plane $z=0$, and decays at infinity. Hence $w\equiv0$ in the whole diffusion field, $\delta C_{\rm R}(r,z,t)=-\delta C_{\rm O}(r,z,t)$, for any concentrations $C_{\rm O}^*$ and $C_{\rm R}^*$. The functions $\delta C_{\rm O}(t)$ and $\delta C_{\rm R}(t)$ of Eqs.~\eqref{eq:DNO1} and \eqref{eq:DNR1} are auxiliary quantities. They are defined as the uniform surface concentrations of the limiting case of ideal local equilibrium, in which the Nernst relation, Eq.~\eqref{eq:ECOR1}, holds at every point of the disk (infinitely fast kinetics) and the surface concentrations equal $\delta C_{\rm O}(r,0,t)=\delta C_{\rm O}(t)$ and $\delta C_{\rm R}(r,0,t)=\delta C_{\rm R}(t)$ for $r\le a$, the situation of Eqs.~\eqref{eq:responseO} and \eqref{eq:responseR}. In this limit the same current is carried by both species, $\delta I=\delta I_{\rm O}=-\delta I_{\rm R}$ (current continuity), and Eqs.~\eqref{eq:responseO} and \eqref{eq:responseR} give $\delta C_{\rm O}(t)/C_{\rm O}^*=\delta I/(nF G_{\rm O})$ and $\delta C_{\rm R}(t)/C_{\rm R}^*=-\delta I/(nF G_{\rm R})$. For $D_{\rm O}=D_{\rm R}$ the response functions are $G_j=4aDC_j^*\hat\phi$ with the same $\hat\phi$, so that $G_{\rm O}/C_{\rm O}^*=G_{\rm R}/C_{\rm R}^*$ and therefore $\delta C_{\rm R}(t)=-\delta C_{\rm O}(t)$. This relation follows from current continuity under ideal local equilibrium and is not an additional assumption. Equation~\eqref{eq:ECOR1}, the Nernst relation, fixes only the combination $\delta C_{\rm O}(t)/C_{\rm O}^*-\delta C_{\rm R}(t)/C_{\rm R}^*$, and the individual values come from current continuity. For finite kinetics the actual surface concentrations $\delta C_{\rm O}(r,0,t)$ and $\delta C_{\rm R}(r,0,t)$ differ from these auxiliary values and are not uniform. Their differences from $\delta C_{\rm O}(t)$ and $\delta C_{\rm R}(t)$ drive the fluxes in Eqs.~\eqref{eq:DNO1} and \eqref{eq:DNR1} and modify the response functions $G_j$, but Eqs.~\eqref{eq:responseO} and \eqref{eq:responseR} remain valid with the modified $G_j$ (for the coupled problem below, $\hat\phi$ is replaced by $\psi$). Hence
\begin{align}
\delta C_{\rm R}(r,z,t)=-\delta C_{\rm O}(r,z,t),
\qquad
\delta C_{\rm O}(t)=-\delta C_{\rm R}(t)=\frac{nF}{RT}\,\delta E(t)\,\frac{C_{\rm O}^*C_{\rm R}^*}{C_{\rm O}^*+C_{\rm R}^*},
\label{eq:D_dCR}
\end{align}
where Eq.~\eqref{eq:ECOR1} was used for the last equality. The flux continuity in Eqs.~\eqref{eq:DNO} and \eqref{eq:DNR} requires $\bar k_{\rm red}C_{\rm O}^*=\bar k_{\rm ox}C_{\rm R}^*$. Substituting Eq.~\eqref{eq:D_dCR} into Eq.~\eqref{eq:DNO1} gives Eq.~\eqref{eq:D_robin} of the main text;
this is a Robin condition for a single species: the driving force of the diffusive flux into the electrode is the deviation of the surface concentration from its local-equilibrium value $\delta C_{\rm O}(t)$, which is set by the electrode potential. For $C_{\rm O}^*=C_{\rm R}^*$ and $\bar k_{\rm ox}=\bar k_{\rm red}=k^0$ it has $\kappa=2k^0$. The Damk\"ohler number is $\mathrm{Da}=\kappa a/D$, equal to $2k^0a/D$ in the symmetric case. The reduction holds at any frequency and, through Eq.~\eqref{eq:D_dCR}, at any electrode potential.

\subsection*{D.2. Zero-frequency relation}

At zero frequency the linear response is the steady-state response to a small potential step. For the symmetric couple, $C_{\rm O}^*=C_{\rm R}^*=C_b$ (the bulk concentration of each species), the factor $C_{\rm O}^*C_{\rm R}^*/(C_{\rm O}^*+C_{\rm R}^*)$ in Eq.~\eqref{eq:D_dCR} equals $C_b/2$, so that Eq.~\eqref{eq:D_dCR} gives $\delta C_{\rm O}(t)=nFC_b\,\delta E/(2RT)$, and Eq.~\eqref{eq:D_robin} then gives $\delta I=\chi(\mathrm{Da})\,Y_{\infty,{\rm rev}}\,\delta E$, with $\chi=I/I_{\rm S}$ the steady-state current in units of Saito's current and $Y_{\infty,{\rm rev}}$ of Eq.~\eqref{eq:AdSaito}. With $R_\infty=1/Y_{\infty,{\rm rev}}$ and $R_{\rm ct}/R_\infty=4/(\pi\mathrm{Da})$ [from Eq.~\eqref{eq:Rct}, $R_{\rm ct}Y_{\infty,{\rm rev}}=2D/(\pi k^0a)$], this gives Eq.~\eqref{eq:Rdc}. The function $\chi(\mathrm{Da})=\psi(\mathrm{Da},0)$ is the solution of D.3 at $x=0$, where $\sqrt{k^2+x}=k$.

\subsection*{D.3. Coupled problem at finite frequency}

\textbf{D.3.1 The problem.} We use dimensionless units: lengths in units of $a$, time in units of $a^2/D$, and the Laplace variable $x=sa^2/D=i\Omega$ with $\Omega=\omega a^2/D$. For a small sinusoidal perturbation we describe the species of D.1 by $c(r,z)=\delta C_{\rm O}(r,z,t)/\delta C_{\rm O}(t)$, the concentration deviation in units of the local equilibrium deviation $\delta C_{\rm O}(t)$ of Eq.~\eqref{eq:D_dCR}; thus $c=1$ corresponds to ideal local equilibrium and $c\to0$ far from the electrode. The function $c$ obeys $(\nabla^2-x)c=0$ in $z>0$. We denote by $u(r)=c(r,0)$ the surface concentration and by $q(r)=-\partial_zc|_{z=0}$ the flux density into the electrode, in units of $D|\delta C_{\rm O}(t)|/a$. At the plane $z=0$ the flux vanishes outside the disk, $q(r)=0$ for $r>1$, and on the disk the Robin condition, Eq.~\eqref{eq:D_robin}, gives
\begin{align}
q(r)=\mathrm{Da}\,[1-u(r)],\qquad r<1 .
\label{eq:D_robin_dl}
\end{align}
The current flowing over the disk in units of Saito's current, $I_{\rm S}=4nFDa|\delta C_{\rm O}(t)|$, is $\psi(\mathrm{Da},x)=\delta I/I_{\rm S}$,
\begin{align}
\psi(\mathrm{Da},x)=\frac{1}{4} \int_0^1 2\pi\, r\, dr\, q(r),
\label{eq:psi}
\end{align}
which is the quantity that enters Eq.~\eqref{eq:ZF}:
it is the current amplitude in units of the current that the same perturbation would drive at a disk with uniform surface concentration at zero frequency.

\textbf{D.3.2 Numerical method.} The solution of $(\nabla^2-x)c=0$ in $z>0$ that vanishes far from the electrode is \cite{Duffy_08}
\begin{align}
c(r,z)=\int_0^\infty A(k)\,J_0(kr)\,e^{-z\sqrt{k^2+x}}\,dk ,
\label{eq:D_rep}
\end{align}
which gives at $z=0$ the surface concentration and the flux density into the electrode,
\begin{align}
u(r)=\int_0^\infty A(k)\,J_0(kr)\,dk,
\qquad
q(r)=\int_0^\infty\sqrt{k^2+x}\,A(k)\,J_0(kr)\,dk ,
\end{align}
with $q(r)=0$ for $r>1$. The second relation says that $\tilde q(k)=\sqrt{k^2+x}\,A(k)/k$ is the Hankel transform of order zero of $q$,
$q(r)=\int_0^\infty\tilde q(k)\,J_0(kr)\,k\,dk$ and $\tilde q(k)=\int_0^1q(r)J_0(kr)\,r\,dr$ [the transform pair is $f(r)=\int_0^\infty\tilde f(k)J_0(kr)\,k\,dk$ and $\tilde f(k)=\int_0^\infty f(r)J_0(kr)\,r\,dr$]. Hence $A(k)=k\tilde q(k)/\sqrt{k^2+x}$, and
\begin{align}
u(r)=\int_0^\infty\frac{\tilde q(k)\,J_0(kr)\,k\,dk}{\sqrt{k^2+x}} .
\label{eq:D_u}
\end{align}
Equation~\eqref{eq:D_robin_dl} therefore becomes an integral equation for $q$ on the disk, $q(r)/\mathrm{Da}+u(r)=1$.
We solve it by expanding the flux density in $N$ basis functions,
\begin{align}
q(r)=\sum_{n=0}^{N-1}b_n\,g_n(r),\qquad g_n(r)=P_{2n+1}(y),\quad y=\sqrt{1-r^2}\ \mbox{ for } r\leq 1,
\label{eq:D_basis}
\end{align}
where $P_l(y)$ are the Legendre polynomials \cite{NIST_10}. The odd polynomials $P_{2n+1}(y)$ vanish at the edge, $y=0$, as $\sqrt{1-r^2}$ does, the form of a bounded flux density near the edge of the disk; the $N$ functions span $y$ times the polynomials of degree $N-1$ in $r^2$.
The Hankel transforms of $g_n(r)$ can be expressed in terms of
the spherical Bessel functions as shown below.
With $r\,dr=-y\,dy$, $\tilde g_n(k)=\int_0^1g_n(r)J_0(kr)\,r\,dr=\int_0^1y\,P_{2n+1}(y)\,J_0(k\sqrt{1-y^2})\,dy$. The expansion $J_0(kr)=\sum_{l=0}^\infty(4l+1)\,\kappa_l\,j_{2l}(k)\,P_{2l}(y)$ \cite[Eq.~(10.60.10)]{NIST_10}, with $\kappa_l=(2l)!/[2^{2l}(l!)^2]$ ($\kappa_0=1$, $\kappa_l=\kappa_{l-1}(2l-1)/(2l)$) and $j_l(k)$ the spherical Bessel functions, and the orthogonality $\int_0^1P_{n}(y)P_{m}(y)\,dy=\delta_{nm}/(2n+1)$ give
\begin{align}
\int_0^1P_{2l}(y)\,J_0(k\sqrt{1-y^2})\,dy=\kappa_l\,j_{2l}(k),
\label{eq:D_PJ}
\end{align}
and the recurrence $y\,P_{2n+1}(y)=[2(n+1)P_{2(n+1)}(y)+(2n+1)P_{2n}(y)]/(4n+3)$ then gives
\begin{align}
\tilde g_n(k)&=\frac{2(n+1)\,\kappa_{n+1}\,j_{2(n+1)}(k)+(2n+1)\,\kappa_n\,j_{2n}(k)}{4n+3}\\
&= (2n+1)\kappa_n\,j_{2n+1}(k)/k ,
\label{eq:D_gt}
\end{align}
where $\kappa_{n+1}=\kappa_n (2n+1)/[2(n+1)]$ and the recurrence $j_{2n}(k)+j_{2(n+1)}(k)=(4n+3)j_{2n+1}(k)/k$ are used.
Multiplication of the integral equation by $g_m(r)$ and integration over the disk
($\int_0^\infty \cdots \,r\,dr$, with $g_m(r)=0$ for $r>1$)
gives an $N\times N$ linear system, in which the term with $u$ is evaluated as,
\begin{align}
\int_0^1 r\, dr\, g_m(r)\int_0^\infty\frac{\tilde q(k)\,J_0(kr)\,k\,dk}{\sqrt{k^2+x}}&=
\int_0^\infty\frac{\tilde q(k)\,k\,dk}{\sqrt{k^2+x}}\int_0^1 r\, dr\, g_m(r)\,J_0(kr)\\
&=\int_0^\infty\frac{\tilde q(k)\tilde g_m(k)\,k\,dk}{\sqrt{k^2+x}}.
\end{align}
Using $\tilde q(k)=\sum_{n=0}^{N-1}b_n\,\tilde g_n(k)$ and $\int_0^1g_m(r)\,r\,dr=\int_0^1y\,P_{2m+1}(y)\,dy=\delta_{m0}/3$
(from the orthogonal relation and $P_{1}(y)=y$), we obtain,
\begin{align}
\sum_{n=0}^{N-1}\left[\frac{\Gamma_{mn}}{\mathrm{Da}}+W_{mn}(x)\right]b_n=\frac13\delta_{m0},
\qquad
\Gamma_{mn}=\int_0^1g_m(r) g_n(r)\,r\,dr,
\quad
W_{mn}(x)=\int_0^\infty\frac{\tilde g_m(k)\,\tilde g_n(k)\,k\,dk}{\sqrt{k^2+x}},
\label{eq:D_psi}
\end{align}
and the current is $\psi=\pi b_0/6$ using Eqs. (\ref{eq:psi}) and (\ref{eq:D_basis}) with the same relation. The overlap integrals are $\Gamma_{mn}=\int_0^1P_{2m+1}(y)P_{2n+1}(y)\,y\,dy$ (with $r\,dr=-y\,dy$). With the recurrence for $y\,P_{2n+1}(y)$ above they reduce to integrals of Legendre polynomials of opposite parity, which are given by Rayleigh's formula \cite{Rayleigh_1870} (see also \cite[Eq.~7.221.1]{GR_1994}), $\int_0^1P_{2m+1}(y)P_{2p}(y)\,dy=(-1)^{m+p}(2m+1)\kappa_m\kappa_p/[(2m+1-2p)(2m+2p+2)]$. The result is
\begin{align}
\Gamma_{mn}=\frac{(-1)^{m+n+1}(2m+1)(2n+1)\,\kappa_m\kappa_n}{2(m+n+1)(m+n+2)\,[4(m-n)^2-1]},
\label{eq:D_Gamma}
\end{align}
for example $\Gamma_{nn}=(2n+1)\kappa_n^2/[4(n+1)]$.
The integrals $W_{mn}$ over $k$ converge rapidly (the integrand decays as $k^{-4}$) and are evaluated as follows. 

For $x=0$ (steady state), $W_{mn}(0)=\int_0^\infty\tilde g_m\tilde g_n\,dk$ is available in closed form. With Eq.~\eqref{eq:D_gt}, $W_{mn}(0)=(2m+1)(2n+1)\kappa_m\kappa_n\int_0^\infty j_{2m+1}(k)\,j_{2n+1}(k)\,k^{-2}\,dk$, and with $j_l(k)=(\pi/2k)^{1/2}J_{l+1/2}(k)$ the Weber--Schafheitlin integral \cite[Sec.~13.4]{Watson_44} gives $\int_0^\infty j_{2m+1}j_{2n+1}\,k^{-2}\,dk=(\pi/8)\,\Gamma(m+n+\tfrac12)/[\Gamma(m+n+\tfrac72)\,\Gamma(m-n+2)\,\Gamma(n-m+2)]$, which vanishes for $|m-n|\ge2$ because $1/\Gamma(2-|m-n|)=0$. The matrix $W(0)$ is therefore tridiagonal,
\begin{align}
W_{mn}(0)=\frac{\pi\,(2m+1)(2n+1)\,\kappa_m\kappa_n}{(2m+2n+1)(2m+2n+3)(2m+2n+5)}\times
\begin{cases}1, & m=n,\\ 1/2, & |m-n|=1,\\ 0, & |m-n|\ge2.\end{cases}
\label{eq:D_W0}
\end{align}

For $x\neq0$ we write $W_{mn}(x)=W_{mn}(0)+\int_0^\infty\tilde g_m\tilde g_n\,[k/\sqrt{k^2+x}-1]\,dk$, whose integrand decays as $k^{-6}$, so that the neglected tail is negligible; the integral is computed by composite 16-point Gauss--Legendre quadrature on $0\le k\le K$ with $K=\max(300,\,40|\lambda|+100)$, $\lambda=\sqrt{x}$; the panel edges are $0,e_0,1.6e_0,1.6^2e_0,\ldots$ with $e_0=\max(|\lambda|,1)/8$ up to the last edge below 1, followed by the integers $1,2,\ldots,K$ (for $|\lambda|\ge8$ the panels are the unit intervals). The linear system is solved directly. We used $N=1600$ for Table~\ref{tab:ss}, and $N=200$ and 61 logarithmically spaced frequencies between $\omega a^2/D=10^{-2}$ and $10^3$ for Figs.~\ref{fig:Gerr} and \ref{fig:fd}.

\textbf{D.3.3 Convergence.} The solution is converged in the number of basis functions. For $\mathrm{Da}=1$, 6, 20, and 100 and the 61 frequencies of Figs.~\ref{fig:Gerr} and \ref{fig:fd}, $\psi$ changes by at most $1.9\times10^{-5}$ (relative) when $N$ is increased from 100 to 200, and the change is reduced by about a factor of four for each doubling of $N$ (for example, it is $4\times10^{-6}$ to $5\times10^{-6}$ from $N=200$ to 400 for $\mathrm{Da}=100$ at $\Omega=12$ and $10^3$). The steady-state current $\chi=\psi(\mathrm{Da},0)$ changes by at most $2\times10^{-5}$ from $N=100$ to 400 and by less than $2\times10^{-6}$ from $N=400$ to 1600 for $\mathrm{Da}\le10^3$. The results of Table~\ref{tab:ss} ($N=1600$) and Figs.~\ref{fig:Gerr} and \ref{fig:fd} ($N=200$) are therefore exact to the digits given.

\begin{figure*}[t]
\centering
\includegraphics[width=0.85\textwidth]{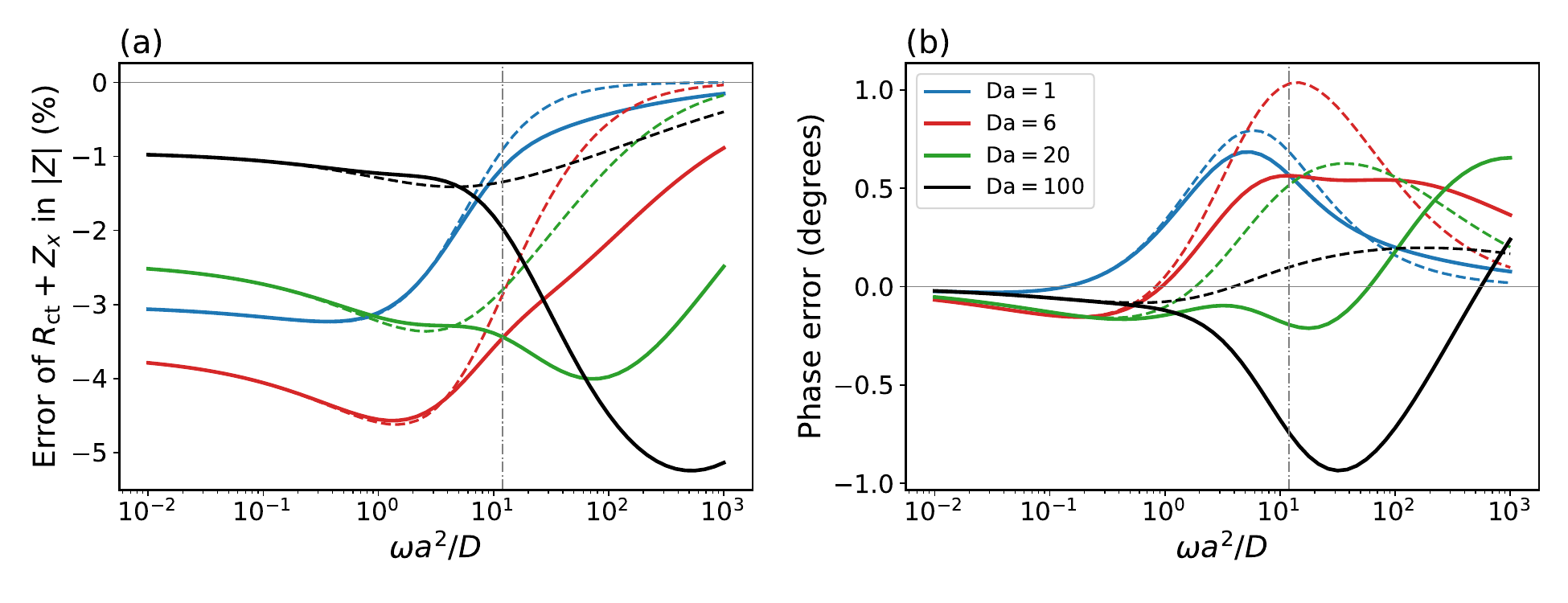}
\caption{Error of the Randles sum $R_{\rm ct}+Z_x$ relative to the coupled impedance $Z_{\rm F}$, Eq.~\eqref{eq:ZF}, as a function of the dimensionless frequency $\omega a^2/D$. Solid curves: Pad\'e $Z_{\rm W}$, Eq.~\eqref{eq:Z}; dashed curves: converged $Z_\infty=R_\infty/\hat\phi(i\Omega)$ (Appendix~C). (a) Modulus $|Z|$. (b) Phase. The dash-dotted line marks $\omega a^2/D=12$. $\mathrm{Da}=1$, 6, 20, and 100 correspond to $k^0=6.3\times10^{-5}$, $3.8\times10^{-4}$, $1.3\times10^{-3}$, and $6.3\times10^{-3}~{\rm m~s^{-1}}$ for $a=5~\mu$m and $D=6.33\times10^{-10}~{\rm m^2~s^{-1}}$.}
\label{fig:fd}
\end{figure*}

\subsection*{D.4. Unequal diffusion coefficients}

\textbf{D.4.1 What is tested.} Equation~\eqref{eq:zw_reversible_pade} adds the responses of the two species as if each of them had a uniform surface concentration on the disk. For $D_{\rm O}=D_{\rm R}$ this is exact (D.1). For $D_{\rm O}\neq D_{\rm R}$ it is an approximation, which we test against the coupled solution.

\textbf{D.4.2 Coupled solution.} In the coupled problem the same flux density $q(r)$ enters the electrode as oxidized species and leaves it as reduced species (flux continuity). Each species obeys its own diffusion equation, with $x_j=i\omega a^2/D_j$ ($j={\rm O},{\rm R}$). The Nernst condition requires that the combination $\delta C_{\rm O}(r,0)/C_{\rm O}^*-\delta C_{\rm R}(r,0)/C_{\rm R}^*$ of the surface concentration deviations has the same value, $(nF/RT)\,\delta E$, everywhere on the disk, because the electrode potential is uniform. As in D.3.2, the surface concentrations follow from the flux density through the half-space response of each species, Eq.~\eqref{eq:D_u} with $x\to x_j$, where $\tilde q(k)=\int_0^1q(r)J_0(kr)\,r\,dr$ is the Hankel transform of $q$ and lengths are in units of $a$; the factor $a/D_j$ converts the flux density into a concentration. For the oxidized species, which is depleted at the electrode, $\delta C_{\rm O}(r,0)=-(a/D_{\rm O})\int_0^\infty\tilde q(k)J_0(kr)\,k\,dk/\sqrt{k^2+x_{\rm O}}$, and for the reduced species, which is produced there, $\delta C_{\rm R}(r,0)=+(a/D_{\rm R})\int_0^\infty\tilde q(k)J_0(kr)\,k\,dk/\sqrt{k^2+x_{\rm R}}$. The Nernst condition thus becomes an integral equation for $q$ alone,
\begin{align}
\sum_{j={\rm O},{\rm R}}\frac{a}{D_jC_j^*}\int_0^\infty\frac{\tilde q(k)\,J_0(kr)\,k\,dk}{\sqrt{k^2+x_j}}=-\frac{nF}{RT}\,\delta E,
\qquad 0\le r\le1 .
\label{eq:D_ie}
\end{align}

We expand the flux density in $N$ basis functions, $q(r)=-[nF\,\delta E/(aRT)]\sum_nb_n\,h_n(r)$ with
\begin{align*}
h_n(r)=\frac{1}{\kappa_n}\,\frac{P_{2n}(y)}{y},
\qquad y=\sqrt{1-r^2},
\qquad \kappa_n=\frac{(2n)!}{2^{2n}(n!)^2},
\end{align*}
where $P_{2n}(y)$ are the Legendre polynomials \cite{NIST_10}. The functions $h_n$, built from the even Legendre polynomials, are the natural basis for a uniform surface concentration, because the flux density at a disk with uniform surface concentration diverges as $1/\sqrt{1-r^2}$ at the edge. Their Hankel transforms are the spherical Bessel functions $j_{2n}(k)$: with $r\,dr=-y\,dy$ and Eq.~\eqref{eq:D_PJ}, $\tilde h_m(k)=\int_0^1h_m(r)J_0(kr)\,r\,dr=\kappa_m^{-1}\int_0^1P_{2m}(y)\,J_0(k\sqrt{1-y^2})\,dy=j_{2m}(k)$. Multiplying Eq.~\eqref{eq:D_ie} by $h_m(r)\,r$ and integrating over the disk ($0\le r\le1$) gives $\mathbf M\mathbf b=\mathbf e_0$ with
\begin{align}
\mathbf M=\frac{\mathbf W(x_{\rm O})}{D_{\rm O}C_{\rm O}^*}+\frac{\mathbf W(x_{\rm R})}{D_{\rm R}C_{\rm R}^*},
\qquad
W_{mn}(x)=\int_0^\infty\frac{j_{2m}(k)\,j_{2n}(k)\,k\,dk}{\sqrt{k^2+x}},
\end{align}
where the lengths in $\mathbf W$ are in units of $a$. Because $\int_0^1h_n(r)\,r\,dr=\kappa_n^{-1}\int_0^1P_{2n}(y)\,dy=\delta_{n0}$, the anodic current $\delta I=-nF\,2\pi a^2\int_0^1q(r)\,r\,dr$ equals $2\pi n^2F^2a\,(\delta E/RT)\,b_0$ with $b_0=[\mathbf M^{-1}]_{00}$, and the impedance $Z_{\rm F}=\delta E/\delta I$ is
\begin{align}
Z=\frac{RT}{2\pi n^2F^2a}\,\frac{1}{[\mathbf M^{-1}]_{00}} .
\end{align}
Because $\int_0^\infty j_{2m}(k)j_{2n}(k)\,dk=\pi\delta_{mn}/[2(4n+1)]$, the matrix elements are evaluated as $W_{mn}(x)=\pi\delta_{mn}/[2(4n+1)]+\int_0^\infty j_{2m}(k)j_{2n}(k)\,[k/\sqrt{k^2+x}-1]\,dk$, whose integrand decays as $k^{-4}$; the integral is computed by Gauss--Legendre quadrature with the panels of D.3.2, with $K=\max(400,40|\sqrt{x}|)$. These are integrals of the same kind as those of D.3.2, with the basis functions $h_n$ instead of $g_n$. For a single species $\mathbf M$ reduces to $\mathbf W(x)/(DC)$, the impedance to $Z_j=RT/(2\pi n^2F^2aD_jC_j^*[\mathbf W(x_j)^{-1}]_{00}) = RT/(n^2F^2G_j)$ with $G_j=4aD_jC_j^*\hat\phi(x_j)$, and the response function of Appendix~C is $\hat\phi(x)=(\pi/2)[\mathbf W(x)^{-1}]_{00}$; for $x=0$, $W_{mn}=\pi\delta_{mn}/[2(4n+1)]$, so that $\hat\phi(0)=1$. Equation~\eqref{eq:zw_reversible_pade}, in contrast, corresponds to the sum $Z_{\rm O}+Z_{\rm R}$ of the two single-species impedances, with $\hat\phi$ the converged response function. The two coincide exactly for $D_{\rm O}=D_{\rm R}$.

\textbf{D.4.3 Result.} We used the converged uniform-concentration responses in Eq.~\eqref{eq:zw_reversible_pade}, so that only the closure is tested. At 16 frequencies between $\omega a^2/D_{\rm O}=0.01$ and 1000 ($N=50$; Fig.~\ref{fig:D_asym}(b)), the maximum error of $|Z|$, attained at $\omega a^2/D_{\rm O}=10$ in all cases, is 0.002, 0.017, 0.096, and 0.305\% for $D_{\rm R}/D_{\rm O}=0.9$, 0.75, 0.5, and 0.25 (with $C_{\rm R}^*=C_{\rm O}^*$), and below 0.01\% for $C_{\rm R}^*/C_{\rm O}^*=0.1$ and 10 at $D_{\rm R}/D_{\rm O}=0.75$ (Fig.~\ref{fig:D_asym}). The additional error of about 6\% that appears when the Pad\'e response is used instead comes from its high-frequency amplitude (Appendix~C), not from the closure.

\begin{figure}[h]
\centering
\includegraphics[width=0.95\linewidth]{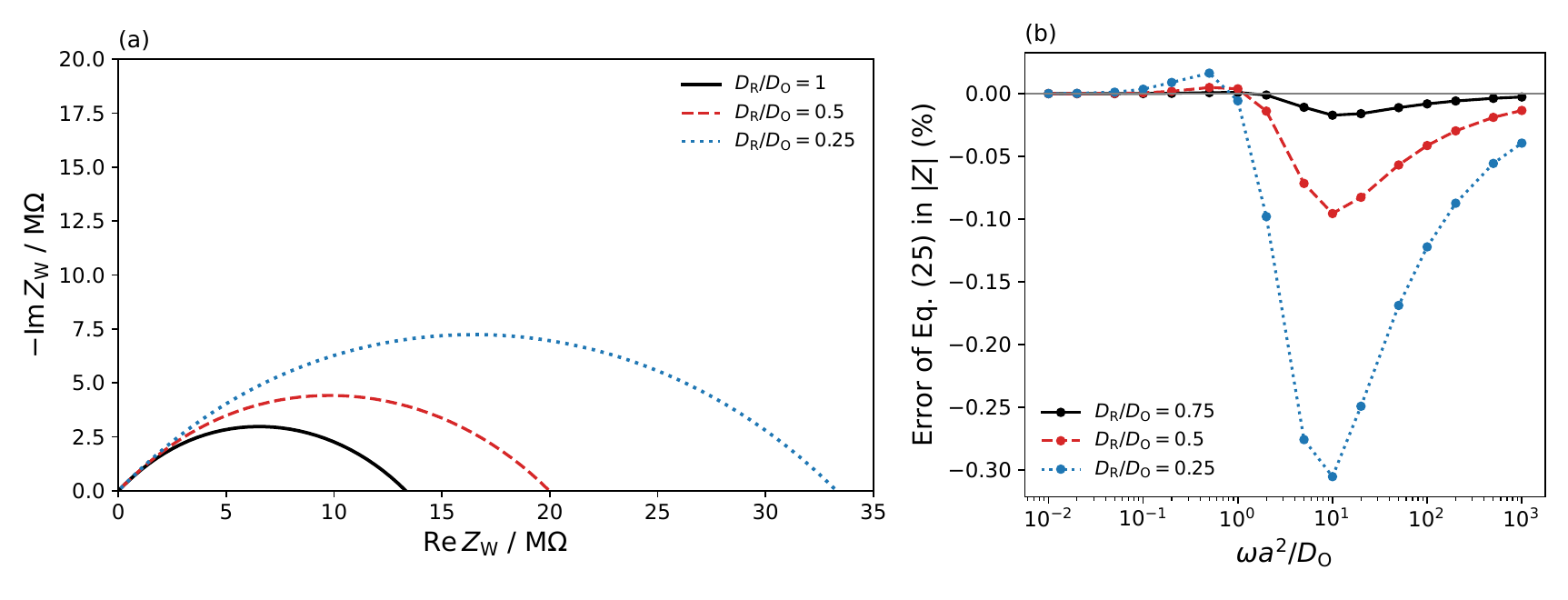}
\caption{(a) Diffusion impedance, Eq.~\eqref{eq:zw_reversible_pade}, in the complex plane for $D_{\rm R}/D_{\rm O}=1$, 0.5, and 0.25 ($a=10~\mu$m, $D_{\rm O}=10^{-9}~{\rm m^2~s^{-1}}$, $C_{\rm O}^{*}=C_{\rm R}^{*}=1~{\rm mol~m^{-3}}$). (b) Error of Eq.~\eqref{eq:zw_reversible_pade} relative to the coupled solution, with the converged uniform-concentration responses used for the closure.}
\label{fig:D_asym}
\end{figure}

\endgroup

% Acknowledgments / funding: none at present; insert here if needed.

\section*{Declaration of competing interest}
The authors declare that they have no known competing financial interests or personal relationships that could have appeared to influence the work reported in this paper.

\section*{Data availability}
The digitized data reanalyzed in Sec.~\ref{sec:Compex} were extracted from the published figure of Ref.~\cite{Gabrielli_06}; readers seeking the original experimental data should contact the corresponding author of that work. All other data supporting this study (the coefficients of the response function, the construction of the Pad\'{e} approximant, the numerical methods of Appendices~C and D, and the calculated impedance spectra) are provided within the article and its appendices, or are available from the corresponding author upon reasonable request.

\section*{Declaration of generative AI and AI-assisted technologies in the manuscript preparation process}
During the preparation of this work one of the authors used Claude Sonnet 5.5 and Claude Opus 5.5 (Anthropic) in order to carry out and check numerical calculations and the associated code, to check derivations, and to revise the text. After using this tool, the authors reviewed and edited the content as needed and take full responsibility for the content of the published article.

\printcredits

%\bibliographystyle{cas-model2-names}
\bibliographystyle{elsarticle-num}

%\bibliography{IS_Saito_v2}
  %your .bib file



\end{document}